\documentclass[aps,prl,twocolumn,superscriptaddress]{revtex4-2}
\usepackage{amssymb}
\usepackage{booktabs}
\usepackage{physics}
\usepackage{graphicx}
\usepackage{amsmath}
\usepackage{amsthm}
\usepackage{amsfonts}
\usepackage[T1]{fontenc}
\usepackage{tikz}
\usepackage{array}
\usepackage{multirow}
\usepackage{color}
\usepackage{esint}
\usepackage{bm}
\usepackage{color}
\definecolor{LinkColor}{rgb}{0.75,0.0,0.2}
\usepackage{bbm}
\usepackage{hyperref}
\hypersetup{
	pdfauthor={good guys},
	pdftitle={good title},
	colorlinks=true,
	citecolor=LinkColor,
	linkcolor=LinkColor,
	urlcolor=LinkColor,
}
\usepackage{babel}
\usepackage{titlesec}

\begin{document}
\title{Symmetry Breaking Dynamics in Quantum Many-Body Systems}
	
\author{Hui Yu}
\affiliation{Beijing National Laboratory for Condensed Matter Physics and Institute of Physics, Chinese Academy of Sciences, Beijing 100910, China}

\author{Zi-Xiang Li}
\email{zixiangli@iphy.ac.cn}

\affiliation{Beijing National Laboratory for Condensed Matter Physics and Institute of Physics, Chinese Academy of Sciences, Beijing 100910, China}
\affiliation{University of Chinese Academy of Sciences, Beijing 100049, China}

\author{Shi-Xin Zhang}
\email{shixinzhang@iphy.ac.cn}

\affiliation{Beijing National Laboratory for Condensed Matter Physics and Institute of Physics, Chinese Academy of Sciences, Beijing 100910, China}

\date{\today}

\begin{abstract}
Entanglement asymmetry has emerged as a powerful tool for characterizing symmetry breaking in quantum many-body systems. In this Letter, we explore how symmetry is dynamically broken through the lens of entanglement asymmetry in two distinct scenarios: a non-symmetric Hamiltonian quench and a non-symmetric random quantum circuit, with a particular focus on U(1) symmetry. In the former case, symmetry remains broken in the subsystem at late times, whereas in the latter case, the symmetry is initially broken and subsequently restored, consistent with the principles of quantum thermalization. Notably, the growth of entanglement asymmetry exhibits unexpected overshooting behavior at early times in both contexts, contrasting with the behavior of charge variance. We also consider dynamics of non-symmetric initial states under the symmetry-breaking evolution. Due to the competition of symmetry-breaking in both the initial state and Hamiltonian, the early-time entanglement asymmetry can increase and decrease, while quantum Mpemba effects remain evident despite the weak symmetry-breaking in both settings.
\end{abstract}

\maketitle

\textit{Introduction.---} 
Symmetry breaking is a ubiquitous phenomenon across all branches of physics. A well-known example is the Higgs mechanism \cite{bernstein1974spontaneous} in particle physics, where the vacuum state of the universe causes different particles to acquire mass, spontaneously breaking the electroweak symmetry. This type of symmetry breaking, which occurs without external influences, is referred to as spontaneous symmetry breaking. In contrast, a symmetry can also be explicitly broken when the Hamiltonian describing the system directly breaks the symmetry. How symmetry breaks dynamically in this case is an interesting fundamental question to explore.

Symmetry properties are also closely related to the concept of quantum thermalization \cite{deutsch1991quantum,srednicki1994chaos,d2016quantum,rigol2012thermalization,deutsch2018eigenstate} for generic quantum many-body systems.
In general, when a closed quantum system evolves with a chaotic Hamiltonian, the reduced density matrix of a small subsystem $a$ thermalizes to the equilibrium finite-temperature state: $\rho_{a} \propto e^{-\beta \hat{H}_{a}}$ where $\hat{H}_{a}$ is the Hamiltonian of the subsystem and $\beta$ is a Lagrangian multiplier determined by the initial energy density. Symmetry is restored at later times for symmetric Hamiltonian $\hat{H}_{a}$, since $[\hat{Q}_{a}, \rho_{a}]=0$ where $\hat{Q}_a$ represents the corresponding symmetry generator. However, if $\hat{H}_{a}$ does not respect the symmetry, the reduced density matrix $\rho_{a}$ at late times is non-commuting with $\hat{Q}_{a}$. In this case, symmetry breaking persists even if the system begins in a symmetric state. It is important to distinguish this explicit breaking from spontaneous symmetry breaking (SSB). In the thermodynamic limit, SSB enables a low-temperature symmetric initial state to evolve dynamically into a symmetry-broken steady state. In contrast, for finite-size systems, SSB cannot occur, and the equilibrium state must retain the symmetry of the Hamiltonian. 

Apart from the richness of the late-time behavior, early-time dynamics have also garnered significant attention. For example, the Mpemba effect \cite{mpemba1969cool}, which claims that hot water freezes faster than cold water, has been widely explored in both classical and quantum contexts \cite{lu2017nonequilibrium,lasanta2017hotter,kumar2020exponentially,klich2019mpemba,teza2023relaxation,bechhoefer2021fresh,malhotra2024double,kumar2022anomalous,manikandan2021equidistant,chatterjee2023quantum,aharony2024inverse,wang2024mpemba,nava2019lindblad,carollo2021exponentially,chatterjee2024multiple,ivander2023hyperacceleration}. Recently, quantum Mpemba effect (QME) is reported in integrable systems and chaotic systems \cite{ares2023entanglement,liu2024symmetry,turkeshi2024quantum}. QME refers to the phenomenon where, during the relaxation toward a steady state value, the time evolution curves of a physical quantity for different initial conditions cross each other. For instance, U(1)-symmetry restoration occurs more rapidly for more asymmetric initial states under the U(1)-symmetric Hamiltonian quench \cite{fagotti2014relaxation,essler2016quench,doyon2020lecture,fagotti2014conservation,bertini2015pre,vidmar2016generalized,alba2021generalized,calabrese2016quantum,polkovnikov83nonequilibrium,bastianello2022introduction,ares2023lack}. This finding was subsequently explored in various other settings \cite{klobas2024translation,rylands2024microscopic,yamashika2024entanglement,khor2024confinement,rylands2024dynamical,ares2024entanglement,murciano2024entanglement,liu2024quantum,chang2024imaginary,russotto2024non} and experimentally realized on a trapped-ion quantum simulator \cite{joshi2024observing}. 

Previous studies \cite{ares2023entanglement,liu2024symmetry} have primarily focused on characterizing symmetry restoration when an asymmetric initial state evolves under a symmetric Hamiltonian or random circuit. In contrast, this Letter examines the dynamical aspects of symmetry breaking, exploring the behavior of symmetric and asymmetric initial states under non-symmetric evolution \cite{rossi2023long}. In addition, due to experimental limitations, symmetric evolutions are often affected by noises and defects, resulting in non-symmetric contributions as well. In such cases, can symmetry restoration still occur, or does symmetry breaking become more pronounced over time? Additionally, how does QME behave in the presence of symmetry-breaking interactions? Addressing these questions offers a more comprehensive understanding of symmetry and symmetry breaking in quantum many-body systems.

In this Letter, we investigate and compare the dynamics of symmetry breaking with two distinct models: a non-symmetric random circuit \cite{fisher2023random} and a non-symmetric  Hamiltonian evolution, each with different symmetric and asymmetric initial states. To characterize the extent of symmetry breaking in subsystem $a$, we employ the metric of entanglement asymmetry (EA) \cite{ares2023entanglement}, which has been extensively utilized as a measure of symmetry breaking in quantum field theories \cite{capizzi2024universal,chen2024renyi,capizzi2023entanglement} and out-of-equilibrium many-body systems \cite{rylands2024microscopic,khor2024confinement,ares2025entanglementasymmetrydynamicsrandom}. EA is defined as
\begin{eqnarray}
    \Delta S_{a}  = S(\rho_{a, Q})   - S(\rho_{a}).
\end{eqnarray}
Here, $S(\rho_{a})$ denotes the standard Von Neumann entropy of subsystem $a$, and $\rho_{a, Q} = \sum_{q \in \mathbb{Z}} \Pi_{q} \rho_{a} \Pi_{q}$ where $\hat{Q}_{a} = \sum_{i \in a} \sigma_{i}^{z}$ in case of U(1) symmetry and $\Pi_{q}$ is the projector onto eigenspace of $\hat{Q}_{a}$ with charge $q$. Consequently, $\rho_{a, Q}$ is block diagonal in the eigenbasis of $\hat{Q}_{a}$. The EA satisfies two key properties: (1) $\Delta S_{a} \geq 0$ since the EA is the relative entropy between $\rho_{a, Q}$ and $\rho_{a}$. (2) $\Delta S_{a} = 0$ if and only if $\rho_{a, Q}=\rho_{a}$. In random circuit settings, $\mathbb{E} [\Delta S_{a}]$ is employed as the circuit-averaged value of $\Delta S_{a}$. Note that the symmetry for subsystem mixed states investigated here corresponds to the weak symmetry in Refs. \cite{lessa2024strong,sala2024spontaneous}. In parallel with the analysis of EA, we also compute the charge variance (CV) $\sigma^{2}_{Q}= \langle \hat{Q}^{2} \rangle - \langle \hat{Q} \rangle^{2}$, where $\hat{Q} = \sum_{i=1}^{L} \sigma_{i}^{z}$. This quantity serves as a measure of charge fluctuations within the system, offering a complementary perspective on symmetry breaking.

\begin{figure}
\centering
\includegraphics[width=0.96\linewidth, keepaspectratio]{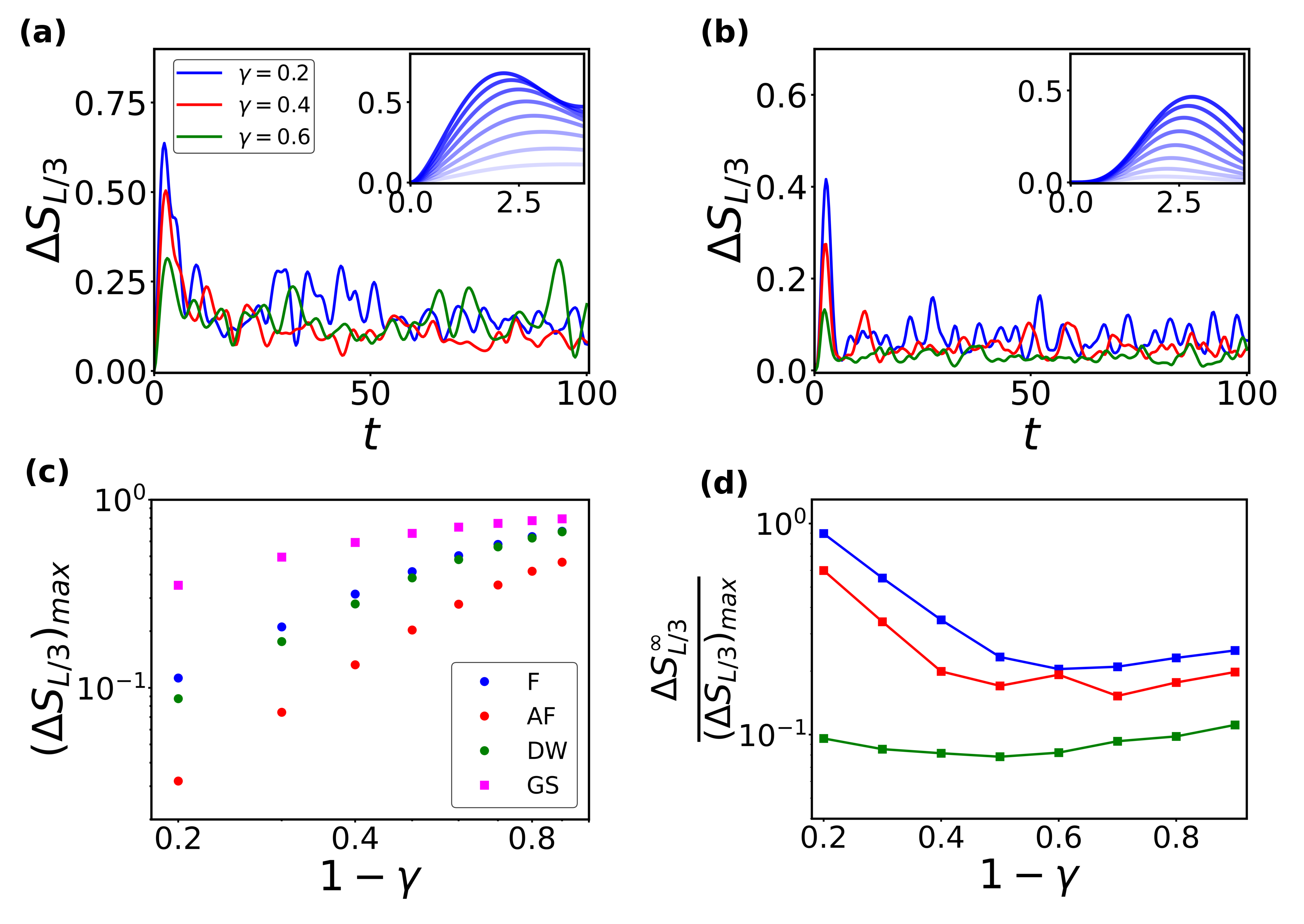}
\caption{EA as a function of time with (a) ferromagnetic and (b) antiferromagnetic states for different values of $\gamma$ under $H_{1}$. The insets show the peak of EA at different values of $\gamma$. From bottom to top: $\gamma=0.8,0.7,0.6,0.5,0.4,0.3,0.2,0.1$. Panels (c) and (d) show the peak value of EA, $(\Delta S_{L/3})_{max}$, and the ratio of the late-time EA, $\Delta S_{L/3}^{\infty}$, to $(\Delta S_{L/3})_{max}$ as a function of  $1-\gamma$ for various initial states under $H_{1}$. GS denotes the value of EA calculated from the ground state of $H_{1}$.}
\label{fig:Ham_one}
\end{figure}

For a non-symmetric Hamiltonian evolution, we find that U(1) symmetry can not be restored in a subsystem which can be explained by the late-time reduced density matrix relaxing to the form $e^{-\beta \hat{H}_{a}}$, where $\hat{H}_{a}$ explicitly includes symmetry-breaking terms. In this scenario, the EA shows nontrivial overshooting at early times, characterized by a peak in EA that significantly exceeds its late-time saturation value. This behavior contrasts with other symmetry-breaking measures like charge variance and mirrors the thermal overshooting of the classical Mpemba effect \cite{klich2019mpemba}, where systems transiently exceed their equilibrium temperature. Furthermore, the QME originated from symmetric evolution disappears when the strength of symmetry breaking in the evolution exceeds some thresholds.

In the case of non-symmetric random circuits, we show that U(1) symmetry for a small subsystem can still be restored regardless of the initial states. As a result, EA also exhibits overshooting at early times. Additionally, QME appears at early times unless all U(1)-symmetric gates are replaced by random Haar gates, where EA dynamics are the same for different U(1)-asymmetric initial states.

\begin{figure}
\centering
\includegraphics[width=0.47\textwidth, keepaspectratio]{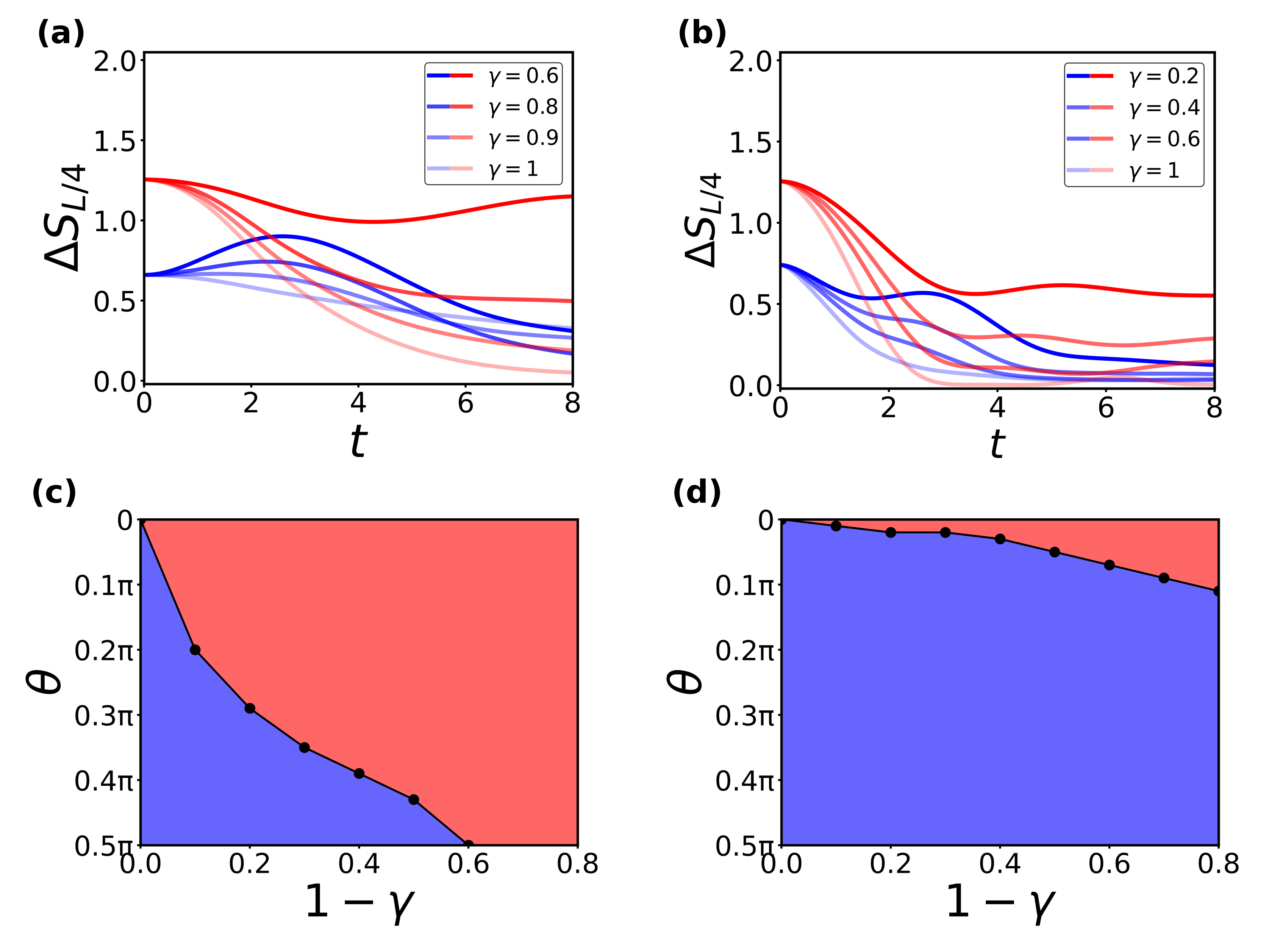}
\caption{EA dynamics for (a) tilted ferromagnetic states and (b) tilted antiferromagnetic states with varying $\gamma$. The blue curves correspond to $\theta=0.2\pi$, and the red curves represent $\theta=0.5\pi$. Panels (c) and (d) depict the dependence of early-time EA dynamics on $\theta$ and $1-\gamma$ for ferromagnetic and antiferromagnetic states, respectively. When the parameter is in the red region, EA can exceed the initial value, while in the blue region, EA firstly decreases and never grows higher than the initial value. All black dots are obtained through numerical simulation. All calculations are based on $H_{1}$.}
\label{fig:Ham_three}
\end{figure}

\textit{Setup.---} To study dynamics in these systems, we consider three initial states: the ferromagnetic state $\vert 000...0 \rangle$, the antiferromagnetic state $\vert 0101..1 \rangle$ and the domain-wall state $\vert 000..111 \rangle$, where the domain wall is positioned at the center of the chain. To incorporate the effect of symmetry breaking in the initial state, we introduce tilted ferromagnetic states \cite{ares2023entanglement,liu2024symmetry}, defined as 
\begin{eqnarray}
    \vert \psi_{i} (\theta)\rangle =  e^{-i\frac{\theta}{2} \sum_{j} \sigma_{j}^{y}} \vert 000...0 \rangle
    \label{eq: tilted ferromagnetic}
\end{eqnarray}
where $\sigma_{j}^{y}$ is the Pauli-$y$ matrix on the $j$-th qubit, and $\theta$ is a tuning parameter controlling the strength of symmetry breaking in the initial state. When $\theta=0$, Eq.~\eqref{eq: tilted ferromagnetic} is U(1)-symmetric with zero EA. As $\theta$ increases, the EA grows, reaching its maximum value at $\theta=\pi/2$. The tilted antiferromagnetic and tilted domain wall states are constructed in a similar manner. 

The random circuit architecture consists of two-qubit U(1)-symmetric gates and Haar-random gates arranged in a brick-wall pattern. The U(1)-symmetric gates have a block-diagonal matrix structure, with each block independently sampled from the Haar measure \cite{li2023d,hearth2023unitary,li2023designs}. The effect of symmetry breaking depends on the density (doped probability) of random Haar gates without U(1) symmetry, denoted as $P_{\text{Haar}}$. The time unit in the circuit is defined by the application of two consecutive layers of gates. 
$\mathbb{E} [\Delta S_{a}]$ is computed by averaging over $5000$ circuit configurations.

We also investigate Hamiltonian dynamics where the state $\vert \psi_{i} (\theta) \rangle$ undergoes unitary evolution by $e^{-iHt}\vert \psi_{i} (\theta) \rangle$, and the Hamiltonian is 
\begin{eqnarray}
H = & -\frac{1}{4} \sum_{j=1}^{L} \Big[ \sigma_j^x \sigma_{j+1}^x + \gamma \sigma_j^y \sigma_{j+1}^y + \Delta_{1} \sigma_j^z \sigma_{j+1}^z \Big]  \label{eq:Ham} \\
& -\Delta_2 \sum_{j=1}^{L} \Big[ \sigma_j^x \sigma_{j+2}^x + \sigma_j^y \sigma_{j+2}^y + \sigma_j^z \sigma_{j+2}^z \Big] \notag.
\end{eqnarray}
Here, $L$ denotes the total system size, while $\Delta_{1}$ and $\Delta_{2}$ represent the coefficients for nearest-neighbor and next-nearest-neighbor interactions, respectively. $\Delta_{2}$ introduces non-integrability, and $\gamma$ controls the strength of symmetry breaking. Periodic boundary conditions are imposed in both contexts.

\textit{U(1)-Symmetric Initial States with U(1) Non-Symmetric Hamiltonian.---} All numerical simulations are performed using the {\sf TensorCircuit-NG} package \cite{zhang2023tensorcircuit}. Here, we investigate the dynamics of symmetry breaking under an integrable Hamiltonian $H_{1}$ with $\Delta_{1}=0.4$ and $\Delta_{2}=0$, and a non-integrable Hamiltonian $H_{2}$ with $\Delta_{1}=0.4$ and $\Delta_{2}=0.05$, with system size $L=12$ sites. As revealed in Fig.~\ref{fig:Ham_one}(a) and (b), EA for various different Hamiltonian symmetry-breaking $\gamma$ exhibit peaks at early times that are much larger than steady values. Furthermore, the peak value of the EA, $(\Delta S_{L/3})_{max}$, is found to be correlated with the strength of symmetry breaking, $1-\gamma$, for different symmetric initial states as shown in Fig.~\ref{fig:Ham_one}(c) where EA of the ground state of $H_{1}$ follows the same trend. Notably, the peak heights nearly coincide between the ferromagnetic and domain wall states, as the early-time peak primarily depends on the local configurations of the initial state. Moreover, the finite-size scaling analysis demonstrates the existence of this peak in the thermodynamic limit (see SM).

\begin{table}[ht]
\centering
\resizebox{0.49\textwidth}{!}{ 
\begin{tabular}{cccc} 
\toprule
& \textbf{Ferromagnetic} & \textbf{Domain Wall} & \textbf{Antiferromagnetic} \\
\midrule
$EA\ (early\ time)$ & \parbox{3cm}{crossing for small $1-\gamma$} & \parbox{3cm}{crossing for small $1-\gamma$} &  \parbox{3cm}{crossing for small $1-\gamma$} \\
\hline
$CV\ (early\ time)$ & \parbox{3cm}{crossing for  $\gamma \neq 1$} & no crossing & no crossing \\
\hline
$EA\ (late\ time)$ & \tikz[baseline]{\draw[->, line width=0.35mm] (-0.1,-0.1) -- (0.2,0.2);} & \tikz[baseline]{\draw[->, line width=0.35mm] (-0.1,-0.1) -- (0.2,0.2);} & \tikz[baseline]{\draw[->, line width=0.35mm] (-0.1,-0.1) -- (0.2,0.2);} \\
\hline
$CV\ (late\ time)$ & \tikz[baseline]{\draw[->, line width=0.35mm] (-0.2,0.2) -- (0.1,-0.1);}  & \tikz[baseline]{\draw[->, line width=0.35mm] (-0.1,-0.1) -- (0.2,0.2);} & \tikz[baseline]{\draw[->, line width=0.35mm] (-0.1,-0.1) -- (0.2,0.2);} \\
\bottomrule
\end{tabular}
}
\caption{The early- and late-time behavior of EA and CV under the evolution of $H_{1}$ or $H_{2}$ ($0.5 \leq \gamma \leq 1$). Crossing in EA (CV) means when the time-evolution curves of EA (CV) for states with larger $\theta$ intersect with those for smaller $\theta$ at early-times. The right-up (right-down) arrow indicates that the late-time value is increasing (decreasing) with increasing tilted angle $\theta$.}
\label{Table:EA and CV}
\end{table}

By analyzing Fig.~\ref{fig:Ham_one}, we identify that the late-time EA $\Delta S_{L/3}^{\infty}$ oscillates and does not approach zero. This is because the reduced density matrix of subsystem $a$ evolves towards a canonical ensemble $e^{-\beta \hat{H_{a}}}$, where $\hat{H_{a}}$ has the same form as $\hat{H}$ in Eq.~\eqref{eq:Ham}, but acts solely on subsystem $a$. Since $\hat{H_{a}}$ includes symmetry breaking terms, $[\rho_{a},\hat{Q_{a}}]\neq0$, leading to a non-vanishing EA at long times. In Fig.~\ref{fig:Ham_one}(d), we calculate the ratio of $\Delta S_{L/3}^{\infty}$ to $(\Delta S_{L/3})_{max}$ with varying $\gamma$. The late-time EA, $\Delta S_{L/3}^{\infty}$, is obtained by averaging $\Delta S_{L/3}$ over $2000$ time points between $t_{1}=2000$ and $t_{2}=40000$. The results further confirm the overshooting behavior as the late-time saturating EA value is much lower than the early-time peak value. This phenomenon stems from the competition between symmetry breaking and subsystem decoherence. Initially, the non-symmetric Hamiltonian dynamically generates asymmetry, causing EA to grow as symmetry breaking predominates over decoherence. Later, symmetry breaking saturates while decoherence becomes the sole governing factor for EA evolution, driving EA toward a steady-state value. The early-time peak emerges from the intricate interplay between these competing effects. On the contrary, the CV dynamics in this setting shows no evident overshooting pattern but instead directly grows to the saturating values.

\begin{figure*}
\begin{center}
\includegraphics[width=0.88\linewidth]{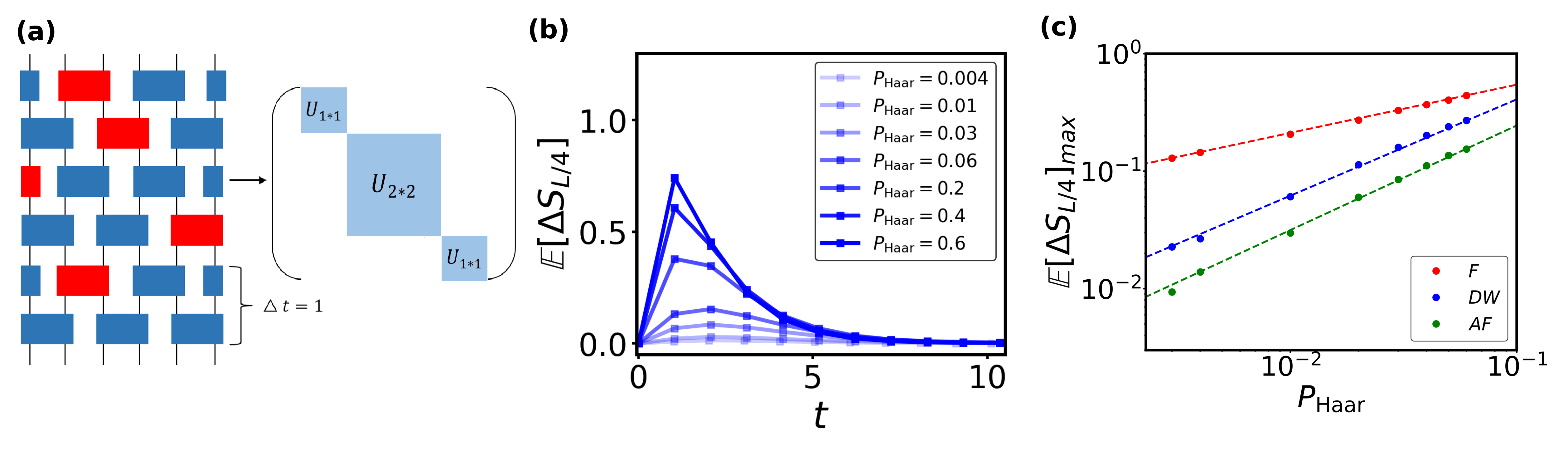}
\caption{(a) Schematic illustration of a non-symmetric random circuit with 6 qubits. Gates are arranged in the even-odd brick-wall pattern. The blue and red rectangles represent U(1)-symmetric and random Haar gates, respectively. The basis for the U(1)-symmetric gate is listed in the following order: $\vert 00 \rangle$, $\vert 01 \rangle$, $\vert 10 \rangle$ and $\vert 11 \rangle$. (b) The circuit-averaged EA, $\mathbb{E} [\Delta S_{L/4}]$, as a function of time with the antiferromagnetic initial state at different values of $P_{\text{Haar}}$. (c) The peak value, $\mathbb{E} [\Delta S_{L/4}]_{max}$, as a function of $P_{\text{Haar}}$. All three curves follow a power law $y=ax^{b}$. F: Ferromagnetic state ($a=1.4$, $b=0.4$); DW: Domain Wall state ($a=2.7$, $b=0.8$); AF: Antiferromagnetic state ($a=1.9$, $b=0.9$).}
\label{fig:circuit_one}
\end{center}
\end{figure*}

\textit{U(1)-Asymmetric Initial States with U(1) Non-Symmetric Hamiltonian.---}
Next we investigate dynamics with U(1)-asymmetric initial states under $H_{1}$ where EA dynamics depends on both symmetry-breaking parameters, $\theta$ and $\gamma$. $\theta$ describes the symmetry breaking in the initial state while $\gamma$ describes the symmetry breaking in the Hamiltonian. The interplay between these two parameters results in distinct behaviors in the EA dynamics. This is illustrated in the schematic figures with varying $\theta$ and $\gamma$ in Fig.~\ref{fig:Ham_three}(c) and (d). The colors highlight the tendency in EA at early times. Blue regime indicates that $\Delta S_{L/4}(t)$ never exceeds its initial value for early times, while red regime corresponds to the situations where EA can grow larger than its initial value at early times. It is clearly reflected in Fig.~\ref{fig:Ham_three}(a), the initial growth of EA at $\theta=0.2\pi$ and $\gamma=0.8$, $0.6$ aligns with the red region. For a fixed $\gamma$, EA grows with weaker asymmetric effects (small $\theta$) in the initial states or for a fixed $\theta$, EA increases with stronger symmetry breaking effects (large $1-\gamma$) in the Hamiltonian. Consequently, the early-time behavior of EA serves as a witness to compare the symmetry-breaking strength hosting in the quantum state and the Hamiltonian. 

Another key feature of the early-time dynamics is the emergence of QME, as shown in Fig.~\ref{fig:Ham_three}(a) for the symmetric case $\gamma=1$. The origin of this QME lies in the relative small $ZZ$ term and gapless nature in the Hamiltonian \cite{rylands2024dynamical}. QME persists for ferromagnetic (antiferromagnetic) states when $0.8 \leq \gamma \leq 1$ ($0.4 \leq \gamma \leq 1$).
We also report relevant results for non-integrable Hamiltonian $H_{2}$ in the SM, and the results remain qualitatively consistent with cases of $H_{1}$, demonstrating the universal applicability of conclusions in this Letter for Hamiltonian evolutions. 

Our simulation on the other symmetry-breaking measure, charge variance, reveals that QME can also emerge for CV with initial tilted ferromagnetic states, but only in cases of non-symmetric evolution. The reversed monotonicity of CV with respect to $\theta$ can persist even at late times. Table.~\ref{Table:EA and CV} summarizes the early- and late-time behavior of EA and CV for different initial states. The distinction shows the richness in characterizing symmetry breaking strength and patterns.

\textit{U(1)-Symmetric (Asymmetric) States with U(1) Non-Symmetric Random Circuit.---} A schematic diagram of the circuit architecture is shown in Fig.~\ref{fig:circuit_one}(a). The circuit under consideration consists of 16 qubits. We evaluate the EA at different $P_{\text{Haar}}$, using an antiferromagnetic initial state. We observe that EAs approach zero at late times, as illustrated in Fig.~\ref{fig:circuit_one}(b). This behavior can be understood in the context of quantum thermalization and information scrambling \cite{chen2024subsystem,hayden2007black,sekino2008fast,lashkari2013towards}, where the reduced density matrix of the subsystem is a fully mixed state for the random circuit cases, as long as the subsystem size does not exceed half of the total system. Additionally, for all probabilities chosen in Fig.~\ref{fig:circuit_one}(b), EAs reach their maximum after only a few layers of unitaries. The rate of symmetry restoration also depends on the initial state. In the SM, we find that symmetry restoration occurs more quickly for antiferromagnetic or domain wall states than for ferromagnetic states, due to the larger Hilbert space sector of the initial states in the former cases. In Fig.~\ref{fig:circuit_one}(c), we reveal that the peak of the circuit-averaged EA, $\mathbb{E} [\Delta S_{L/4}]_{max}$ follows a power-law with respect to $P_{\text{Haar}}$ for small $P_{\text{Haar}}$. 

\begin{figure}
\centering
\includegraphics[width=0.9\linewidth, keepaspectratio]{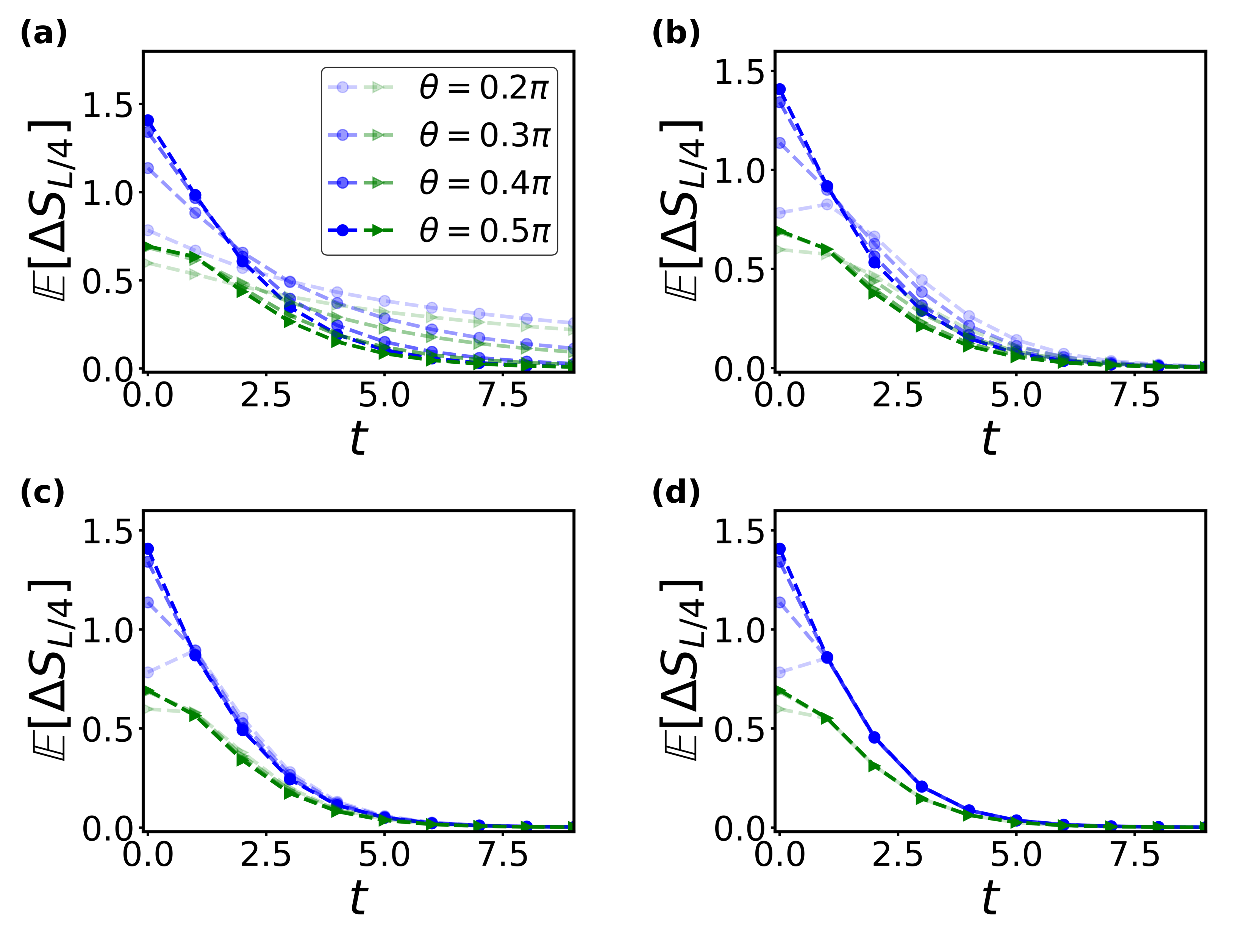}
\caption{The circuit-averaged EA, $\mathbb{E} [\Delta S_{L/4}]$, as a function of time for different values of $P_{\text{Haar}}$. Blue: U(1) EA. Green:  $Z_{2}$ EA. Panels (a)-(d) correspond to different values of $P_{\text{Haar}}$ (a) $P_{\text{Haar}}=0$, (b) $P_{\text{Haar}}=0.3$, (c) $P_{\text{Haar}}=0.7$, and (d) $P_{\text{Haar}}=1$, respectively.}
\label{fig:cirucit_two}
\end{figure}

Next, we examine the dynamics from U(1)-asymmetric initial states, i.e. a tilted ferromagnetic state. We compute the EA for both U(1) symmetry with $\hat{Q_{a}} = \sum_{i \in a} \sigma_{i}^{z}$ and $Z_{2}$ symmetry with $\hat{Q_{a}} = \prod_{i \in a} \sigma_{i}^{z}$. As depicted in Fig.~\ref{fig:cirucit_two}(a), for $P_{\text{Haar}}=0$, we clearly notice the emergence of QME in U(1) case. Surprisingly, we also find that the QME appears in the $Z_{2}$ probe, which does not contradict previous study \cite{liu2024symmetry} suggesting the absence of QME in $Z_{2}$-symmetric circuits. Even though U(1)-symmetric gates are also $Z_{2}$ symmetric, there is no off-diagonal coupling between $\vert 00 \rangle$ and $\vert 11 \rangle$, leading to different thermalization rates between two $Z_{2}$ charge sectors ($Q_{a}=\pm 1$), and thus resulting in QME. As we replace a portion of U(1)-symmetric gates with random Haar gates, QME remains evident with a finite number of random Haar gates.
However, when the circuit consists entirely of random Haar gates, all charge sectors thermalize at the same rate after circuit averaging, and QME disappears. In this case, the overshooting mechanism becomes apparent: EA curves for all initial states (parametrized by $\theta$) converge to a common finite value after a single gate layer, then decay toward zero at late times. Overshooting occurs when this transient value exceeds the initial EA of certain states - particularly symmetric initial states.

\textit{Conclusions and discussions.---} In this Letter, we present a comprehensive study of subsystem symmetry breaking within two frameworks: a non-symmetric Hamiltonian evolution and a non-symmetric random circuit. Our simulation reveals that U(1) symmetry is always restored in the non-symmetric random circuit case, regardless of the initial states or the density of symmetry-breaking random Haar gates $P_{\text{Haar}}$. On the contrary, subsystem U(1) symmetry remains broken in the case of a U(1) non-symmetric Hamiltonian.

In addition to the late-time results, the early-time dynamics of EA shows a universal and surprising feature of overshooting. Specifically, the initial growth of EA can reach a peak significantly higher than its late-time steady value. This behavior is unexpected and is distinct from the growth of entanglement or charge variance, another measure of symmetry-breaking, where the value increases monotonically to its saturating level without any evident overshooting.  Furthermore, for asymmetric initial states evolved under non-symmetric Hamiltonians, the distinct and rich early-time dynamics of EA (increase versus decrease) allow for a direct comparison of the symmetry-breaking extent in both the state and the Hamiltonian.

There are several promising directions for further exploration. For instance, studying the dynamics of symmetry breaking in a non-unitary random circuit with mid-circuit measurements 
\cite{hosur2016chaos,skinner2019measurement,choi2020quantum,li2019measurement,jian2021measurement,minato2022fate,dhar2016measurement,vijay2020measurement,bao2020theory,gullans2020dynamical,zabalo2020critical,ippoliti2021entanglement,szyniszewski2019entanglement,jian2020measurement,liu2024entanglement,liu2024noise,liu2023universal}, could offer valuable insights. Additionally, examining the effect of symmetry breaking in Hamiltonians that avoid thermalization such as many-body localization systems \cite{liu2024quantum,alet2018many,abanin2019colloquium,pal2010many,nandkishore2015many,imbrie2017local,altman2015universal,huse2014phenomenology,lukin2019probing,morningstar2022avalanches} can provide a more unified picture on the understanding of symmetry-breaking dynamics.

\textit{Acknowledgement.---}  HY is supported by the International Young Scientist Fellowship of Institute of Physics Chinese Academy of Sciences (No.202407). ZXL is supported by the National Natural Science Foundation of China under the Grant No.12347107 and Grant No.12474146. SXZ is supported by a start-up grant at IOP-CAS.

\textit{Data availability.}  Numerical data for this manuscript are publicly accessible in Ref. \cite{data-availbale}.


\begin{thebibliography}{91}%
\makeatletter
\providecommand \@ifxundefined [1]{%
 \@ifx{#1\undefined}
}%
\providecommand \@ifnum [1]{%
 \ifnum #1\expandafter \@firstoftwo
 \else \expandafter \@secondoftwo
 \fi
}%
\providecommand \@ifx [1]{%
 \ifx #1\expandafter \@firstoftwo
 \else \expandafter \@secondoftwo
 \fi
}%
\providecommand \natexlab [1]{#1}%
\providecommand \enquote  [1]{``#1''}%
\providecommand \bibnamefont  [1]{#1}%
\providecommand \bibfnamefont [1]{#1}%
\providecommand \citenamefont [1]{#1}%
\providecommand \href@noop [0]{\@secondoftwo}%
\providecommand \href [0]{\begingroup \@sanitize@url \@href}%
\providecommand \@href[1]{\@@startlink{#1}\@@href}%
\providecommand \@@href[1]{\endgroup#1\@@endlink}%
\providecommand \@sanitize@url [0]{\catcode `\\12\catcode `\$12\catcode `\&12\catcode `\#12\catcode `\^12\catcode `\_12\catcode `\%12\relax}%
\providecommand \@@startlink[1]{}%
\providecommand \@@endlink[0]{}%
\providecommand \url  [0]{\begingroup\@sanitize@url \@url }%
\providecommand \@url [1]{\endgroup\@href {#1}{\urlprefix }}%
\providecommand \urlprefix  [0]{URL }%
\providecommand \Eprint [0]{\href }%
\providecommand \doibase [0]{https://doi.org/}%
\providecommand \selectlanguage [0]{\@gobble}%
\providecommand \bibinfo  [0]{\@secondoftwo}%
\providecommand \bibfield  [0]{\@secondoftwo}%
\providecommand \translation [1]{[#1]}%
\providecommand \BibitemOpen [0]{}%
\providecommand \bibitemStop [0]{}%
\providecommand \bibitemNoStop [0]{.\EOS\space}%
\providecommand \EOS [0]{\spacefactor3000\relax}%
\providecommand \BibitemShut  [1]{\csname bibitem#1\endcsname}%
\let\auto@bib@innerbib\@empty
\bibitem [{\citenamefont {Bernstein}(1974)}]{bernstein1974spontaneous}%
  \BibitemOpen
  \bibfield  {author} {\bibinfo {author} {\bibfnamefont {J.}~\bibnamefont {Bernstein}},\ }\href@noop {} {\bibfield  {journal} {\bibinfo  {journal} {Reviews of modern physics}\ }\textbf {\bibinfo {volume} {46}},\ \bibinfo {pages} {7} (\bibinfo {year} {1974})}\BibitemShut {NoStop}%
\bibitem [{\citenamefont {Deutsch}(1991)}]{deutsch1991quantum}%
  \BibitemOpen
  \bibfield  {author} {\bibinfo {author} {\bibfnamefont {J.~M.}\ \bibnamefont {Deutsch}},\ }\href@noop {} {\bibfield  {journal} {\bibinfo  {journal} {Physical review A}\ }\textbf {\bibinfo {volume} {43}},\ \bibinfo {pages} {2046} (\bibinfo {year} {1991})}\BibitemShut {NoStop}%
\bibitem [{\citenamefont {Srednicki}(1994)}]{srednicki1994chaos}%
  \BibitemOpen
  \bibfield  {author} {\bibinfo {author} {\bibfnamefont {M.}~\bibnamefont {Srednicki}},\ }\href@noop {} {\bibfield  {journal} {\bibinfo  {journal} {Physical review E}\ }\textbf {\bibinfo {volume} {50}},\ \bibinfo {pages} {888} (\bibinfo {year} {1994})}\BibitemShut {NoStop}%
\bibitem [{\citenamefont {D'Alessio}\ \emph {et~al.}(2016)\citenamefont {D'Alessio}, \citenamefont {Kafri}, \citenamefont {Polkovnikov},\ and\ \citenamefont {Rigol}}]{d2016quantum}%
  \BibitemOpen
  \bibfield  {author} {\bibinfo {author} {\bibfnamefont {L.}~\bibnamefont {D'Alessio}}, \bibinfo {author} {\bibfnamefont {Y.}~\bibnamefont {Kafri}}, \bibinfo {author} {\bibfnamefont {A.}~\bibnamefont {Polkovnikov}},\ and\ \bibinfo {author} {\bibfnamefont {M.}~\bibnamefont {Rigol}},\ }\href@noop {} {\bibfield  {journal} {\bibinfo  {journal} {Advances in Physics}\ }\textbf {\bibinfo {volume} {65}},\ \bibinfo {pages} {239} (\bibinfo {year} {2016})}\BibitemShut {NoStop}%
\bibitem [{\citenamefont {Rigol}\ \emph {et~al.}(2012)\citenamefont {Rigol}, \citenamefont {Dunjko},\ and\ \citenamefont {Olshanii}}]{rigol2012thermalization}%
  \BibitemOpen
  \bibfield  {author} {\bibinfo {author} {\bibfnamefont {M.}~\bibnamefont {Rigol}}, \bibinfo {author} {\bibfnamefont {V.}~\bibnamefont {Dunjko}},\ and\ \bibinfo {author} {\bibfnamefont {M.}~\bibnamefont {Olshanii}},\ }\href@noop {} {\bibfield  {journal} {\bibinfo  {journal} {Nature}\ }\textbf {\bibinfo {volume} {481}},\ \bibinfo {pages} {224} (\bibinfo {year} {2012})}\BibitemShut {NoStop}%
\bibitem [{\citenamefont {Deutsch}(2018)}]{deutsch2018eigenstate}%
  \BibitemOpen
  \bibfield  {author} {\bibinfo {author} {\bibfnamefont {J.~M.}\ \bibnamefont {Deutsch}},\ }\href@noop {} {\bibfield  {journal} {\bibinfo  {journal} {Reports on Progress in Physics}\ }\textbf {\bibinfo {volume} {81}},\ \bibinfo {pages} {082001} (\bibinfo {year} {2018})}\BibitemShut {NoStop}%
\bibitem [{\citenamefont {Mpemba}\ and\ \citenamefont {Osborne}(1969)}]{mpemba1969cool}%
  \BibitemOpen
  \bibfield  {author} {\bibinfo {author} {\bibfnamefont {E.~B.}\ \bibnamefont {Mpemba}}\ and\ \bibinfo {author} {\bibfnamefont {D.~G.}\ \bibnamefont {Osborne}},\ }\href@noop {} {\bibfield  {journal} {\bibinfo  {journal} {Physics Education}\ }\textbf {\bibinfo {volume} {4}},\ \bibinfo {pages} {172} (\bibinfo {year} {1969})}\BibitemShut {NoStop}%
\bibitem [{\citenamefont {Lu}\ and\ \citenamefont {Raz}(2017)}]{lu2017nonequilibrium}%
  \BibitemOpen
  \bibfield  {author} {\bibinfo {author} {\bibfnamefont {Z.}~\bibnamefont {Lu}}\ and\ \bibinfo {author} {\bibfnamefont {O.}~\bibnamefont {Raz}},\ }\href@noop {} {\bibfield  {journal} {\bibinfo  {journal} {Proceedings of the National Academy of Sciences}\ }\textbf {\bibinfo {volume} {114}},\ \bibinfo {pages} {5083} (\bibinfo {year} {2017})}\BibitemShut {NoStop}%
\bibitem [{\citenamefont {Lasanta}\ \emph {et~al.}(2017)\citenamefont {Lasanta}, \citenamefont {Vega~Reyes}, \citenamefont {Prados},\ and\ \citenamefont {Santos}}]{lasanta2017hotter}%
  \BibitemOpen
  \bibfield  {author} {\bibinfo {author} {\bibfnamefont {A.}~\bibnamefont {Lasanta}}, \bibinfo {author} {\bibfnamefont {F.}~\bibnamefont {Vega~Reyes}}, \bibinfo {author} {\bibfnamefont {A.}~\bibnamefont {Prados}},\ and\ \bibinfo {author} {\bibfnamefont {A.}~\bibnamefont {Santos}},\ }\href@noop {} {\bibfield  {journal} {\bibinfo  {journal} {Physical review letters}\ }\textbf {\bibinfo {volume} {119}},\ \bibinfo {pages} {148001} (\bibinfo {year} {2017})}\BibitemShut {NoStop}%
\bibitem [{\citenamefont {Kumar}\ and\ \citenamefont {Bechhoefer}(2020)}]{kumar2020exponentially}%
  \BibitemOpen
  \bibfield  {author} {\bibinfo {author} {\bibfnamefont {A.}~\bibnamefont {Kumar}}\ and\ \bibinfo {author} {\bibfnamefont {J.}~\bibnamefont {Bechhoefer}},\ }\href@noop {} {\bibfield  {journal} {\bibinfo  {journal} {Nature}\ }\textbf {\bibinfo {volume} {584}},\ \bibinfo {pages} {64} (\bibinfo {year} {2020})}\BibitemShut {NoStop}%
\bibitem [{\citenamefont {Klich}\ \emph {et~al.}(2019)\citenamefont {Klich}, \citenamefont {Raz}, \citenamefont {Hirschberg},\ and\ \citenamefont {Vucelja}}]{klich2019mpemba}%
  \BibitemOpen
  \bibfield  {author} {\bibinfo {author} {\bibfnamefont {I.}~\bibnamefont {Klich}}, \bibinfo {author} {\bibfnamefont {O.}~\bibnamefont {Raz}}, \bibinfo {author} {\bibfnamefont {O.}~\bibnamefont {Hirschberg}},\ and\ \bibinfo {author} {\bibfnamefont {M.}~\bibnamefont {Vucelja}},\ }\href@noop {} {\bibfield  {journal} {\bibinfo  {journal} {Physical Review X}\ }\textbf {\bibinfo {volume} {9}},\ \bibinfo {pages} {021060} (\bibinfo {year} {2019})}\BibitemShut {NoStop}%
\bibitem [{\citenamefont {Teza}\ \emph {et~al.}(2023)\citenamefont {Teza}, \citenamefont {Yaacoby},\ and\ \citenamefont {Raz}}]{teza2023relaxation}%
  \BibitemOpen
  \bibfield  {author} {\bibinfo {author} {\bibfnamefont {G.}~\bibnamefont {Teza}}, \bibinfo {author} {\bibfnamefont {R.}~\bibnamefont {Yaacoby}},\ and\ \bibinfo {author} {\bibfnamefont {O.}~\bibnamefont {Raz}},\ }\href@noop {} {\bibfield  {journal} {\bibinfo  {journal} {Physical review letters}\ }\textbf {\bibinfo {volume} {131}},\ \bibinfo {pages} {017101} (\bibinfo {year} {2023})}\BibitemShut {NoStop}%
\bibitem [{\citenamefont {Bechhoefer}\ \emph {et~al.}(2021)\citenamefont {Bechhoefer}, \citenamefont {Kumar},\ and\ \citenamefont {Ch{\'e}trite}}]{bechhoefer2021fresh}%
  \BibitemOpen
  \bibfield  {author} {\bibinfo {author} {\bibfnamefont {J.}~\bibnamefont {Bechhoefer}}, \bibinfo {author} {\bibfnamefont {A.}~\bibnamefont {Kumar}},\ and\ \bibinfo {author} {\bibfnamefont {R.}~\bibnamefont {Ch{\'e}trite}},\ }\href@noop {} {\bibfield  {journal} {\bibinfo  {journal} {Nature Reviews Physics}\ }\textbf {\bibinfo {volume} {3}},\ \bibinfo {pages} {534} (\bibinfo {year} {2021})}\BibitemShut {NoStop}%
\bibitem [{\citenamefont {Malhotra}\ and\ \citenamefont {L{\"o}wen}(2024)}]{malhotra2024double}%
  \BibitemOpen
  \bibfield  {author} {\bibinfo {author} {\bibfnamefont {I.}~\bibnamefont {Malhotra}}\ and\ \bibinfo {author} {\bibfnamefont {H.}~\bibnamefont {L{\"o}wen}},\ }\href@noop {} {\bibfield  {journal} {\bibinfo  {journal} {The Journal of Chemical Physics}\ }\textbf {\bibinfo {volume} {161}} (\bibinfo {year} {2024})}\BibitemShut {NoStop}%
\bibitem [{\citenamefont {Kumar}\ \emph {et~al.}(2022)\citenamefont {Kumar}, \citenamefont {Ch{\'e}trite},\ and\ \citenamefont {Bechhoefer}}]{kumar2022anomalous}%
  \BibitemOpen
  \bibfield  {author} {\bibinfo {author} {\bibfnamefont {A.}~\bibnamefont {Kumar}}, \bibinfo {author} {\bibfnamefont {R.}~\bibnamefont {Ch{\'e}trite}},\ and\ \bibinfo {author} {\bibfnamefont {J.}~\bibnamefont {Bechhoefer}},\ }\href@noop {} {\bibfield  {journal} {\bibinfo  {journal} {Proceedings of the National Academy of Sciences}\ }\textbf {\bibinfo {volume} {119}},\ \bibinfo {pages} {e2118484119} (\bibinfo {year} {2022})}\BibitemShut {NoStop}%
\bibitem [{\citenamefont {Manikandan}(2021)}]{manikandan2021equidistant}%
  \BibitemOpen
  \bibfield  {author} {\bibinfo {author} {\bibfnamefont {S.~K.}\ \bibnamefont {Manikandan}},\ }\href@noop {} {\bibfield  {journal} {\bibinfo  {journal} {Physical Review Research}\ }\textbf {\bibinfo {volume} {3}},\ \bibinfo {pages} {043108} (\bibinfo {year} {2021})}\BibitemShut {NoStop}%
\bibitem [{\citenamefont {Chatterjee}\ \emph {et~al.}(2023)\citenamefont {Chatterjee}, \citenamefont {Takada},\ and\ \citenamefont {Hayakawa}}]{chatterjee2023quantum}%
  \BibitemOpen
  \bibfield  {author} {\bibinfo {author} {\bibfnamefont {A.~K.}\ \bibnamefont {Chatterjee}}, \bibinfo {author} {\bibfnamefont {S.}~\bibnamefont {Takada}},\ and\ \bibinfo {author} {\bibfnamefont {H.}~\bibnamefont {Hayakawa}},\ }\href@noop {} {\bibfield  {journal} {\bibinfo  {journal} {Physical Review Letters}\ }\textbf {\bibinfo {volume} {131}},\ \bibinfo {pages} {080402} (\bibinfo {year} {2023})}\BibitemShut {NoStop}%
\bibitem [{\citenamefont {Aharony~Shapira}\ \emph {et~al.}(2024)\citenamefont {Aharony~Shapira}, \citenamefont {Shapira}, \citenamefont {Markov}, \citenamefont {Teza}, \citenamefont {Akerman}, \citenamefont {Raz},\ and\ \citenamefont {Ozeri}}]{aharony2024inverse}%
  \BibitemOpen
  \bibfield  {author} {\bibinfo {author} {\bibfnamefont {S.}~\bibnamefont {Aharony~Shapira}}, \bibinfo {author} {\bibfnamefont {Y.}~\bibnamefont {Shapira}}, \bibinfo {author} {\bibfnamefont {J.}~\bibnamefont {Markov}}, \bibinfo {author} {\bibfnamefont {G.}~\bibnamefont {Teza}}, \bibinfo {author} {\bibfnamefont {N.}~\bibnamefont {Akerman}}, \bibinfo {author} {\bibfnamefont {O.}~\bibnamefont {Raz}},\ and\ \bibinfo {author} {\bibfnamefont {R.}~\bibnamefont {Ozeri}},\ }\href@noop {} {\bibfield  {journal} {\bibinfo  {journal} {Physical Review Letters}\ }\textbf {\bibinfo {volume} {133}},\ \bibinfo {pages} {010403} (\bibinfo {year} {2024})}\BibitemShut {NoStop}%
\bibitem [{\citenamefont {Wang}\ and\ \citenamefont {Wang}(2024)}]{wang2024mpemba}%
  \BibitemOpen
  \bibfield  {author} {\bibinfo {author} {\bibfnamefont {X.}~\bibnamefont {Wang}}\ and\ \bibinfo {author} {\bibfnamefont {J.}~\bibnamefont {Wang}},\ }\href@noop {} {\bibfield  {journal} {\bibinfo  {journal} {Physical Review Research}\ }\textbf {\bibinfo {volume} {6}},\ \bibinfo {pages} {033330} (\bibinfo {year} {2024})}\BibitemShut {NoStop}%
\bibitem [{\citenamefont {Nava}\ and\ \citenamefont {Fabrizio}(2019)}]{nava2019lindblad}%
  \BibitemOpen
  \bibfield  {author} {\bibinfo {author} {\bibfnamefont {A.}~\bibnamefont {Nava}}\ and\ \bibinfo {author} {\bibfnamefont {M.}~\bibnamefont {Fabrizio}},\ }\href@noop {} {\bibfield  {journal} {\bibinfo  {journal} {Physical Review B}\ }\textbf {\bibinfo {volume} {100}},\ \bibinfo {pages} {125102} (\bibinfo {year} {2019})}\BibitemShut {NoStop}%
\bibitem [{\citenamefont {Carollo}\ \emph {et~al.}(2021)\citenamefont {Carollo}, \citenamefont {Lasanta},\ and\ \citenamefont {Lesanovsky}}]{carollo2021exponentially}%
  \BibitemOpen
  \bibfield  {author} {\bibinfo {author} {\bibfnamefont {F.}~\bibnamefont {Carollo}}, \bibinfo {author} {\bibfnamefont {A.}~\bibnamefont {Lasanta}},\ and\ \bibinfo {author} {\bibfnamefont {I.}~\bibnamefont {Lesanovsky}},\ }\href@noop {} {\bibfield  {journal} {\bibinfo  {journal} {Physical Review Letters}\ }\textbf {\bibinfo {volume} {127}},\ \bibinfo {pages} {060401} (\bibinfo {year} {2021})}\BibitemShut {NoStop}%
\bibitem [{\citenamefont {Chatterjee}\ \emph {et~al.}(2024)\citenamefont {Chatterjee}, \citenamefont {Takada},\ and\ \citenamefont {Hayakawa}}]{chatterjee2024multiple}%
  \BibitemOpen
  \bibfield  {author} {\bibinfo {author} {\bibfnamefont {A.~K.}\ \bibnamefont {Chatterjee}}, \bibinfo {author} {\bibfnamefont {S.}~\bibnamefont {Takada}},\ and\ \bibinfo {author} {\bibfnamefont {H.}~\bibnamefont {Hayakawa}},\ }\href@noop {} {\bibfield  {journal} {\bibinfo  {journal} {Physical Review A}\ }\textbf {\bibinfo {volume} {110}},\ \bibinfo {pages} {022213} (\bibinfo {year} {2024})}\BibitemShut {NoStop}%
\bibitem [{\citenamefont {Ivander}\ \emph {et~al.}(2023)\citenamefont {Ivander}, \citenamefont {Anto-Sztrikacs},\ and\ \citenamefont {Segal}}]{ivander2023hyperacceleration}%
  \BibitemOpen
  \bibfield  {author} {\bibinfo {author} {\bibfnamefont {F.}~\bibnamefont {Ivander}}, \bibinfo {author} {\bibfnamefont {N.}~\bibnamefont {Anto-Sztrikacs}},\ and\ \bibinfo {author} {\bibfnamefont {D.}~\bibnamefont {Segal}},\ }\href@noop {} {\bibfield  {journal} {\bibinfo  {journal} {Physical Review E}\ }\textbf {\bibinfo {volume} {108}},\ \bibinfo {pages} {014130} (\bibinfo {year} {2023})}\BibitemShut {NoStop}%
\bibitem [{\citenamefont {Ares}\ \emph {et~al.}(2023{\natexlab{a}})\citenamefont {Ares}, \citenamefont {Murciano},\ and\ \citenamefont {Calabrese}}]{ares2023entanglement}%
  \BibitemOpen
  \bibfield  {author} {\bibinfo {author} {\bibfnamefont {F.}~\bibnamefont {Ares}}, \bibinfo {author} {\bibfnamefont {S.}~\bibnamefont {Murciano}},\ and\ \bibinfo {author} {\bibfnamefont {P.}~\bibnamefont {Calabrese}},\ }\href@noop {} {\bibfield  {journal} {\bibinfo  {journal} {Nature Communications}\ }\textbf {\bibinfo {volume} {14}},\ \bibinfo {pages} {2036} (\bibinfo {year} {2023}{\natexlab{a}})}\BibitemShut {NoStop}%
\bibitem [{\citenamefont {Liu}\ \emph {et~al.}(2024{\natexlab{a}})\citenamefont {Liu}, \citenamefont {Zhang}, \citenamefont {Yin},\ and\ \citenamefont {Zhang}}]{liu2024symmetry}%
  \BibitemOpen
  \bibfield  {author} {\bibinfo {author} {\bibfnamefont {S.}~\bibnamefont {Liu}}, \bibinfo {author} {\bibfnamefont {H.-K.}\ \bibnamefont {Zhang}}, \bibinfo {author} {\bibfnamefont {S.}~\bibnamefont {Yin}},\ and\ \bibinfo {author} {\bibfnamefont {S.-X.}\ \bibnamefont {Zhang}},\ }\href@noop {} {\bibfield  {journal} {\bibinfo  {journal} {Physical Review Letters}\ }\textbf {\bibinfo {volume} {133}},\ \bibinfo {pages} {140405} (\bibinfo {year} {2024}{\natexlab{a}})}\BibitemShut {NoStop}%
\bibitem [{\citenamefont {Turkeshi}\ \emph {et~al.}(2024)\citenamefont {Turkeshi}, \citenamefont {Calabrese},\ and\ \citenamefont {De~Luca}}]{turkeshi2024quantum}%
  \BibitemOpen
  \bibfield  {author} {\bibinfo {author} {\bibfnamefont {X.}~\bibnamefont {Turkeshi}}, \bibinfo {author} {\bibfnamefont {P.}~\bibnamefont {Calabrese}},\ and\ \bibinfo {author} {\bibfnamefont {A.}~\bibnamefont {De~Luca}},\ }\href@noop {} {\bibfield  {journal} {\bibinfo  {journal} {arXiv preprint arXiv:2405.14514}\ } (\bibinfo {year} {2024})}\BibitemShut {NoStop}%
\bibitem [{\citenamefont {Fagotti}\ \emph {et~al.}(2014)\citenamefont {Fagotti}, \citenamefont {Collura}, \citenamefont {Essler},\ and\ \citenamefont {Calabrese}}]{fagotti2014relaxation}%
  \BibitemOpen
  \bibfield  {author} {\bibinfo {author} {\bibfnamefont {M.}~\bibnamefont {Fagotti}}, \bibinfo {author} {\bibfnamefont {M.}~\bibnamefont {Collura}}, \bibinfo {author} {\bibfnamefont {F.~H.}\ \bibnamefont {Essler}},\ and\ \bibinfo {author} {\bibfnamefont {P.}~\bibnamefont {Calabrese}},\ }\href@noop {} {\bibfield  {journal} {\bibinfo  {journal} {Physical Review B}\ }\textbf {\bibinfo {volume} {89}},\ \bibinfo {pages} {125101} (\bibinfo {year} {2014})}\BibitemShut {NoStop}%
\bibitem [{\citenamefont {Essler}\ and\ \citenamefont {Fagotti}(2016)}]{essler2016quench}%
  \BibitemOpen
  \bibfield  {author} {\bibinfo {author} {\bibfnamefont {F.~H.}\ \bibnamefont {Essler}}\ and\ \bibinfo {author} {\bibfnamefont {M.}~\bibnamefont {Fagotti}},\ }\href@noop {} {\bibfield  {journal} {\bibinfo  {journal} {Journal of Statistical Mechanics: Theory and Experiment}\ }\textbf {\bibinfo {volume} {2016}},\ \bibinfo {pages} {064002} (\bibinfo {year} {2016})}\BibitemShut {NoStop}%
\bibitem [{\citenamefont {Doyon}(2020)}]{doyon2020lecture}%
  \BibitemOpen
  \bibfield  {author} {\bibinfo {author} {\bibfnamefont {B.}~\bibnamefont {Doyon}},\ }\href@noop {} {\bibfield  {journal} {\bibinfo  {journal} {SciPost Physics Lecture Notes}\ ,\ \bibinfo {pages} {018}} (\bibinfo {year} {2020})}\BibitemShut {NoStop}%
\bibitem [{\citenamefont {Fagotti}(2014)}]{fagotti2014conservation}%
  \BibitemOpen
  \bibfield  {author} {\bibinfo {author} {\bibfnamefont {M.}~\bibnamefont {Fagotti}},\ }\href@noop {} {\bibfield  {journal} {\bibinfo  {journal} {Journal of Statistical Mechanics: Theory and Experiment}\ }\textbf {\bibinfo {volume} {2014}},\ \bibinfo {pages} {P03016} (\bibinfo {year} {2014})}\BibitemShut {NoStop}%
\bibitem [{\citenamefont {Bertini}\ and\ \citenamefont {Fagotti}(2015)}]{bertini2015pre}%
  \BibitemOpen
  \bibfield  {author} {\bibinfo {author} {\bibfnamefont {B.}~\bibnamefont {Bertini}}\ and\ \bibinfo {author} {\bibfnamefont {M.}~\bibnamefont {Fagotti}},\ }\href@noop {} {\bibfield  {journal} {\bibinfo  {journal} {Journal of Statistical Mechanics: Theory and Experiment}\ }\textbf {\bibinfo {volume} {2015}},\ \bibinfo {pages} {P07012} (\bibinfo {year} {2015})}\BibitemShut {NoStop}%
\bibitem [{\citenamefont {Vidmar}\ and\ \citenamefont {Rigol}(2016)}]{vidmar2016generalized}%
  \BibitemOpen
  \bibfield  {author} {\bibinfo {author} {\bibfnamefont {L.}~\bibnamefont {Vidmar}}\ and\ \bibinfo {author} {\bibfnamefont {M.}~\bibnamefont {Rigol}},\ }\href@noop {} {\bibfield  {journal} {\bibinfo  {journal} {Journal of Statistical Mechanics: Theory and Experiment}\ }\textbf {\bibinfo {volume} {2016}},\ \bibinfo {pages} {064007} (\bibinfo {year} {2016})}\BibitemShut {NoStop}%
\bibitem [{\citenamefont {Alba}\ \emph {et~al.}(2021)\citenamefont {Alba}, \citenamefont {Bertini}, \citenamefont {Fagotti}, \citenamefont {Piroli},\ and\ \citenamefont {Ruggiero}}]{alba2021generalized}%
  \BibitemOpen
  \bibfield  {author} {\bibinfo {author} {\bibfnamefont {V.}~\bibnamefont {Alba}}, \bibinfo {author} {\bibfnamefont {B.}~\bibnamefont {Bertini}}, \bibinfo {author} {\bibfnamefont {M.}~\bibnamefont {Fagotti}}, \bibinfo {author} {\bibfnamefont {L.}~\bibnamefont {Piroli}},\ and\ \bibinfo {author} {\bibfnamefont {P.}~\bibnamefont {Ruggiero}},\ }\href@noop {} {\bibfield  {journal} {\bibinfo  {journal} {Journal of Statistical Mechanics: Theory and Experiment}\ }\textbf {\bibinfo {volume} {2021}},\ \bibinfo {pages} {114004} (\bibinfo {year} {2021})}\BibitemShut {NoStop}%
\bibitem [{\citenamefont {Calabrese}\ \emph {et~al.}(2016)\citenamefont {Calabrese}, \citenamefont {Essler},\ and\ \citenamefont {Mussardo}}]{calabrese2016quantum}%
  \BibitemOpen
  \bibfield  {author} {\bibinfo {author} {\bibfnamefont {P.}~\bibnamefont {Calabrese}}, \bibinfo {author} {\bibfnamefont {F.}~\bibnamefont {Essler}},\ and\ \bibinfo {author} {\bibfnamefont {G.}~\bibnamefont {Mussardo}},\ }\href@noop {} {\bibfield  {journal} {\bibinfo  {journal} {J. Stat. Mech}\ }\textbf {\bibinfo {volume} {2016}},\ \bibinfo {pages} {064001} (\bibinfo {year} {2016})}\BibitemShut {NoStop}%
\bibitem [{\citenamefont {Polkovnikov}\ \emph {et~al.}()\citenamefont {Polkovnikov}, \citenamefont {Sengupta}, \citenamefont {Silva},\ and\ \citenamefont {Vengalatorre}}]{polkovnikov83nonequilibrium}%
  \BibitemOpen
  \bibfield  {author} {\bibinfo {author} {\bibfnamefont {A.}~\bibnamefont {Polkovnikov}}, \bibinfo {author} {\bibfnamefont {K.}~\bibnamefont {Sengupta}}, \bibinfo {author} {\bibfnamefont {A.}~\bibnamefont {Silva}},\ and\ \bibinfo {author} {\bibfnamefont {M.}~\bibnamefont {Vengalatorre}},\ }\href@noop {} {\bibfield  {journal} {\bibinfo  {journal} {Mod. Phys}\ }\textbf {\bibinfo {volume} {83}},\ \bibinfo {pages} {5}}\BibitemShut {NoStop}%
\bibitem [{\citenamefont {Bastianello}\ \emph {et~al.}(2022)\citenamefont {Bastianello}, \citenamefont {Bertini}, \citenamefont {Doyon},\ and\ \citenamefont {Vasseur}}]{bastianello2022introduction}%
  \BibitemOpen
  \bibfield  {author} {\bibinfo {author} {\bibfnamefont {A.}~\bibnamefont {Bastianello}}, \bibinfo {author} {\bibfnamefont {B.}~\bibnamefont {Bertini}}, \bibinfo {author} {\bibfnamefont {B.}~\bibnamefont {Doyon}},\ and\ \bibinfo {author} {\bibfnamefont {R.}~\bibnamefont {Vasseur}},\ }\href@noop {} {\bibfield  {journal} {\bibinfo  {journal} {Journal of Statistical Mechanics: Theory and Experiment}\ }\textbf {\bibinfo {volume} {2022}},\ \bibinfo {pages} {014001} (\bibinfo {year} {2022})}\BibitemShut {NoStop}%
\bibitem [{\citenamefont {Ares}\ \emph {et~al.}(2023{\natexlab{b}})\citenamefont {Ares}, \citenamefont {Murciano}, \citenamefont {Vernier},\ and\ \citenamefont {Calabrese}}]{ares2023lack}%
  \BibitemOpen
  \bibfield  {author} {\bibinfo {author} {\bibfnamefont {F.}~\bibnamefont {Ares}}, \bibinfo {author} {\bibfnamefont {S.}~\bibnamefont {Murciano}}, \bibinfo {author} {\bibfnamefont {E.}~\bibnamefont {Vernier}},\ and\ \bibinfo {author} {\bibfnamefont {P.}~\bibnamefont {Calabrese}},\ }\href@noop {} {\bibfield  {journal} {\bibinfo  {journal} {SciPost Physics}\ }\textbf {\bibinfo {volume} {15}},\ \bibinfo {pages} {089} (\bibinfo {year} {2023}{\natexlab{b}})}\BibitemShut {NoStop}%
\bibitem [{\citenamefont {Klobas}\ \emph {et~al.}(2024)\citenamefont {Klobas}, \citenamefont {Rylands},\ and\ \citenamefont {Bertini}}]{klobas2024translation}%
  \BibitemOpen
  \bibfield  {author} {\bibinfo {author} {\bibfnamefont {K.}~\bibnamefont {Klobas}}, \bibinfo {author} {\bibfnamefont {C.}~\bibnamefont {Rylands}},\ and\ \bibinfo {author} {\bibfnamefont {B.}~\bibnamefont {Bertini}},\ }\href@noop {} {\bibfield  {journal} {\bibinfo  {journal} {arXiv preprint arXiv:2406.04296}\ } (\bibinfo {year} {2024})}\BibitemShut {NoStop}%
\bibitem [{\citenamefont {Rylands}\ \emph {et~al.}(2024{\natexlab{a}})\citenamefont {Rylands}, \citenamefont {Klobas}, \citenamefont {Ares}, \citenamefont {Calabrese}, \citenamefont {Murciano},\ and\ \citenamefont {Bertini}}]{rylands2024microscopic}%
  \BibitemOpen
  \bibfield  {author} {\bibinfo {author} {\bibfnamefont {C.}~\bibnamefont {Rylands}}, \bibinfo {author} {\bibfnamefont {K.}~\bibnamefont {Klobas}}, \bibinfo {author} {\bibfnamefont {F.}~\bibnamefont {Ares}}, \bibinfo {author} {\bibfnamefont {P.}~\bibnamefont {Calabrese}}, \bibinfo {author} {\bibfnamefont {S.}~\bibnamefont {Murciano}},\ and\ \bibinfo {author} {\bibfnamefont {B.}~\bibnamefont {Bertini}},\ }\href@noop {} {\bibfield  {journal} {\bibinfo  {journal} {Physical Review Letters}\ }\textbf {\bibinfo {volume} {133}},\ \bibinfo {pages} {010401} (\bibinfo {year} {2024}{\natexlab{a}})}\BibitemShut {NoStop}%
\bibitem [{\citenamefont {Yamashika}\ \emph {et~al.}(2024)\citenamefont {Yamashika}, \citenamefont {Ares},\ and\ \citenamefont {Calabrese}}]{yamashika2024entanglement}%
  \BibitemOpen
  \bibfield  {author} {\bibinfo {author} {\bibfnamefont {S.}~\bibnamefont {Yamashika}}, \bibinfo {author} {\bibfnamefont {F.}~\bibnamefont {Ares}},\ and\ \bibinfo {author} {\bibfnamefont {P.}~\bibnamefont {Calabrese}},\ }\href@noop {} {\bibfield  {journal} {\bibinfo  {journal} {Physical Review B}\ }\textbf {\bibinfo {volume} {110}},\ \bibinfo {pages} {085126} (\bibinfo {year} {2024})}\BibitemShut {NoStop}%
\bibitem [{\citenamefont {Khor}\ \emph {et~al.}(2024)\citenamefont {Khor}, \citenamefont {K{\"u}rk{\c{c}}{\"u}oglu}, \citenamefont {Hobbs}, \citenamefont {Perdue},\ and\ \citenamefont {Klich}}]{khor2024confinement}%
  \BibitemOpen
  \bibfield  {author} {\bibinfo {author} {\bibfnamefont {B.~J.}\ \bibnamefont {Khor}}, \bibinfo {author} {\bibfnamefont {D.}~\bibnamefont {K{\"u}rk{\c{c}}{\"u}oglu}}, \bibinfo {author} {\bibfnamefont {T.}~\bibnamefont {Hobbs}}, \bibinfo {author} {\bibfnamefont {G.}~\bibnamefont {Perdue}},\ and\ \bibinfo {author} {\bibfnamefont {I.}~\bibnamefont {Klich}},\ }\href@noop {} {\bibfield  {journal} {\bibinfo  {journal} {Quantum}\ }\textbf {\bibinfo {volume} {8}},\ \bibinfo {pages} {1462} (\bibinfo {year} {2024})}\BibitemShut {NoStop}%
\bibitem [{\citenamefont {Rylands}\ \emph {et~al.}(2024{\natexlab{b}})\citenamefont {Rylands}, \citenamefont {Vernier},\ and\ \citenamefont {Calabrese}}]{rylands2024dynamical}%
  \BibitemOpen
  \bibfield  {author} {\bibinfo {author} {\bibfnamefont {C.}~\bibnamefont {Rylands}}, \bibinfo {author} {\bibfnamefont {E.}~\bibnamefont {Vernier}},\ and\ \bibinfo {author} {\bibfnamefont {P.}~\bibnamefont {Calabrese}},\ }\href@noop {} {\bibfield  {journal} {\bibinfo  {journal} {Journal of Statistical Mechanics: Theory and Experiment}\ }\textbf {\bibinfo {volume} {2024}},\ \bibinfo {pages} {123102} (\bibinfo {year} {2024}{\natexlab{b}})}\BibitemShut {NoStop}%
\bibitem [{\citenamefont {Ares}\ \emph {et~al.}(2024)\citenamefont {Ares}, \citenamefont {Murciano}, \citenamefont {Piroli},\ and\ \citenamefont {Calabrese}}]{ares2024entanglement}%
  \BibitemOpen
  \bibfield  {author} {\bibinfo {author} {\bibfnamefont {F.}~\bibnamefont {Ares}}, \bibinfo {author} {\bibfnamefont {S.}~\bibnamefont {Murciano}}, \bibinfo {author} {\bibfnamefont {L.}~\bibnamefont {Piroli}},\ and\ \bibinfo {author} {\bibfnamefont {P.}~\bibnamefont {Calabrese}},\ }\href@noop {} {\bibfield  {journal} {\bibinfo  {journal} {Physical Review D}\ }\textbf {\bibinfo {volume} {110}},\ \bibinfo {pages} {L061901} (\bibinfo {year} {2024})}\BibitemShut {NoStop}%
\bibitem [{\citenamefont {Murciano}\ \emph {et~al.}(2024)\citenamefont {Murciano}, \citenamefont {Ares}, \citenamefont {Klich},\ and\ \citenamefont {Calabrese}}]{murciano2024entanglement}%
  \BibitemOpen
  \bibfield  {author} {\bibinfo {author} {\bibfnamefont {S.}~\bibnamefont {Murciano}}, \bibinfo {author} {\bibfnamefont {F.}~\bibnamefont {Ares}}, \bibinfo {author} {\bibfnamefont {I.}~\bibnamefont {Klich}},\ and\ \bibinfo {author} {\bibfnamefont {P.}~\bibnamefont {Calabrese}},\ }\href@noop {} {\bibfield  {journal} {\bibinfo  {journal} {Journal of Statistical Mechanics: Theory and Experiment}\ }\textbf {\bibinfo {volume} {2024}},\ \bibinfo {pages} {013103} (\bibinfo {year} {2024})}\BibitemShut {NoStop}%
\bibitem [{\citenamefont {Liu}\ \emph {et~al.}(2024{\natexlab{b}})\citenamefont {Liu}, \citenamefont {Zhang}, \citenamefont {Yin}, \citenamefont {Zhang},\ and\ \citenamefont {Yao}}]{liu2024quantum}%
  \BibitemOpen
  \bibfield  {author} {\bibinfo {author} {\bibfnamefont {S.}~\bibnamefont {Liu}}, \bibinfo {author} {\bibfnamefont {H.-K.}\ \bibnamefont {Zhang}}, \bibinfo {author} {\bibfnamefont {S.}~\bibnamefont {Yin}}, \bibinfo {author} {\bibfnamefont {S.-X.}\ \bibnamefont {Zhang}},\ and\ \bibinfo {author} {\bibfnamefont {H.}~\bibnamefont {Yao}},\ }\href@noop {} {\bibfield  {journal} {\bibinfo  {journal} {arXiv preprint arXiv:2408.07750}\ } (\bibinfo {year} {2024}{\natexlab{b}})}\BibitemShut {NoStop}%
\bibitem [{\citenamefont {Chang}\ \emph {et~al.}(2024)\citenamefont {Chang}, \citenamefont {Yin}, \citenamefont {Zhang},\ and\ \citenamefont {Li}}]{chang2024imaginary}%
  \BibitemOpen
  \bibfield  {author} {\bibinfo {author} {\bibfnamefont {W.-X.}\ \bibnamefont {Chang}}, \bibinfo {author} {\bibfnamefont {S.}~\bibnamefont {Yin}}, \bibinfo {author} {\bibfnamefont {S.-X.}\ \bibnamefont {Zhang}},\ and\ \bibinfo {author} {\bibfnamefont {Z.-X.}\ \bibnamefont {Li}},\ }\href@noop {} {\bibfield  {journal} {\bibinfo  {journal} {arXiv preprint arXiv:2409.06547}\ } (\bibinfo {year} {2024})}\BibitemShut {NoStop}%
\bibitem [{\citenamefont {Russotto}\ \emph {et~al.}(2024)\citenamefont {Russotto}, \citenamefont {Ares},\ and\ \citenamefont {Calabrese}}]{russotto2024non}%
  \BibitemOpen
  \bibfield  {author} {\bibinfo {author} {\bibfnamefont {A.}~\bibnamefont {Russotto}}, \bibinfo {author} {\bibfnamefont {F.}~\bibnamefont {Ares}},\ and\ \bibinfo {author} {\bibfnamefont {P.}~\bibnamefont {Calabrese}},\ }\href@noop {} {\bibfield  {journal} {\bibinfo  {journal} {arXiv preprint arXiv:2411.13337}\ } (\bibinfo {year} {2024})}\BibitemShut {NoStop}%
\bibitem [{\citenamefont {Joshi}\ \emph {et~al.}(2024)\citenamefont {Joshi}, \citenamefont {Franke}, \citenamefont {Rath}, \citenamefont {Ares}, \citenamefont {Murciano}, \citenamefont {Kranzl}, \citenamefont {Blatt}, \citenamefont {Zoller}, \citenamefont {Vermersch}, \citenamefont {Calabrese} \emph {et~al.}}]{joshi2024observing}%
  \BibitemOpen
  \bibfield  {author} {\bibinfo {author} {\bibfnamefont {L.~K.}\ \bibnamefont {Joshi}}, \bibinfo {author} {\bibfnamefont {J.}~\bibnamefont {Franke}}, \bibinfo {author} {\bibfnamefont {A.}~\bibnamefont {Rath}}, \bibinfo {author} {\bibfnamefont {F.}~\bibnamefont {Ares}}, \bibinfo {author} {\bibfnamefont {S.}~\bibnamefont {Murciano}}, \bibinfo {author} {\bibfnamefont {F.}~\bibnamefont {Kranzl}}, \bibinfo {author} {\bibfnamefont {R.}~\bibnamefont {Blatt}}, \bibinfo {author} {\bibfnamefont {P.}~\bibnamefont {Zoller}}, \bibinfo {author} {\bibfnamefont {B.}~\bibnamefont {Vermersch}}, \bibinfo {author} {\bibfnamefont {P.}~\bibnamefont {Calabrese}}, \emph {et~al.},\ }\href@noop {} {\bibfield  {journal} {\bibinfo  {journal} {Physical Review Letters}\ }\textbf {\bibinfo {volume} {133}},\ \bibinfo {pages} {010402} (\bibinfo {year} {2024})}\BibitemShut {NoStop}%
\bibitem [{\citenamefont {Rossi}\ \emph {et~al.}(2023)\citenamefont {Rossi}, \citenamefont {Barbiero}, \citenamefont {Budich},\ and\ \citenamefont {Dolcini}}]{rossi2023long}%
  \BibitemOpen
  \bibfield  {author} {\bibinfo {author} {\bibfnamefont {L.}~\bibnamefont {Rossi}}, \bibinfo {author} {\bibfnamefont {L.}~\bibnamefont {Barbiero}}, \bibinfo {author} {\bibfnamefont {J.~C.}\ \bibnamefont {Budich}},\ and\ \bibinfo {author} {\bibfnamefont {F.}~\bibnamefont {Dolcini}},\ }\href@noop {} {\bibfield  {journal} {\bibinfo  {journal} {Physical Review B}\ }\textbf {\bibinfo {volume} {108}},\ \bibinfo {pages} {155420} (\bibinfo {year} {2023})}\BibitemShut {NoStop}%
\bibitem [{\citenamefont {Fisher}\ \emph {et~al.}(2023)\citenamefont {Fisher}, \citenamefont {Khemani}, \citenamefont {Nahum},\ and\ \citenamefont {Vijay}}]{fisher2023random}%
  \BibitemOpen
  \bibfield  {author} {\bibinfo {author} {\bibfnamefont {M.~P.}\ \bibnamefont {Fisher}}, \bibinfo {author} {\bibfnamefont {V.}~\bibnamefont {Khemani}}, \bibinfo {author} {\bibfnamefont {A.}~\bibnamefont {Nahum}},\ and\ \bibinfo {author} {\bibfnamefont {S.}~\bibnamefont {Vijay}},\ }\href@noop {} {\bibfield  {journal} {\bibinfo  {journal} {Annual Review of Condensed Matter Physics}\ }\textbf {\bibinfo {volume} {14}},\ \bibinfo {pages} {335} (\bibinfo {year} {2023})}\BibitemShut {NoStop}%
\bibitem [{\citenamefont {Capizzi}\ and\ \citenamefont {Vitale}(2024)}]{capizzi2024universal}%
  \BibitemOpen
  \bibfield  {author} {\bibinfo {author} {\bibfnamefont {L.}~\bibnamefont {Capizzi}}\ and\ \bibinfo {author} {\bibfnamefont {V.}~\bibnamefont {Vitale}},\ }\href@noop {} {\bibfield  {journal} {\bibinfo  {journal} {Journal of Physics A: Mathematical and Theoretical}\ }\textbf {\bibinfo {volume} {57}},\ \bibinfo {pages} {45LT01} (\bibinfo {year} {2024})}\BibitemShut {NoStop}%
\bibitem [{\citenamefont {Chen}\ and\ \citenamefont {Chen}(2024)}]{chen2024renyi}%
  \BibitemOpen
  \bibfield  {author} {\bibinfo {author} {\bibfnamefont {M.}~\bibnamefont {Chen}}\ and\ \bibinfo {author} {\bibfnamefont {H.-H.}\ \bibnamefont {Chen}},\ }\href@noop {} {\bibfield  {journal} {\bibinfo  {journal} {Physical Review D}\ }\textbf {\bibinfo {volume} {109}},\ \bibinfo {pages} {065009} (\bibinfo {year} {2024})}\BibitemShut {NoStop}%
\bibitem [{\citenamefont {Capizzi}\ and\ \citenamefont {Mazzoni}(2023)}]{capizzi2023entanglement}%
  \BibitemOpen
  \bibfield  {author} {\bibinfo {author} {\bibfnamefont {L.}~\bibnamefont {Capizzi}}\ and\ \bibinfo {author} {\bibfnamefont {M.}~\bibnamefont {Mazzoni}},\ }\href@noop {} {\bibfield  {journal} {\bibinfo  {journal} {Journal of High Energy Physics}\ }\textbf {\bibinfo {volume} {2023}},\ \bibinfo {pages} {1} (\bibinfo {year} {2023})}\BibitemShut {NoStop}%
\bibitem [{\citenamefont {Ares}\ \emph {et~al.}(2025)\citenamefont {Ares}, \citenamefont {Murciano}, \citenamefont {Calabrese},\ and\ \citenamefont {Piroli}}]{ares2025entanglementasymmetrydynamicsrandom}%
  \BibitemOpen
  \bibfield  {author} {\bibinfo {author} {\bibfnamefont {F.}~\bibnamefont {Ares}}, \bibinfo {author} {\bibfnamefont {S.}~\bibnamefont {Murciano}}, \bibinfo {author} {\bibfnamefont {P.}~\bibnamefont {Calabrese}},\ and\ \bibinfo {author} {\bibfnamefont {L.}~\bibnamefont {Piroli}},\ }\href@noop {} {\bibfield  {journal} {\bibinfo  {journal} {arXiv preprint arXiv:2501.12459}\ } (\bibinfo {year} {2025})}\BibitemShut {NoStop}%
\bibitem [{\citenamefont {Lessa}\ \emph {et~al.}(2024)\citenamefont {Lessa}, \citenamefont {Ma}, \citenamefont {Zhang}, \citenamefont {Bi}, \citenamefont {Cheng},\ and\ \citenamefont {Wang}}]{lessa2024strong}%
  \BibitemOpen
  \bibfield  {author} {\bibinfo {author} {\bibfnamefont {L.~A.}\ \bibnamefont {Lessa}}, \bibinfo {author} {\bibfnamefont {R.}~\bibnamefont {Ma}}, \bibinfo {author} {\bibfnamefont {J.-H.}\ \bibnamefont {Zhang}}, \bibinfo {author} {\bibfnamefont {Z.}~\bibnamefont {Bi}}, \bibinfo {author} {\bibfnamefont {M.}~\bibnamefont {Cheng}},\ and\ \bibinfo {author} {\bibfnamefont {C.}~\bibnamefont {Wang}},\ }\href@noop {} {\bibfield  {journal} {\bibinfo  {journal} {arXiv preprint arXiv:2405.03639}\ } (\bibinfo {year} {2024})}\BibitemShut {NoStop}%
\bibitem [{\citenamefont {Sala}\ \emph {et~al.}(2024)\citenamefont {Sala}, \citenamefont {Gopalakrishnan}, \citenamefont {Oshikawa},\ and\ \citenamefont {You}}]{sala2024spontaneous}%
  \BibitemOpen
  \bibfield  {author} {\bibinfo {author} {\bibfnamefont {P.}~\bibnamefont {Sala}}, \bibinfo {author} {\bibfnamefont {S.}~\bibnamefont {Gopalakrishnan}}, \bibinfo {author} {\bibfnamefont {M.}~\bibnamefont {Oshikawa}},\ and\ \bibinfo {author} {\bibfnamefont {Y.}~\bibnamefont {You}},\ }\href@noop {} {\bibfield  {journal} {\bibinfo  {journal} {Physical Review B}\ }\textbf {\bibinfo {volume} {110}},\ \bibinfo {pages} {155150} (\bibinfo {year} {2024})}\BibitemShut {NoStop}%
\bibitem [{\citenamefont {Li}\ \emph {et~al.}(2023{\natexlab{a}})\citenamefont {Li}, \citenamefont {Zheng}, \citenamefont {Wang}, \citenamefont {Jiang}, \citenamefont {Liu},\ and\ \citenamefont {Liu}}]{li2023d}%
  \BibitemOpen
  \bibfield  {author} {\bibinfo {author} {\bibfnamefont {Z.}~\bibnamefont {Li}}, \bibinfo {author} {\bibfnamefont {H.}~\bibnamefont {Zheng}}, \bibinfo {author} {\bibfnamefont {Y.}~\bibnamefont {Wang}}, \bibinfo {author} {\bibfnamefont {L.}~\bibnamefont {Jiang}}, \bibinfo {author} {\bibfnamefont {Z.-W.}\ \bibnamefont {Liu}},\ and\ \bibinfo {author} {\bibfnamefont {J.}~\bibnamefont {Liu}},\ }\href@noop {} {\bibfield  {journal} {\bibinfo  {journal} {arXiv preprint arXiv:2309.16556}\ } (\bibinfo {year} {2023}{\natexlab{a}})}\BibitemShut {NoStop}%
\bibitem [{\citenamefont {Hearth}\ \emph {et~al.}(2023)\citenamefont {Hearth}, \citenamefont {Flynn}, \citenamefont {Chandran},\ and\ \citenamefont {Laumann}}]{hearth2023unitary}%
  \BibitemOpen
  \bibfield  {author} {\bibinfo {author} {\bibfnamefont {S.~N.}\ \bibnamefont {Hearth}}, \bibinfo {author} {\bibfnamefont {M.~O.}\ \bibnamefont {Flynn}}, \bibinfo {author} {\bibfnamefont {A.}~\bibnamefont {Chandran}},\ and\ \bibinfo {author} {\bibfnamefont {C.~R.}\ \bibnamefont {Laumann}},\ }\href@noop {} {\bibfield  {journal} {\bibinfo  {journal} {arXiv preprint arXiv:2306.01035}\ } (\bibinfo {year} {2023})}\BibitemShut {NoStop}%
\bibitem [{\citenamefont {Li}\ \emph {et~al.}(2023{\natexlab{b}})\citenamefont {Li}, \citenamefont {Zheng}, \citenamefont {Liu}, \citenamefont {Jiang},\ and\ \citenamefont {Liu}}]{li2023designs}%
  \BibitemOpen
  \bibfield  {author} {\bibinfo {author} {\bibfnamefont {Z.}~\bibnamefont {Li}}, \bibinfo {author} {\bibfnamefont {H.}~\bibnamefont {Zheng}}, \bibinfo {author} {\bibfnamefont {J.}~\bibnamefont {Liu}}, \bibinfo {author} {\bibfnamefont {L.}~\bibnamefont {Jiang}},\ and\ \bibinfo {author} {\bibfnamefont {Z.-W.}\ \bibnamefont {Liu}},\ }\href@noop {} {\bibfield  {journal} {\bibinfo  {journal} {arXiv preprint arXiv:2309.08155}\ } (\bibinfo {year} {2023}{\natexlab{b}})}\BibitemShut {NoStop}%
\bibitem [{\citenamefont {Zhang}\ \emph {et~al.}(2023)\citenamefont {Zhang}, \citenamefont {Allcock}, \citenamefont {Wan}, \citenamefont {Liu}, \citenamefont {Sun}, \citenamefont {Yu}, \citenamefont {Yang}, \citenamefont {Qiu}, \citenamefont {Ye}, \citenamefont {Chen} \emph {et~al.}}]{zhang2023tensorcircuit}%
  \BibitemOpen
  \bibfield  {author} {\bibinfo {author} {\bibfnamefont {S.-X.}\ \bibnamefont {Zhang}}, \bibinfo {author} {\bibfnamefont {J.}~\bibnamefont {Allcock}}, \bibinfo {author} {\bibfnamefont {Z.-Q.}\ \bibnamefont {Wan}}, \bibinfo {author} {\bibfnamefont {S.}~\bibnamefont {Liu}}, \bibinfo {author} {\bibfnamefont {J.}~\bibnamefont {Sun}}, \bibinfo {author} {\bibfnamefont {H.}~\bibnamefont {Yu}}, \bibinfo {author} {\bibfnamefont {X.-H.}\ \bibnamefont {Yang}}, \bibinfo {author} {\bibfnamefont {J.}~\bibnamefont {Qiu}}, \bibinfo {author} {\bibfnamefont {Z.}~\bibnamefont {Ye}}, \bibinfo {author} {\bibfnamefont {Y.-Q.}\ \bibnamefont {Chen}}, \emph {et~al.},\ }\href@noop {} {\bibfield  {journal} {\bibinfo  {journal} {Quantum}\ }\textbf {\bibinfo {volume} {7}},\ \bibinfo {pages} {912} (\bibinfo {year} {2023})}\BibitemShut {NoStop}%
\bibitem [{\citenamefont {Chen}\ \emph {et~al.}(2024)\citenamefont {Chen}, \citenamefont {Liu},\ and\ \citenamefont {Zhang}}]{chen2024subsystem}%
  \BibitemOpen
  \bibfield  {author} {\bibinfo {author} {\bibfnamefont {Y.-Q.}\ \bibnamefont {Chen}}, \bibinfo {author} {\bibfnamefont {S.}~\bibnamefont {Liu}},\ and\ \bibinfo {author} {\bibfnamefont {S.-X.}\ \bibnamefont {Zhang}},\ }\href@noop {} {\bibfield  {journal} {\bibinfo  {journal} {arXiv preprint arXiv:2405.05076}\ } (\bibinfo {year} {2024})}\BibitemShut {NoStop}%
\bibitem [{\citenamefont {Hayden}\ and\ \citenamefont {Preskill}(2007)}]{hayden2007black}%
  \BibitemOpen
  \bibfield  {author} {\bibinfo {author} {\bibfnamefont {P.}~\bibnamefont {Hayden}}\ and\ \bibinfo {author} {\bibfnamefont {J.}~\bibnamefont {Preskill}},\ }\href@noop {} {\bibfield  {journal} {\bibinfo  {journal} {Journal of high energy physics}\ }\textbf {\bibinfo {volume} {2007}},\ \bibinfo {pages} {120} (\bibinfo {year} {2007})}\BibitemShut {NoStop}%
\bibitem [{\citenamefont {Sekino}\ and\ \citenamefont {Susskind}(2008)}]{sekino2008fast}%
  \BibitemOpen
  \bibfield  {author} {\bibinfo {author} {\bibfnamefont {Y.}~\bibnamefont {Sekino}}\ and\ \bibinfo {author} {\bibfnamefont {L.}~\bibnamefont {Susskind}},\ }\href@noop {} {\bibfield  {journal} {\bibinfo  {journal} {Journal of High Energy Physics}\ }\textbf {\bibinfo {volume} {2008}},\ \bibinfo {pages} {065} (\bibinfo {year} {2008})}\BibitemShut {NoStop}%
\bibitem [{\citenamefont {Lashkari}\ \emph {et~al.}(2013)\citenamefont {Lashkari}, \citenamefont {Stanford}, \citenamefont {Hastings}, \citenamefont {Osborne},\ and\ \citenamefont {Hayden}}]{lashkari2013towards}%
  \BibitemOpen
  \bibfield  {author} {\bibinfo {author} {\bibfnamefont {N.}~\bibnamefont {Lashkari}}, \bibinfo {author} {\bibfnamefont {D.}~\bibnamefont {Stanford}}, \bibinfo {author} {\bibfnamefont {M.}~\bibnamefont {Hastings}}, \bibinfo {author} {\bibfnamefont {T.}~\bibnamefont {Osborne}},\ and\ \bibinfo {author} {\bibfnamefont {P.}~\bibnamefont {Hayden}},\ }\href@noop {} {\bibfield  {journal} {\bibinfo  {journal} {Journal of High Energy Physics}\ }\textbf {\bibinfo {volume} {2013}},\ \bibinfo {pages} {1} (\bibinfo {year} {2013})}\BibitemShut {NoStop}%
\bibitem [{\citenamefont {Hosur}\ \emph {et~al.}(2016)\citenamefont {Hosur}, \citenamefont {Qi}, \citenamefont {Roberts},\ and\ \citenamefont {Yoshida}}]{hosur2016chaos}%
  \BibitemOpen
  \bibfield  {author} {\bibinfo {author} {\bibfnamefont {P.}~\bibnamefont {Hosur}}, \bibinfo {author} {\bibfnamefont {X.-L.}\ \bibnamefont {Qi}}, \bibinfo {author} {\bibfnamefont {D.~A.}\ \bibnamefont {Roberts}},\ and\ \bibinfo {author} {\bibfnamefont {B.}~\bibnamefont {Yoshida}},\ }\href@noop {} {\bibfield  {journal} {\bibinfo  {journal} {Journal of High Energy Physics}\ }\textbf {\bibinfo {volume} {2016}},\ \bibinfo {pages} {1} (\bibinfo {year} {2016})}\BibitemShut {NoStop}%
\bibitem [{\citenamefont {Skinner}\ \emph {et~al.}(2019)\citenamefont {Skinner}, \citenamefont {Ruhman},\ and\ \citenamefont {Nahum}}]{skinner2019measurement}%
  \BibitemOpen
  \bibfield  {author} {\bibinfo {author} {\bibfnamefont {B.}~\bibnamefont {Skinner}}, \bibinfo {author} {\bibfnamefont {J.}~\bibnamefont {Ruhman}},\ and\ \bibinfo {author} {\bibfnamefont {A.}~\bibnamefont {Nahum}},\ }\href@noop {} {\bibfield  {journal} {\bibinfo  {journal} {Physical Review X}\ }\textbf {\bibinfo {volume} {9}},\ \bibinfo {pages} {031009} (\bibinfo {year} {2019})}\BibitemShut {NoStop}%
\bibitem [{\citenamefont {Choi}\ \emph {et~al.}(2020)\citenamefont {Choi}, \citenamefont {Bao}, \citenamefont {Qi},\ and\ \citenamefont {Altman}}]{choi2020quantum}%
  \BibitemOpen
  \bibfield  {author} {\bibinfo {author} {\bibfnamefont {S.}~\bibnamefont {Choi}}, \bibinfo {author} {\bibfnamefont {Y.}~\bibnamefont {Bao}}, \bibinfo {author} {\bibfnamefont {X.-L.}\ \bibnamefont {Qi}},\ and\ \bibinfo {author} {\bibfnamefont {E.}~\bibnamefont {Altman}},\ }\href@noop {} {\bibfield  {journal} {\bibinfo  {journal} {Physical Review Letters}\ }\textbf {\bibinfo {volume} {125}},\ \bibinfo {pages} {030505} (\bibinfo {year} {2020})}\BibitemShut {NoStop}%
\bibitem [{\citenamefont {Li}\ \emph {et~al.}(2019)\citenamefont {Li}, \citenamefont {Chen},\ and\ \citenamefont {Fisher}}]{li2019measurement}%
  \BibitemOpen
  \bibfield  {author} {\bibinfo {author} {\bibfnamefont {Y.}~\bibnamefont {Li}}, \bibinfo {author} {\bibfnamefont {X.}~\bibnamefont {Chen}},\ and\ \bibinfo {author} {\bibfnamefont {M.~P.}\ \bibnamefont {Fisher}},\ }\href@noop {} {\bibfield  {journal} {\bibinfo  {journal} {Physical Review B}\ }\textbf {\bibinfo {volume} {100}},\ \bibinfo {pages} {134306} (\bibinfo {year} {2019})}\BibitemShut {NoStop}%
\bibitem [{\citenamefont {Jian}\ \emph {et~al.}(2021)\citenamefont {Jian}, \citenamefont {Liu}, \citenamefont {Chen}, \citenamefont {Swingle},\ and\ \citenamefont {Zhang}}]{jian2021measurement}%
  \BibitemOpen
  \bibfield  {author} {\bibinfo {author} {\bibfnamefont {S.-K.}\ \bibnamefont {Jian}}, \bibinfo {author} {\bibfnamefont {C.}~\bibnamefont {Liu}}, \bibinfo {author} {\bibfnamefont {X.}~\bibnamefont {Chen}}, \bibinfo {author} {\bibfnamefont {B.}~\bibnamefont {Swingle}},\ and\ \bibinfo {author} {\bibfnamefont {P.}~\bibnamefont {Zhang}},\ }\href@noop {} {\bibfield  {journal} {\bibinfo  {journal} {Physical review letters}\ }\textbf {\bibinfo {volume} {127}},\ \bibinfo {pages} {140601} (\bibinfo {year} {2021})}\BibitemShut {NoStop}%
\bibitem [{\citenamefont {Minato}\ \emph {et~al.}(2022)\citenamefont {Minato}, \citenamefont {Sugimoto}, \citenamefont {Kuwahara},\ and\ \citenamefont {Saito}}]{minato2022fate}%
  \BibitemOpen
  \bibfield  {author} {\bibinfo {author} {\bibfnamefont {T.}~\bibnamefont {Minato}}, \bibinfo {author} {\bibfnamefont {K.}~\bibnamefont {Sugimoto}}, \bibinfo {author} {\bibfnamefont {T.}~\bibnamefont {Kuwahara}},\ and\ \bibinfo {author} {\bibfnamefont {K.}~\bibnamefont {Saito}},\ }\href@noop {} {\bibfield  {journal} {\bibinfo  {journal} {Physical review letters}\ }\textbf {\bibinfo {volume} {128}},\ \bibinfo {pages} {010603} (\bibinfo {year} {2022})}\BibitemShut {NoStop}%
\bibitem [{\citenamefont {Dhar}\ and\ \citenamefont {Dasgupta}(2016)}]{dhar2016measurement}%
  \BibitemOpen
  \bibfield  {author} {\bibinfo {author} {\bibfnamefont {S.}~\bibnamefont {Dhar}}\ and\ \bibinfo {author} {\bibfnamefont {S.}~\bibnamefont {Dasgupta}},\ }\href@noop {} {\bibfield  {journal} {\bibinfo  {journal} {Physical Review A}\ }\textbf {\bibinfo {volume} {93}},\ \bibinfo {pages} {050103} (\bibinfo {year} {2016})}\BibitemShut {NoStop}%
\bibitem [{\citenamefont {Vijay}(2020)}]{vijay2020measurement}%
  \BibitemOpen
  \bibfield  {author} {\bibinfo {author} {\bibfnamefont {S.}~\bibnamefont {Vijay}},\ }\href@noop {} {\bibfield  {journal} {\bibinfo  {journal} {arXiv preprint arXiv:2005.03052}\ } (\bibinfo {year} {2020})}\BibitemShut {NoStop}%
\bibitem [{\citenamefont {Bao}\ \emph {et~al.}(2020)\citenamefont {Bao}, \citenamefont {Choi},\ and\ \citenamefont {Altman}}]{bao2020theory}%
  \BibitemOpen
  \bibfield  {author} {\bibinfo {author} {\bibfnamefont {Y.}~\bibnamefont {Bao}}, \bibinfo {author} {\bibfnamefont {S.}~\bibnamefont {Choi}},\ and\ \bibinfo {author} {\bibfnamefont {E.}~\bibnamefont {Altman}},\ }\href@noop {} {\bibfield  {journal} {\bibinfo  {journal} {Physical Review B}\ }\textbf {\bibinfo {volume} {101}},\ \bibinfo {pages} {104301} (\bibinfo {year} {2020})}\BibitemShut {NoStop}%
\bibitem [{\citenamefont {Gullans}\ and\ \citenamefont {Huse}(2020)}]{gullans2020dynamical}%
  \BibitemOpen
  \bibfield  {author} {\bibinfo {author} {\bibfnamefont {M.~J.}\ \bibnamefont {Gullans}}\ and\ \bibinfo {author} {\bibfnamefont {D.~A.}\ \bibnamefont {Huse}},\ }\href@noop {} {\bibfield  {journal} {\bibinfo  {journal} {Physical Review X}\ }\textbf {\bibinfo {volume} {10}},\ \bibinfo {pages} {041020} (\bibinfo {year} {2020})}\BibitemShut {NoStop}%
\bibitem [{\citenamefont {Zabalo}\ \emph {et~al.}(2020)\citenamefont {Zabalo}, \citenamefont {Gullans}, \citenamefont {Wilson}, \citenamefont {Gopalakrishnan}, \citenamefont {Huse},\ and\ \citenamefont {Pixley}}]{zabalo2020critical}%
  \BibitemOpen
  \bibfield  {author} {\bibinfo {author} {\bibfnamefont {A.}~\bibnamefont {Zabalo}}, \bibinfo {author} {\bibfnamefont {M.~J.}\ \bibnamefont {Gullans}}, \bibinfo {author} {\bibfnamefont {J.~H.}\ \bibnamefont {Wilson}}, \bibinfo {author} {\bibfnamefont {S.}~\bibnamefont {Gopalakrishnan}}, \bibinfo {author} {\bibfnamefont {D.~A.}\ \bibnamefont {Huse}},\ and\ \bibinfo {author} {\bibfnamefont {J.}~\bibnamefont {Pixley}},\ }\href@noop {} {\bibfield  {journal} {\bibinfo  {journal} {Physical Review B}\ }\textbf {\bibinfo {volume} {101}},\ \bibinfo {pages} {060301} (\bibinfo {year} {2020})}\BibitemShut {NoStop}%
\bibitem [{\citenamefont {Ippoliti}\ \emph {et~al.}(2021)\citenamefont {Ippoliti}, \citenamefont {Gullans}, \citenamefont {Gopalakrishnan}, \citenamefont {Huse},\ and\ \citenamefont {Khemani}}]{ippoliti2021entanglement}%
  \BibitemOpen
  \bibfield  {author} {\bibinfo {author} {\bibfnamefont {M.}~\bibnamefont {Ippoliti}}, \bibinfo {author} {\bibfnamefont {M.~J.}\ \bibnamefont {Gullans}}, \bibinfo {author} {\bibfnamefont {S.}~\bibnamefont {Gopalakrishnan}}, \bibinfo {author} {\bibfnamefont {D.~A.}\ \bibnamefont {Huse}},\ and\ \bibinfo {author} {\bibfnamefont {V.}~\bibnamefont {Khemani}},\ }\href@noop {} {\bibfield  {journal} {\bibinfo  {journal} {Physical Review X}\ }\textbf {\bibinfo {volume} {11}},\ \bibinfo {pages} {011030} (\bibinfo {year} {2021})}\BibitemShut {NoStop}%
\bibitem [{\citenamefont {Szyniszewski}\ \emph {et~al.}(2019)\citenamefont {Szyniszewski}, \citenamefont {Romito},\ and\ \citenamefont {Schomerus}}]{szyniszewski2019entanglement}%
  \BibitemOpen
  \bibfield  {author} {\bibinfo {author} {\bibfnamefont {M.}~\bibnamefont {Szyniszewski}}, \bibinfo {author} {\bibfnamefont {A.}~\bibnamefont {Romito}},\ and\ \bibinfo {author} {\bibfnamefont {H.}~\bibnamefont {Schomerus}},\ }\href@noop {} {\bibfield  {journal} {\bibinfo  {journal} {Physical Review B}\ }\textbf {\bibinfo {volume} {100}},\ \bibinfo {pages} {064204} (\bibinfo {year} {2019})}\BibitemShut {NoStop}%
\bibitem [{\citenamefont {Jian}\ \emph {et~al.}(2020)\citenamefont {Jian}, \citenamefont {You}, \citenamefont {Vasseur},\ and\ \citenamefont {Ludwig}}]{jian2020measurement}%
  \BibitemOpen
  \bibfield  {author} {\bibinfo {author} {\bibfnamefont {C.-M.}\ \bibnamefont {Jian}}, \bibinfo {author} {\bibfnamefont {Y.-Z.}\ \bibnamefont {You}}, \bibinfo {author} {\bibfnamefont {R.}~\bibnamefont {Vasseur}},\ and\ \bibinfo {author} {\bibfnamefont {A.~W.}\ \bibnamefont {Ludwig}},\ }\href@noop {} {\bibfield  {journal} {\bibinfo  {journal} {Physical Review B}\ }\textbf {\bibinfo {volume} {101}},\ \bibinfo {pages} {104302} (\bibinfo {year} {2020})}\BibitemShut {NoStop}%
\bibitem [{\citenamefont {Liu}\ \emph {et~al.}(2024{\natexlab{c}})\citenamefont {Liu}, \citenamefont {Li}, \citenamefont {Zhang},\ and\ \citenamefont {Jian}}]{liu2024entanglement}%
  \BibitemOpen
  \bibfield  {author} {\bibinfo {author} {\bibfnamefont {S.}~\bibnamefont {Liu}}, \bibinfo {author} {\bibfnamefont {M.-R.}\ \bibnamefont {Li}}, \bibinfo {author} {\bibfnamefont {S.-X.}\ \bibnamefont {Zhang}},\ and\ \bibinfo {author} {\bibfnamefont {S.-K.}\ \bibnamefont {Jian}},\ }\href@noop {} {\bibfield  {journal} {\bibinfo  {journal} {Physical Review Letters}\ }\textbf {\bibinfo {volume} {132}},\ \bibinfo {pages} {240402} (\bibinfo {year} {2024}{\natexlab{c}})}\BibitemShut {NoStop}%
\bibitem [{\citenamefont {Liu}\ \emph {et~al.}(2024{\natexlab{d}})\citenamefont {Liu}, \citenamefont {Li}, \citenamefont {Zhang}, \citenamefont {Jian},\ and\ \citenamefont {Yao}}]{liu2024noise}%
  \BibitemOpen
  \bibfield  {author} {\bibinfo {author} {\bibfnamefont {S.}~\bibnamefont {Liu}}, \bibinfo {author} {\bibfnamefont {M.-R.}\ \bibnamefont {Li}}, \bibinfo {author} {\bibfnamefont {S.-X.}\ \bibnamefont {Zhang}}, \bibinfo {author} {\bibfnamefont {S.-K.}\ \bibnamefont {Jian}},\ and\ \bibinfo {author} {\bibfnamefont {H.}~\bibnamefont {Yao}},\ }\href@noop {} {\bibfield  {journal} {\bibinfo  {journal} {Physical Review B}\ }\textbf {\bibinfo {volume} {110}},\ \bibinfo {pages} {064323} (\bibinfo {year} {2024}{\natexlab{d}})}\BibitemShut {NoStop}%
\bibitem [{\citenamefont {Liu}\ \emph {et~al.}(2023)\citenamefont {Liu}, \citenamefont {Li}, \citenamefont {Zhang}, \citenamefont {Jian},\ and\ \citenamefont {Yao}}]{liu2023universal}%
  \BibitemOpen
  \bibfield  {author} {\bibinfo {author} {\bibfnamefont {S.}~\bibnamefont {Liu}}, \bibinfo {author} {\bibfnamefont {M.-R.}\ \bibnamefont {Li}}, \bibinfo {author} {\bibfnamefont {S.-X.}\ \bibnamefont {Zhang}}, \bibinfo {author} {\bibfnamefont {S.-K.}\ \bibnamefont {Jian}},\ and\ \bibinfo {author} {\bibfnamefont {H.}~\bibnamefont {Yao}},\ }\href@noop {} {\bibfield  {journal} {\bibinfo  {journal} {Physical Review B}\ }\textbf {\bibinfo {volume} {107}},\ \bibinfo {pages} {L201113} (\bibinfo {year} {2023})}\BibitemShut {NoStop}%
\bibitem [{\citenamefont {Alet}\ and\ \citenamefont {Laflorencie}(2018)}]{alet2018many}%
  \BibitemOpen
  \bibfield  {author} {\bibinfo {author} {\bibfnamefont {F.}~\bibnamefont {Alet}}\ and\ \bibinfo {author} {\bibfnamefont {N.}~\bibnamefont {Laflorencie}},\ }\href@noop {} {\bibfield  {journal} {\bibinfo  {journal} {Comptes Rendus Physique}\ }\textbf {\bibinfo {volume} {19}},\ \bibinfo {pages} {498} (\bibinfo {year} {2018})}\BibitemShut {NoStop}%
\bibitem [{\citenamefont {Abanin}\ \emph {et~al.}(2019)\citenamefont {Abanin}, \citenamefont {Altman}, \citenamefont {Bloch},\ and\ \citenamefont {Serbyn}}]{abanin2019colloquium}%
  \BibitemOpen
  \bibfield  {author} {\bibinfo {author} {\bibfnamefont {D.~A.}\ \bibnamefont {Abanin}}, \bibinfo {author} {\bibfnamefont {E.}~\bibnamefont {Altman}}, \bibinfo {author} {\bibfnamefont {I.}~\bibnamefont {Bloch}},\ and\ \bibinfo {author} {\bibfnamefont {M.}~\bibnamefont {Serbyn}},\ }\href@noop {} {\bibfield  {journal} {\bibinfo  {journal} {Reviews of Modern Physics}\ }\textbf {\bibinfo {volume} {91}},\ \bibinfo {pages} {021001} (\bibinfo {year} {2019})}\BibitemShut {NoStop}%
\bibitem [{\citenamefont {Pal}\ and\ \citenamefont {Huse}(2010)}]{pal2010many}%
  \BibitemOpen
  \bibfield  {author} {\bibinfo {author} {\bibfnamefont {A.}~\bibnamefont {Pal}}\ and\ \bibinfo {author} {\bibfnamefont {D.~A.}\ \bibnamefont {Huse}},\ }\href@noop {} {\bibfield  {journal} {\bibinfo  {journal} {Physical Review B—Condensed Matter and Materials Physics}\ }\textbf {\bibinfo {volume} {82}},\ \bibinfo {pages} {174411} (\bibinfo {year} {2010})}\BibitemShut {NoStop}%
\bibitem [{\citenamefont {Nandkishore}\ and\ \citenamefont {Huse}(2015)}]{nandkishore2015many}%
  \BibitemOpen
  \bibfield  {author} {\bibinfo {author} {\bibfnamefont {R.}~\bibnamefont {Nandkishore}}\ and\ \bibinfo {author} {\bibfnamefont {D.~A.}\ \bibnamefont {Huse}},\ }\href@noop {} {\bibfield  {journal} {\bibinfo  {journal} {Annu. Rev. Condens. Matter Phys.}\ }\textbf {\bibinfo {volume} {6}},\ \bibinfo {pages} {15} (\bibinfo {year} {2015})}\BibitemShut {NoStop}%
\bibitem [{\citenamefont {Imbrie}\ \emph {et~al.}(2017)\citenamefont {Imbrie}, \citenamefont {Ros},\ and\ \citenamefont {Scardicchio}}]{imbrie2017local}%
  \BibitemOpen
  \bibfield  {author} {\bibinfo {author} {\bibfnamefont {J.~Z.}\ \bibnamefont {Imbrie}}, \bibinfo {author} {\bibfnamefont {V.}~\bibnamefont {Ros}},\ and\ \bibinfo {author} {\bibfnamefont {A.}~\bibnamefont {Scardicchio}},\ }\href@noop {} {\bibfield  {journal} {\bibinfo  {journal} {Annalen der Physik}\ }\textbf {\bibinfo {volume} {529}},\ \bibinfo {pages} {1600278} (\bibinfo {year} {2017})}\BibitemShut {NoStop}%
\bibitem [{\citenamefont {Altman}\ and\ \citenamefont {Vosk}(2015)}]{altman2015universal}%
  \BibitemOpen
  \bibfield  {author} {\bibinfo {author} {\bibfnamefont {E.}~\bibnamefont {Altman}}\ and\ \bibinfo {author} {\bibfnamefont {R.}~\bibnamefont {Vosk}},\ }\href@noop {} {\bibfield  {journal} {\bibinfo  {journal} {Annu. Rev. Condens. Matter Phys.}\ }\textbf {\bibinfo {volume} {6}},\ \bibinfo {pages} {383} (\bibinfo {year} {2015})}\BibitemShut {NoStop}%
\bibitem [{\citenamefont {Huse}\ \emph {et~al.}(2014)\citenamefont {Huse}, \citenamefont {Nandkishore},\ and\ \citenamefont {Oganesyan}}]{huse2014phenomenology}%
  \BibitemOpen
  \bibfield  {author} {\bibinfo {author} {\bibfnamefont {D.~A.}\ \bibnamefont {Huse}}, \bibinfo {author} {\bibfnamefont {R.}~\bibnamefont {Nandkishore}},\ and\ \bibinfo {author} {\bibfnamefont {V.}~\bibnamefont {Oganesyan}},\ }\href@noop {} {\bibfield  {journal} {\bibinfo  {journal} {Physical Review B}\ }\textbf {\bibinfo {volume} {90}},\ \bibinfo {pages} {174202} (\bibinfo {year} {2014})}\BibitemShut {NoStop}%
\bibitem [{\citenamefont {Lukin}\ \emph {et~al.}(2019)\citenamefont {Lukin}, \citenamefont {Rispoli}, \citenamefont {Schittko}, \citenamefont {Tai}, \citenamefont {Kaufman}, \citenamefont {Choi}, \citenamefont {Khemani}, \citenamefont {L{\'e}onard},\ and\ \citenamefont {Greiner}}]{lukin2019probing}%
  \BibitemOpen
  \bibfield  {author} {\bibinfo {author} {\bibfnamefont {A.}~\bibnamefont {Lukin}}, \bibinfo {author} {\bibfnamefont {M.}~\bibnamefont {Rispoli}}, \bibinfo {author} {\bibfnamefont {R.}~\bibnamefont {Schittko}}, \bibinfo {author} {\bibfnamefont {M.~E.}\ \bibnamefont {Tai}}, \bibinfo {author} {\bibfnamefont {A.~M.}\ \bibnamefont {Kaufman}}, \bibinfo {author} {\bibfnamefont {S.}~\bibnamefont {Choi}}, \bibinfo {author} {\bibfnamefont {V.}~\bibnamefont {Khemani}}, \bibinfo {author} {\bibfnamefont {J.}~\bibnamefont {L{\'e}onard}},\ and\ \bibinfo {author} {\bibfnamefont {M.}~\bibnamefont {Greiner}},\ }\href@noop {} {\bibfield  {journal} {\bibinfo  {journal} {Science}\ }\textbf {\bibinfo {volume} {364}},\ \bibinfo {pages} {256} (\bibinfo {year} {2019})}\BibitemShut {NoStop}%
\bibitem [{\citenamefont {Morningstar}\ \emph {et~al.}(2022)\citenamefont {Morningstar}, \citenamefont {Colmenarez}, \citenamefont {Khemani}, \citenamefont {Luitz},\ and\ \citenamefont {Huse}}]{morningstar2022avalanches}%
  \BibitemOpen
  \bibfield  {author} {\bibinfo {author} {\bibfnamefont {A.}~\bibnamefont {Morningstar}}, \bibinfo {author} {\bibfnamefont {L.}~\bibnamefont {Colmenarez}}, \bibinfo {author} {\bibfnamefont {V.}~\bibnamefont {Khemani}}, \bibinfo {author} {\bibfnamefont {D.~J.}\ \bibnamefont {Luitz}},\ and\ \bibinfo {author} {\bibfnamefont {D.~A.}\ \bibnamefont {Huse}},\ }\href@noop {} {\bibfield  {journal} {\bibinfo  {journal} {Physical Review B}\ }\textbf {\bibinfo {volume} {105}},\ \bibinfo {pages} {174205} (\bibinfo {year} {2022})}\BibitemShut {NoStop}%
\bibitem [{\citenamefont {Yu}\ \emph {et~al.}()\citenamefont {Yu}, \citenamefont {Li},\ and\ \citenamefont {Zhang}}]{data-availbale}%
  \BibitemOpen
  \bibfield  {author} {\bibinfo {author} {\bibfnamefont {H.}~\bibnamefont {Yu}}, \bibinfo {author} {\bibfnamefont {Z.-X.}\ \bibnamefont {Li}},\ and\ \bibinfo {author} {\bibfnamefont {S.-X.}\ \bibnamefont {Zhang}},\ }\href@noop {} {\bibinfo  {journal} {Data available in zenodo for publication}\ }\BibitemShut {NoStop}%
\end{thebibliography}

%


\end{document}


\begin{center}
\textbf{\large Supplemental Material for ``Symmetry Breaking Dynamics in Quantum Many-Body Systems''}
\end{center}

\author{Hui Yu}
\affiliation{Beijing National Laboratory for Condensed Matter Physics and Institute of Physics, Chinese Academy of Sciences, Beijing 100910, China}

\author{Zi-Xiang Li}
\email{zixiangli@iphy.ac.cn}

\affiliation{Beijing National Laboratory for Condensed Matter Physics and Institute of Physics, Chinese Academy of Sciences, Beijing 100910, China}
\affiliation{University of Chinese Academy of Sciences, Beijing 100049, China}

\author{Shi-Xin Zhang}
\email{shixinzhang@iphy.ac.cn}

\affiliation{Beijing National Laboratory for Condensed Matter Physics and Institute of Physics, Chinese Academy of Sciences, Beijing 100910, China}
\maketitle

\tableofcontents
\section{More numerical results for U(1) non-symmetric quantum circuits}

\subsection{Dynamics of EA with other U(1)-symmetric initial states}

\begin{figure}[ht]
\centering
\includegraphics[width=0.7\textwidth, keepaspectratio]{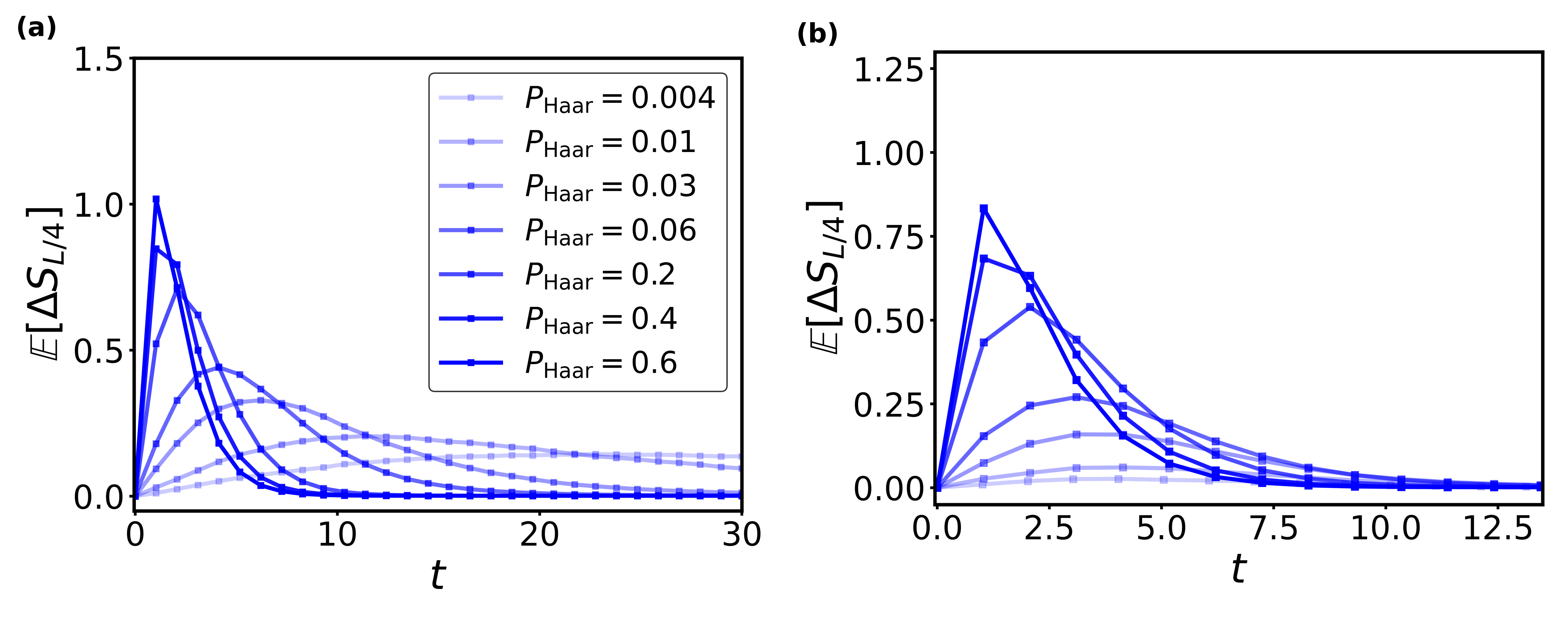}
\caption{The dynamics of  circuit-averaged EA, $\mathbb{E} [\Delta S_{L/4}]$ for different initial states with $L=16$. The varying color intensity represents different densities of random Haar gates. Left: Ferromagnetic state. Right: Domain Wall state.}
\label{fig:smcircuit_one}
\end{figure}

In this section, we present numerical results of the dynamics of entanglement asymmetry with different initial states, using the same setup as described in the main text. Similar to the behavior observed in the main text, EA for the other initial states in Fig. \ref{fig:smcircuit_one} also exhibits an initial peak, which then gradually decays to zero, indicating the restoration of U(1) symmetry. It is most apparent from the case of ferromagnetic states that a smaller density of random Haar gates results in a longer time of symmetry restoration in the subsystem. In addition, the thermalization process occurs more rapidly in domain wall and antiferromagnetic states than ferromagnetic states. This is because, in the limit of symmetry-preserving circuits, the Hilbert space accessible to the antiferromagnetic states is much larger than the ferromagnetic cases.

\subsection{Dynamics of EA with U(1)-asymmetric initial states at different $P_{\text{Haar}}$}
\begin{figure}[ht]
\centering
\includegraphics[width=0.62\textwidth, keepaspectratio]{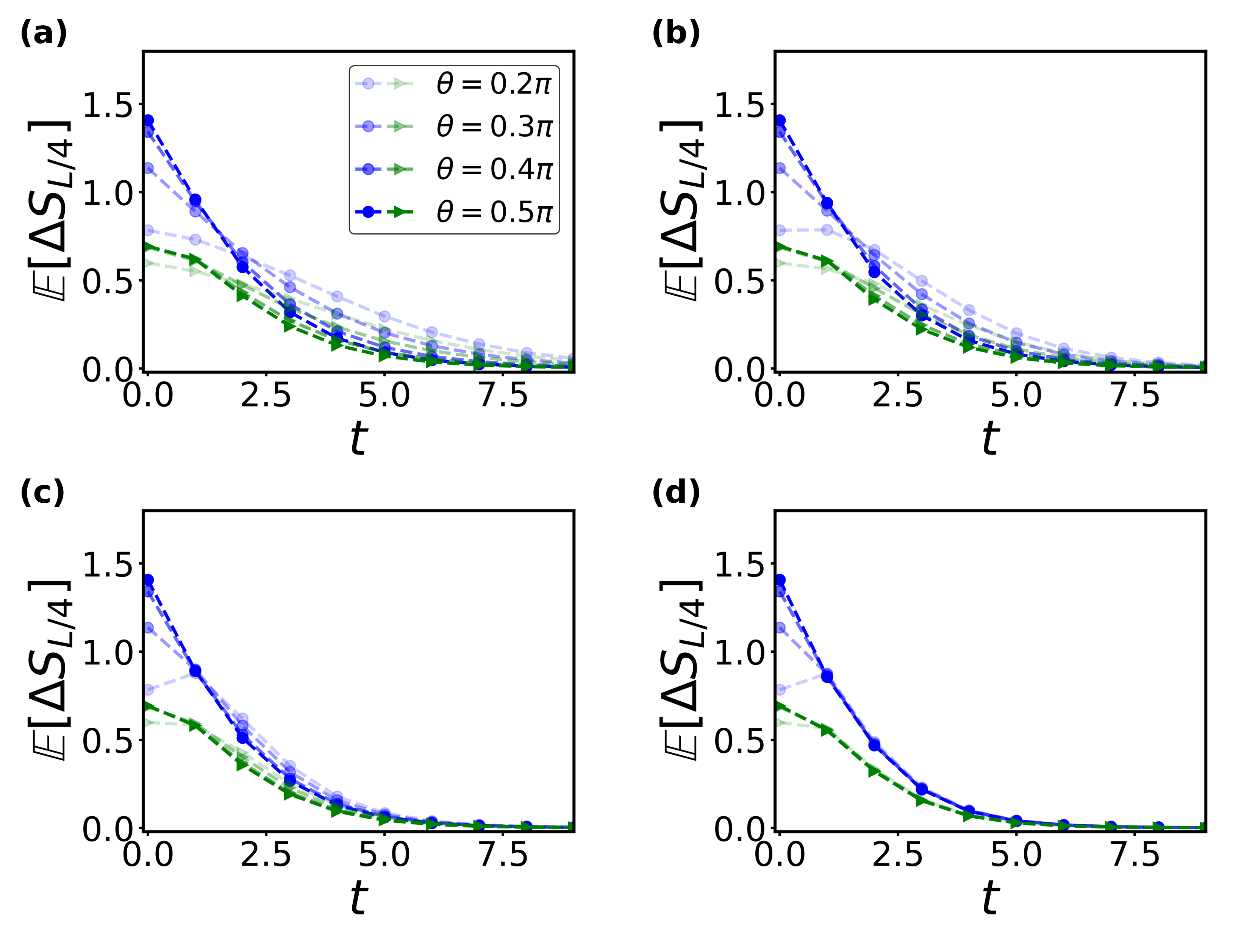}
\caption{The dynamics of circuit-averaged EA, $\mathbb{E} [\Delta S_{L/4}]$, for tilted ferromagnetic initial states is examined for various values of $P_{\text{Haar}}$ with $L=16$. Blue: U(1) EA. Green:  $Z_{2}$ EA. Panels (a)-(d) correspond to the following values of $P_{\text{Haar}}$ (a) $P_{\text{Haar}}=0.1$, (b) $P_{\text{Haar}}=0.2$, (c) $P_{\text{Haar}}=0.5$, and (d) $P_{\text{Haar}}=0.9$.}
\label{fig:smcircuit_two}
\end{figure}

\begin{figure}[ht]
\centering
\includegraphics[width=0.62\textwidth, keepaspectratio]{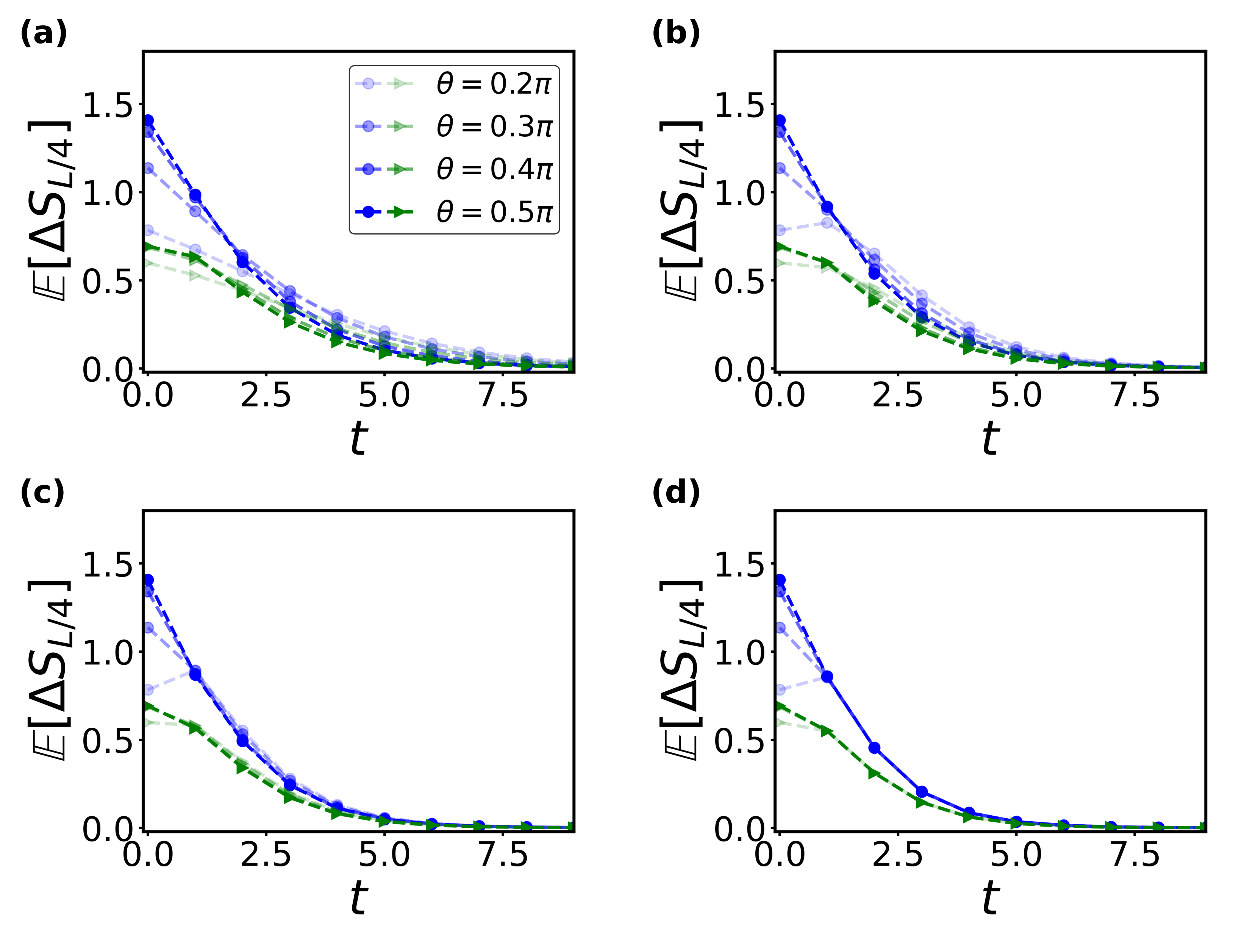}
\caption{The dynamics of circuit-averaged EA, $\mathbb{E} [\Delta S_{L/4}]$, for tilted domain wall initial states is examined for various values of $P_{\text{Haar}}$ with $L=16$. Blue: U(1) EA. Green:  $Z_{2}$ EA. Panels (a)-(d) correspond to the following values of $P_{\text{Haar}}$ (a) $P_{\text{Haar}}=0$, (b) $P_{\text{Haar}}=0.3$, (c) $P_{\text{Haar}}=0.7$, and (d) $P_{\text{Haar}}=1$. }
\label{fig:smcircuit_three}
\end{figure}

\begin{figure}[ht]
\centering
\includegraphics[width=0.62\textwidth, keepaspectratio]{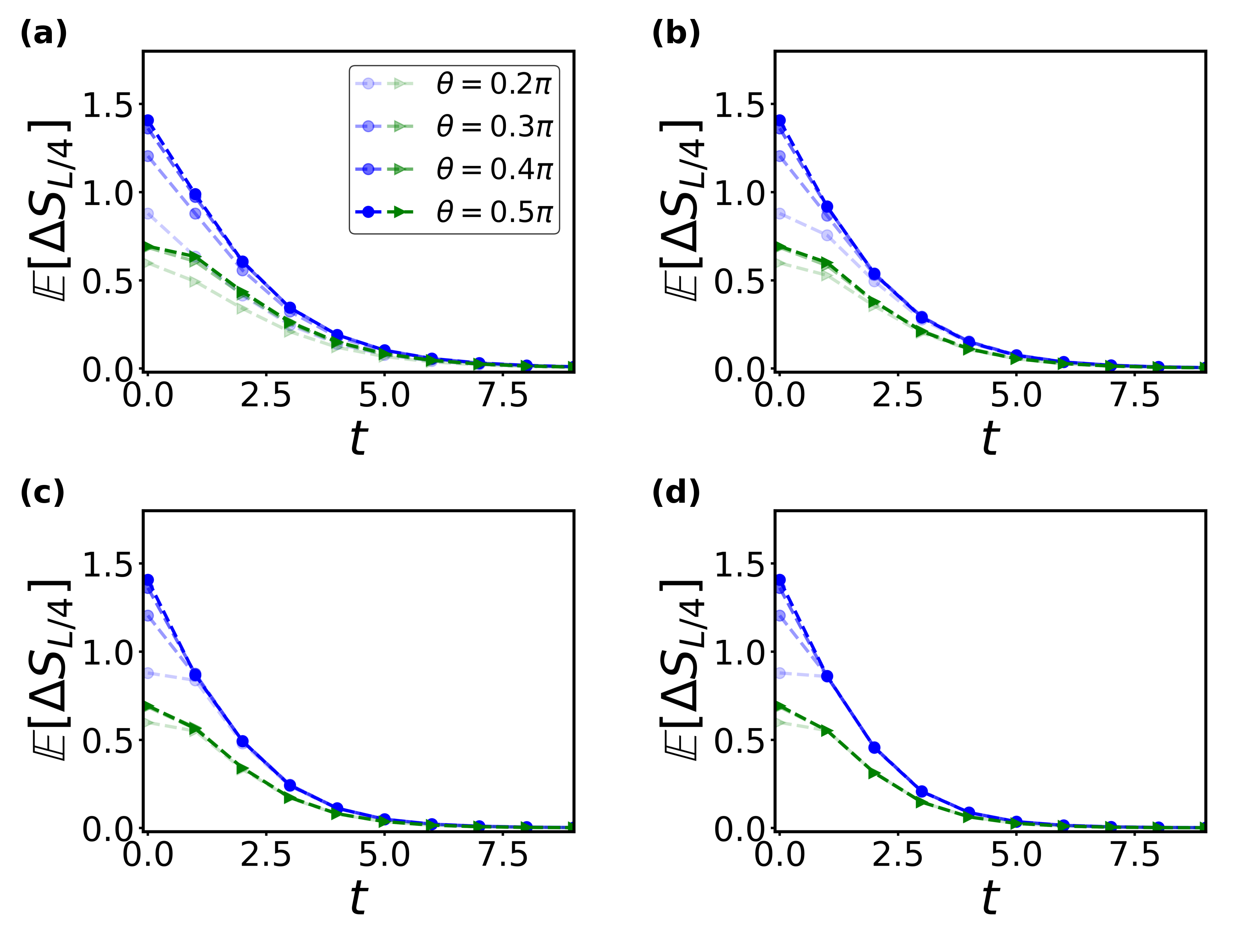}
\caption{The dynamics of circuit-averaged EA, $\mathbb{E} [\Delta S_{L/4}]$, for tilted antiferromagnetic initial states is examined for various values of $P_{\text{Haar}}$ with $L=16$. Blue: U(1) EA. Green:  $Z_{2}$ EA. Panels (a)-(d) correspond to the following values of $P_{\text{Haar}}$ (a) $P_{\text{Haar}}=0$, (b) $P_{\text{Haar}}=0.3$, (c) $P_{\text{Haar}}=0.7$, and (d) $P_{\text{Haar}}=1$.}
\label{fig:smcircuit_four}
\end{figure}

In addition to the $P_{\text{Haar}}$ values used in the main text, we show the behavior of entanglement asymmetry with tilted ferromagnetic state for other $P_{\text{Haar}}$ values in Fig. \ref{fig:smcircuit_two}, and find that QME is indeed present except when the symmetry breaking reaches the maximum $P_{\text{Haar}}=1$. This phenomenon also holds true for tilted domain wall state, as demonstrated in Fig. \ref{fig:smcircuit_three}. On the contrary, as shown in Fig. \ref{fig:smcircuit_four}, QME is absent for tilted antiferromagnetic states across all values of $P_{\text{Haar}}$. Therefore, the emergence of QME is specific to certain initial states and is robust against the effects of random Haar gates on the circuit evolution.

\subsection{Dynamics of CV with U(1)-symmetric(asymmetric) initial states at different $P_{\text{Haar}}$}
\begin{figure}[ht]
\centering
\includegraphics[width=1.02\textwidth, keepaspectratio]{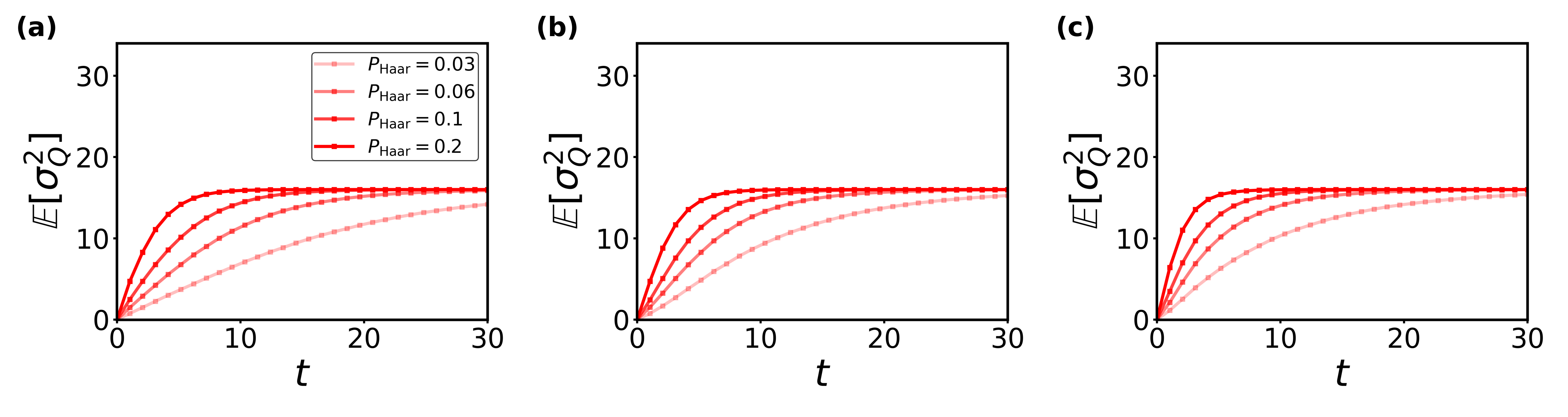}
\caption{The dynamics of circuit-averaged CV, $\mathbb{E} [\sigma^{2}_{Q}]$, for different U(1)-symmetric initial states are examined for various values of $P_{\text{Haar}}$ with $L=16$. From left to right: the initial states are ferromagnetic, domain wall and  antiferromagnetic states, respectively.}
\label{fig:circuitCV_untilt}
\end{figure}

\begin{figure}[ht]
\centering
\includegraphics[width=0.7\textwidth, keepaspectratio]{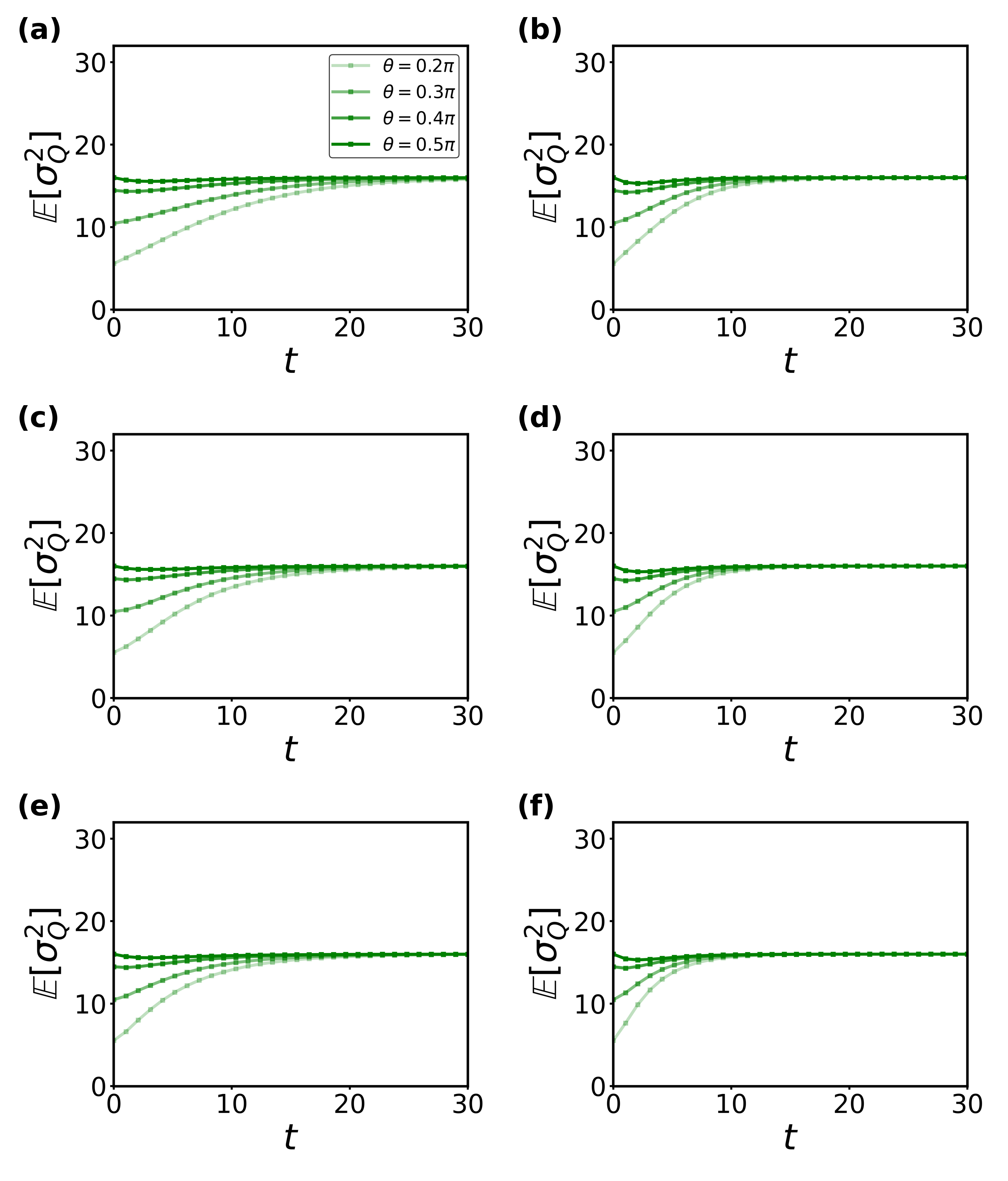}
\caption{Dynamics of circuit averaged CV, $\mathbb{E} [\sigma^{2}_{Q}]$, for different U(1)-asymmetric initial states in a system with $L=16$. The top, middle, and bottom rows correspond to tilted ferromagnetic, tilted domain wall, and tilted antiferromagnetic initial states, respectively. The left and columns show results for $P_{\text{Haar}}=0.05$ and $P_{\text{Haar}}=0.1$, respectively.}
\label{fig:circuitCV_tilt}
\end{figure}

As shown in Fig. \ref{fig:circuitCV_untilt}, the circuit-averaged charge variance $\mathbb{E} [\sigma^{2}_{Q}]$ for all U(1)-symmetric initial states gradually increases until it reaches a plateau. This behavior contrasts with the overshooting dynamics observed in the entanglement asymmetry. Besides, the early-time dynamics of charge variance from different asymmetric initial states show no evidence of QME, as confirmed in Fig. \ref{fig:circuitCV_tilt}. Two observations can be drawn from Figs. \ref{fig:circuitCV_untilt} and \ref{fig:circuitCV_tilt}: (1) The time required to reach the plateau decreases as the density of random Haar gates increases. Furthermore, the charge variance saturates more quickly for the antiferromagnetic state compared to the ferromagnetic state. (2) The charge variance saturates to the same value for all U(1)-symmetric(asymmetric) initial states. This arises because the late-time density matrix for a small subsystem approaches an identity matrix. Here, $\mathbb{E} [\sigma^{2}_{Q}]$ is computed by averaging the charge variance, $\sigma^{2}_{Q}$, over $5000$ circuit configurations.

\subsection{The early and late time behavior of EA and CV in U(1) non-symmetric quantum circuits}
In Table. \ref{table:circuit}, we provide a summary of the behavior of entanglement asymmetry and charge variance at early and late times for U(1)-asymmetric states in the circuit model. We observe that the quantum Mpemba effect occurs only in the early-time dynamics of entanglement asymmetry for tilted ferromagnetic and tilted domain wall states. Moreover, the entanglement asymmetry and charge variance approach distinct values, which are universal across all initial states, at late times.

\begin{table}[ht]
\centering
\resizebox{0.7\textwidth}{!}{ 
\begin{tabular}{cccc} 
\toprule
& \textbf{Ferromagnetic} & \textbf{Domain Wall} & \textbf{Antiferromagnetic} \\
\midrule
$EA\ (early\ time)$ & crossing  & crossing & no crossing \\
\hline
$CV\ (early\ time)$ & no crossing & no crossing & no crossing \\
\hline
$EA\ (late\ time)$ & 0 & 0 & 0 \\
\hline
$CV\ (late\ time)$ & C & C & C \\
\bottomrule
\end{tabular}
}
\caption{The early- and late-time behavior of entanglement asymmetry (EA) and charge variance (CV) under the evolution of random circuit with $0 \leq P_{\text{Haar}} < 1$. Crossing in EA (CV) means when the time-evolution curves of EA (CV) for states with larger $\theta$ intersect with those for smaller $\theta$ at early-times. The constant C can be evaluated as $\mathrm{tr}(\rho Q^{2})-\mathrm{tr}(\rho Q)^{2}$, where $\rho=\frac{I}{2^{L}}$ is the late-time density matrix for the subsystem. }
\label{table:circuit}
\end{table}

\vspace{25cm}

\section{More numerical results of entanglement asymmetry for U(1) non-symmetric Hamiltonians}
\subsection{Dynamics of entanglement asymmetry for various U(1)-symmetric initial states}
\begin{figure}[ht]
\centering
\includegraphics[width=0.8\textwidth, keepaspectratio]{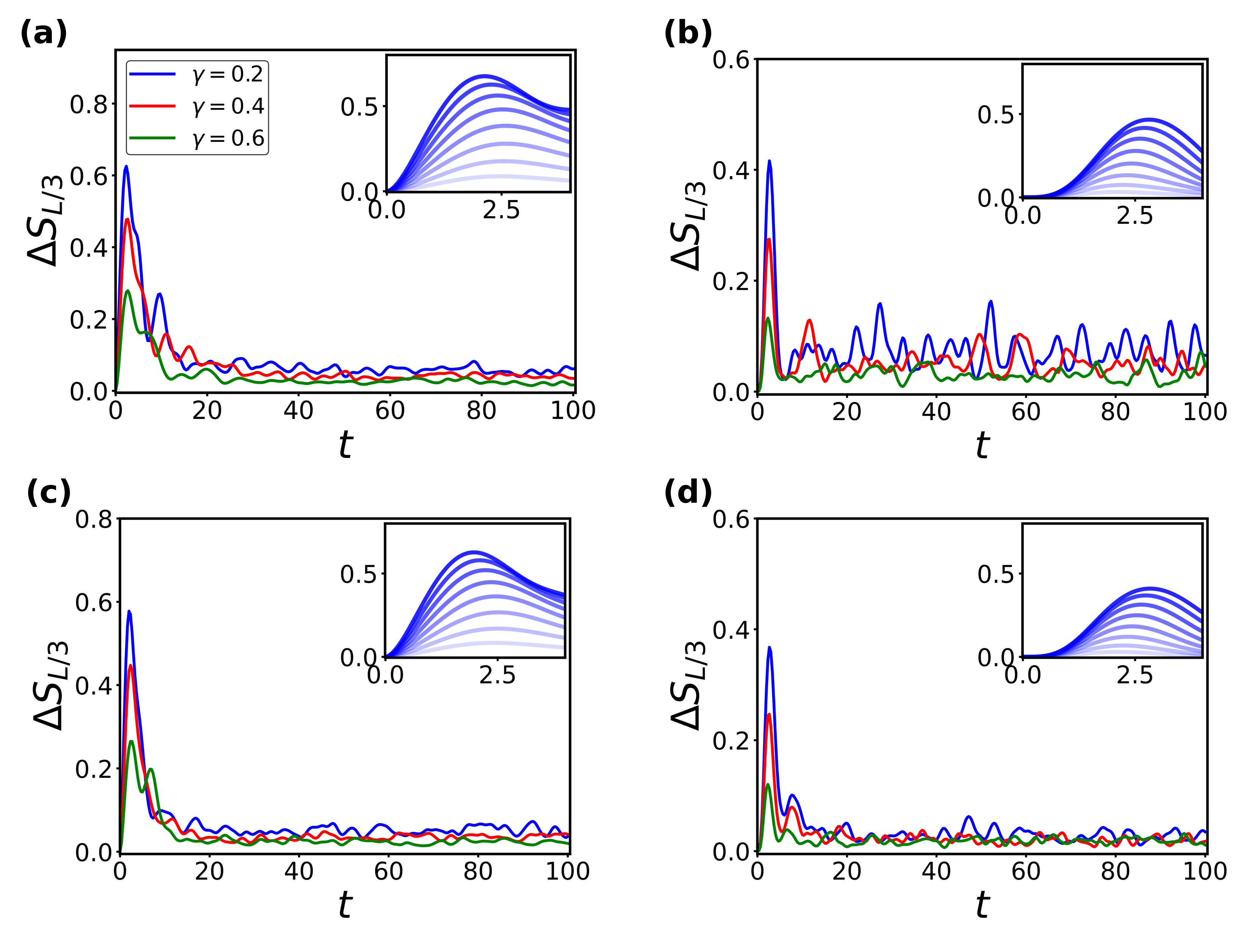}
\caption{EA as a function of time for various initial states and values of $\gamma$, with $L=12$. Panels (a) are based on the $H_{1}$, while panels (b), (c), and (d) correspond to $H_{2}$. Insets zoom in on the peak of EA for different values of $\gamma$ at early times, listed from bottom to top as: 
$\gamma=0.8,0.7,0.6,0.5,0.4,0.3,0.2,0.1$. (a),(c): Domain Wall state. (b): Ferromagnetic state. (d): Antiferromagnetic state.}
\label{fig:Ham_EA}
\end{figure}

\begin{figure}[ht]
\centering
\includegraphics[width=0.8\textwidth, keepaspectratio]{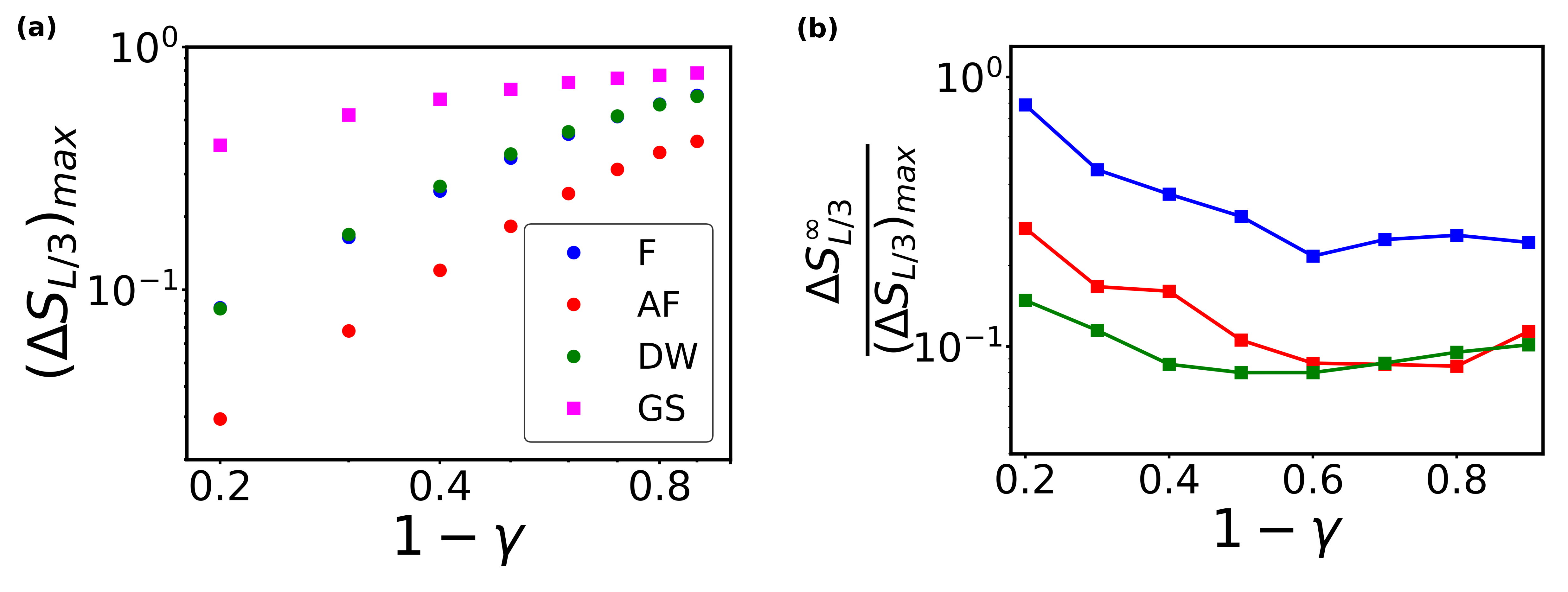}
\caption{Panels (a) and (b) show the peak value of EA, $(\Delta S_{L/3})_{max}$, and the ratio of the late-time EA, $\Delta S_{L/3}^{\infty}$, to $(\Delta S_{L/3})_{max}$ as a function of  $1-\gamma$ for different initial states under $H_{2}$. F: Ferromagnetic state; DW: Domain Wall state; AF: Antiferromagnetic state; GS: Ground State.}
\label{fig:Ham_EAmax}
\end{figure}

In Fig. \ref{fig:Ham_EA}, we show the dynamics of EA across all initial states (ferromagnetic, domain wall, and antiferromagnetic) under the evolution of $H_{2}$. The entanglement asymmetry shows evident overshooting at early times, and the U(1) symmetry remains broken in the subsystem at long times in all cases, consistent with the results described in the main text. As shown in Fig. \ref{fig:Ham_EAmax}, the peak of entanglement asymmetry also correlates with the strength of symmetry breaking, $1-\gamma$, for various initial states in the case of a non-integrable Hamiltonian. EA for ground state of $H_2$ is also shown for comparison. More importantly, the peak value of entanglement asymmetry is significantly higher than its late-time value.  

\subsection{Finite-size scaling of entanglement asymmetry}
\begin{figure}
\centering
\includegraphics[width=0.43\textwidth, keepaspectratio]{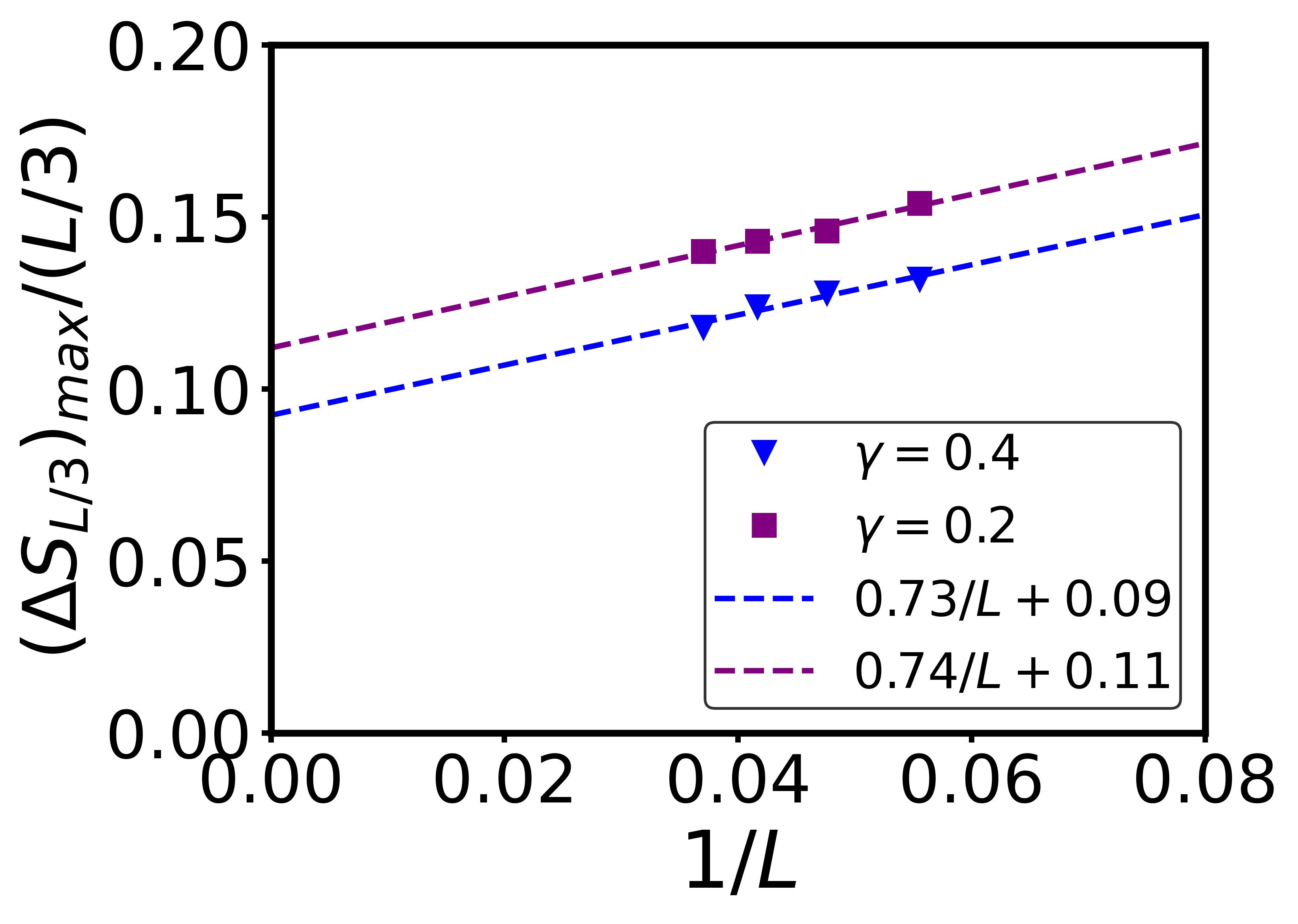}
\caption{Finite-size scaling of entanglement asymmetry peaks for ferromagnetic state at $\gamma=0.2$ and $0.4$. The dots represent numerical results for system sizes $L\in \{18, 21, 24, 27\}$ under Hamiltonian $H_{1}$. The dashed lines indicate a linear fit, demonstrating the peak EA density extrapolates to a non-zero value in the thermodynamic limit.}
\label{fig:HamFS_EA}
\end{figure}

To demonstrate that the overshooting effect persists in the thermodynamic limit, we perform finite size scaling of EA density for ferromagnetic initial states, as depicted in Fig.~\ref{fig:HamFS_EA}. The nonzero intercept of the finite size extrapolation line confirms that the EA density peak survives even in the thermodynamic limit. Our numerical calculations reached the maximum feasible system size of $L=27$ with sparse matrix formulation due to memory constraints. Furthermore, extending these calculations with time-evolving block decimation would be prohibitively challenging, as the overshooting regime coincides with strongly entangled states that exceed practical computational limits. 

\subsection{Dynamics of entanglement entropy}
\begin{figure}[ht]
\centering
\includegraphics[width=1\textwidth, keepaspectratio]{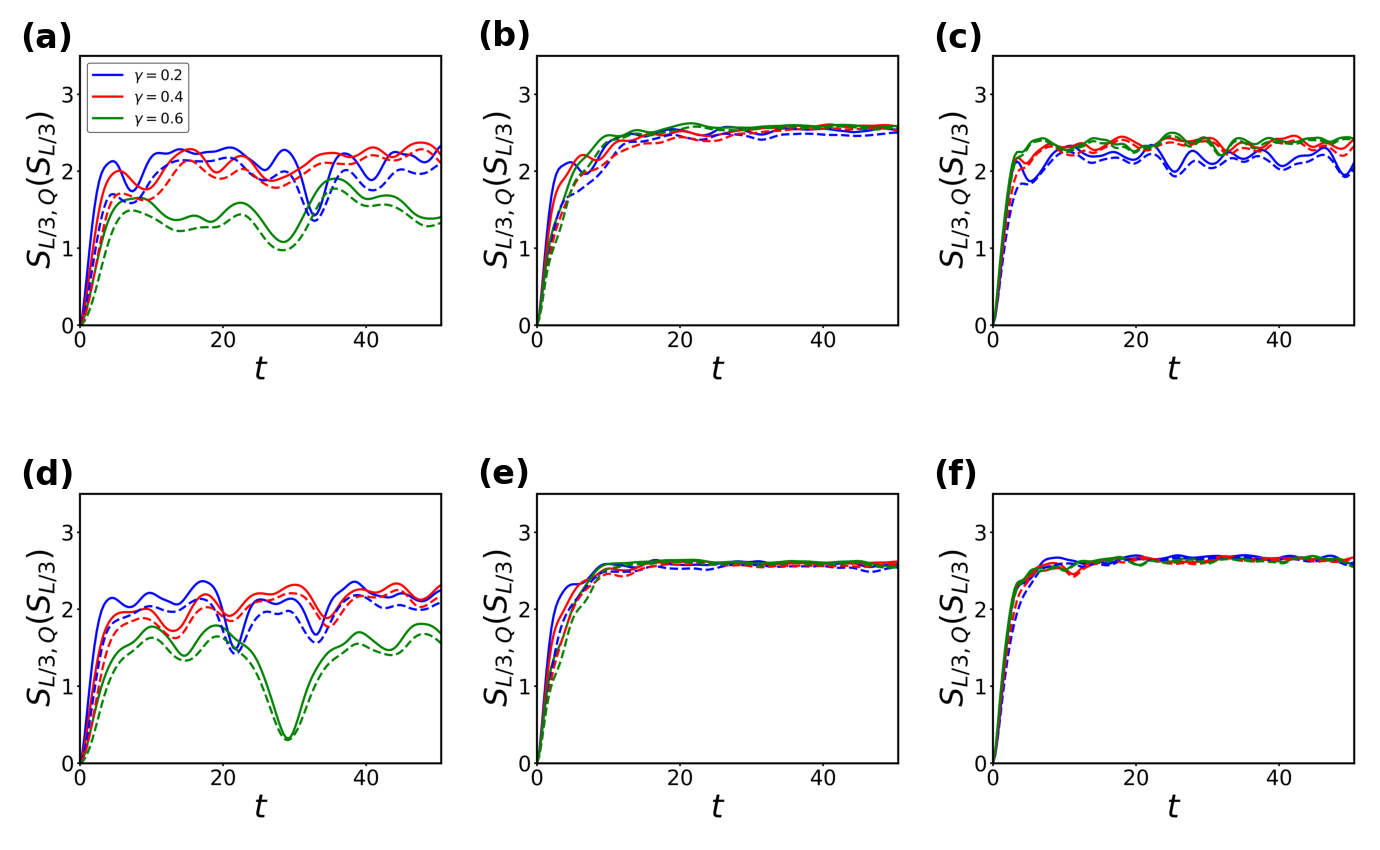}
\caption{Dynamics of each individual term in the expression for the entanglement asymmetry $\Delta S_{L/3}$ with $L=12$ is shown. Solid lines represent $S_{L/3,Q}$, while dash lines correspond to the entanglement entropy $S_{L/3}$. From left to right: initial states are  Ferromagnetic, Domain wall and Antiferromagnetic states, respectively. Panels (a), (b), and (c) are obtained using $H_{1}$, while panels (d), (e), and (f) are based on $H_{2}$.}
\label{fig:Ham_EE}
\end{figure}

To better explain the behavior of entanglement asymmetry, we examine the time evolution of entanglement entropy and explore its connection to entanglement asymmetry in this section. We present both $S_{L/3}$ and $S_{L/3,Q}$ as functions of time, with their difference yielding EA, for various initial states and Hamiltonians. As shown in Fig. \ref{fig:Ham_EE}, we observe that the entanglement entropy grows linearly with time before eventually reaching a plateau. The saturation time of the entanglement entropy is always greater than the time at which the entanglement asymmetry reaches its peak. In most cases, the peak of entanglement asymmetry occurs after $S_{L/3,Q}$ reaches its first peak. This observation is reflected in Fig. \ref{fig:Ham_EE}.

\subsection{Quantum Mpemba effect in U(1) non-symmetric Hamiltonian}
\begin{figure}[ht]
\centering
\includegraphics[width=0.4\textwidth, keepaspectratio]{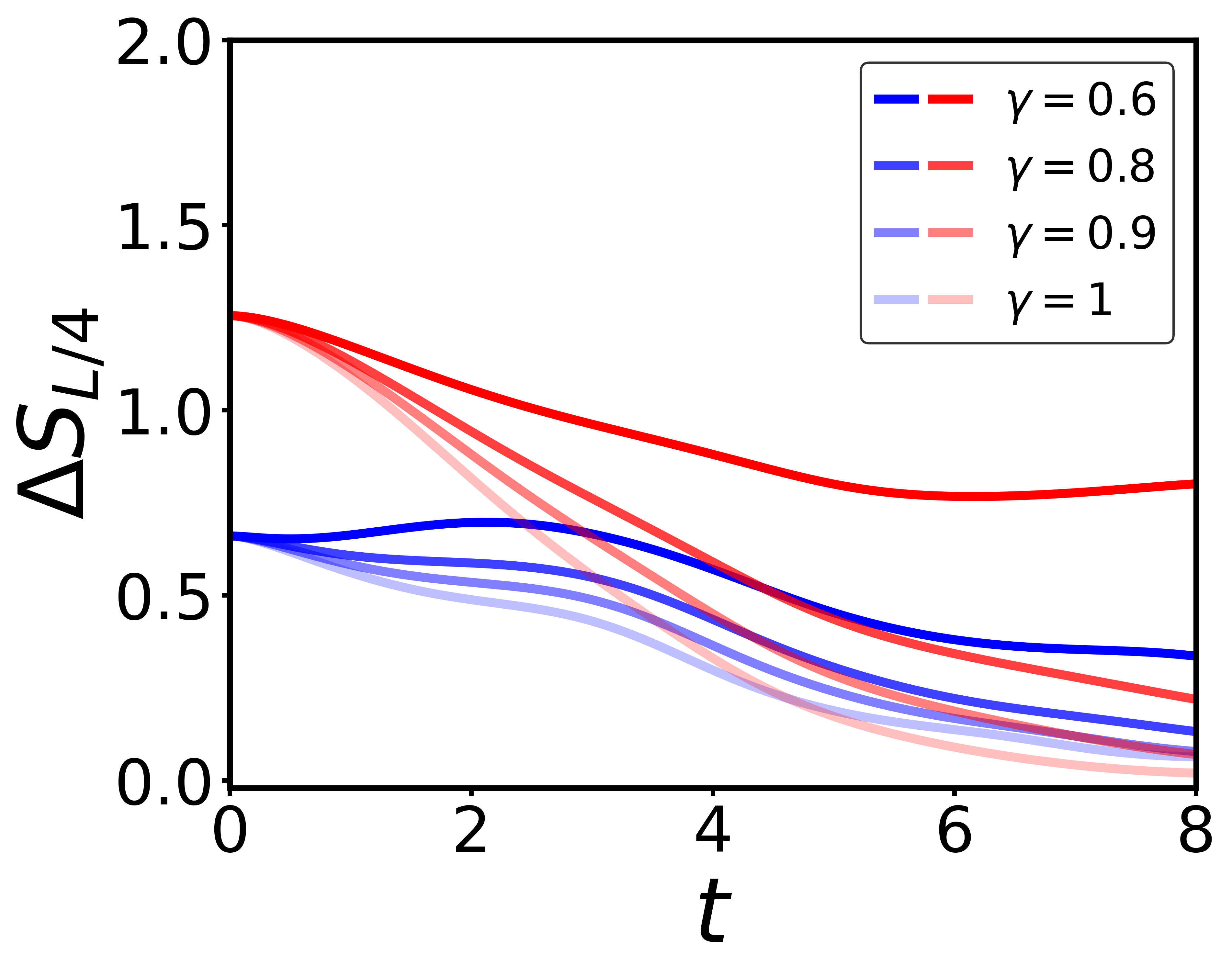}
\caption{EA dynamics at different values of $\gamma$ with $L=12$ are examined for tilted domain wall states. The calculation is based on $H_{1}$. Blue corresponds to $\theta=0.2\pi$, and red corresponds to $\theta=0.5\pi$. The increasing intensity of color reflects a stronger effect of symmetry breaking.}
\label{fig:HamQMB_one}
\end{figure}

\begin{figure}[ht]
\centering
\includegraphics[width=1\textwidth, keepaspectratio]{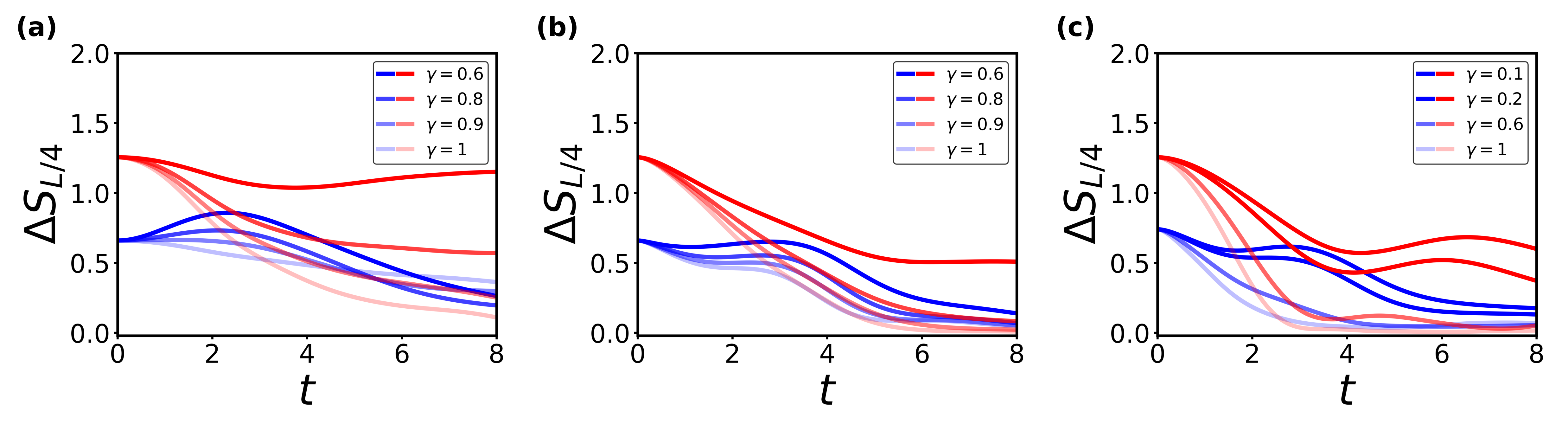}
\caption{The dynamics of entanglement asymmetry are examined for different values of $\gamma$ with $L=12$ across three initial states: (a) tilted ferromagnetic states, (b) tilted domain wall states, and (c) tilted antiferromagnetic states. Panels (a)–(c) are based on the evolution under $H_{2}$. The blue curves correspond to $\theta=0.2\pi$, while the red curves correspond to $\theta=0.5\pi$. The increasing intensity of the color reflects the stronger effect of symmetry breaking.}
\label{fig:HamQMB_two}
\end{figure}

When $\gamma=1$, the emergence of QME is observed in all cases, as shown in Figs. \ref{fig:HamQMB_one} and \ref{fig:HamQMB_two}. Analogous to the discussion in the main text, we find the absence of QME in the non-integrable Hamiltonian across all initial states when $\gamma$ is tuned to above some thresholds. 

\subsection{Kinematics behavior of entanglement asymmetry in U(1) non-symmetric Hamiltonians}
\begin{figure}[ht]
\centering
\includegraphics[width=0.37\textwidth, keepaspectratio]{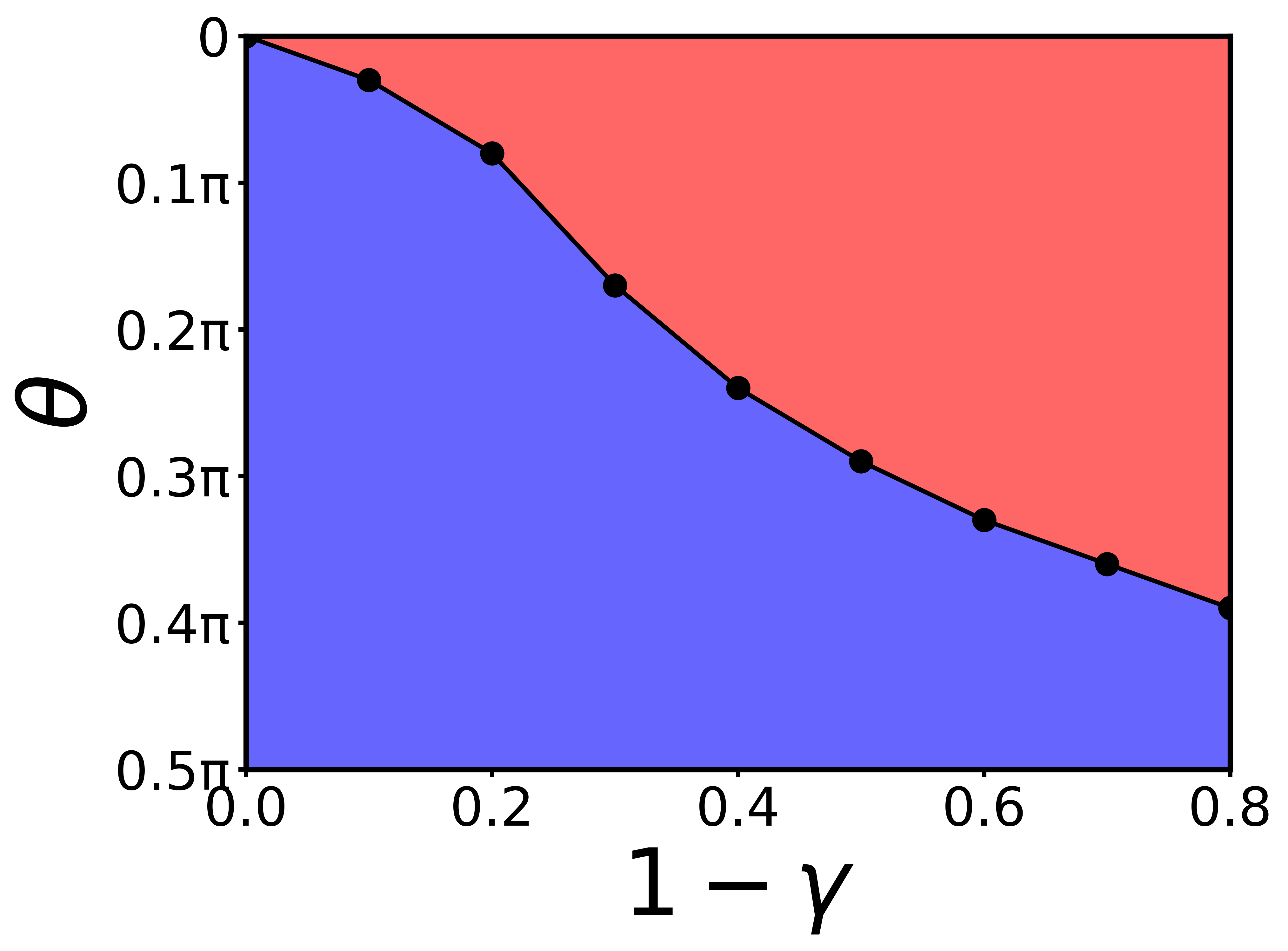}
\caption{A schematic figure illustrating the dependence of EA dynamical patterns on the parameters $\theta$ and $1-\gamma$ for a domain wall state under the evolution of $H_{1}$. Those black dots are obtained through numerical simulation.}
\label{fig:Ham_phasediagram_one}
\end{figure}

\begin{figure}[ht]
\centering
\includegraphics[width=1\textwidth, keepaspectratio]{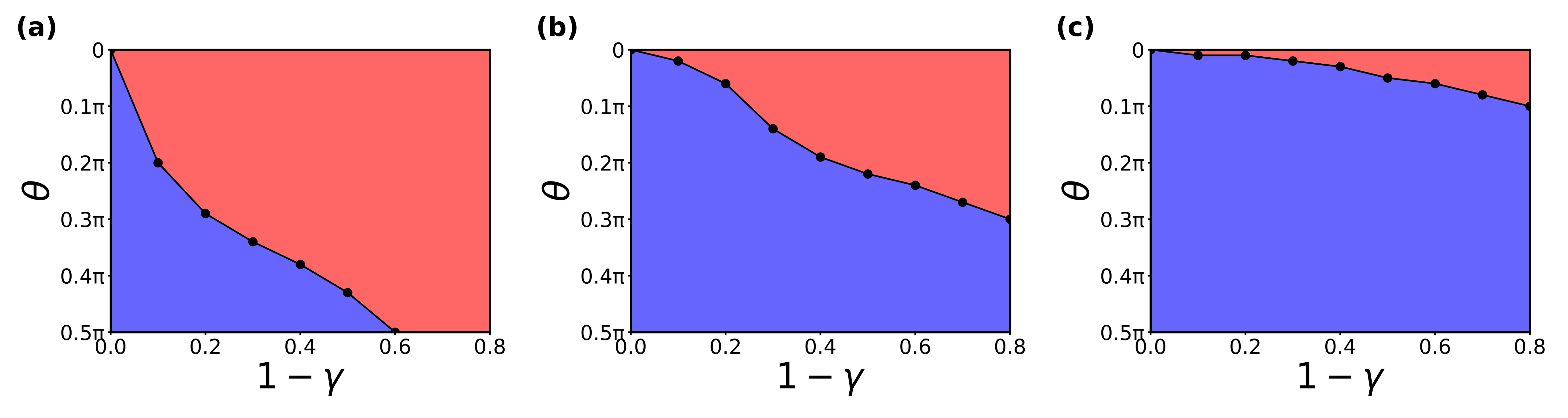}
\caption{Panels (a)-(c) show schematic figures depicting the dependence of EA dynamical patterns on $\theta$ and $1-\gamma$ for different initial states. All panels are based on $H_{2}$. (a) Ferromagnetic state, (b) Domain Wall state, (c) Antiferromagnetic state. Those black dots are obtained through numerical simulation.}
\label{fig:Ham_phasediagram_two}
\end{figure}

\begin{figure}[ht]
\centering
\includegraphics[width=0.9\textwidth, keepaspectratio]{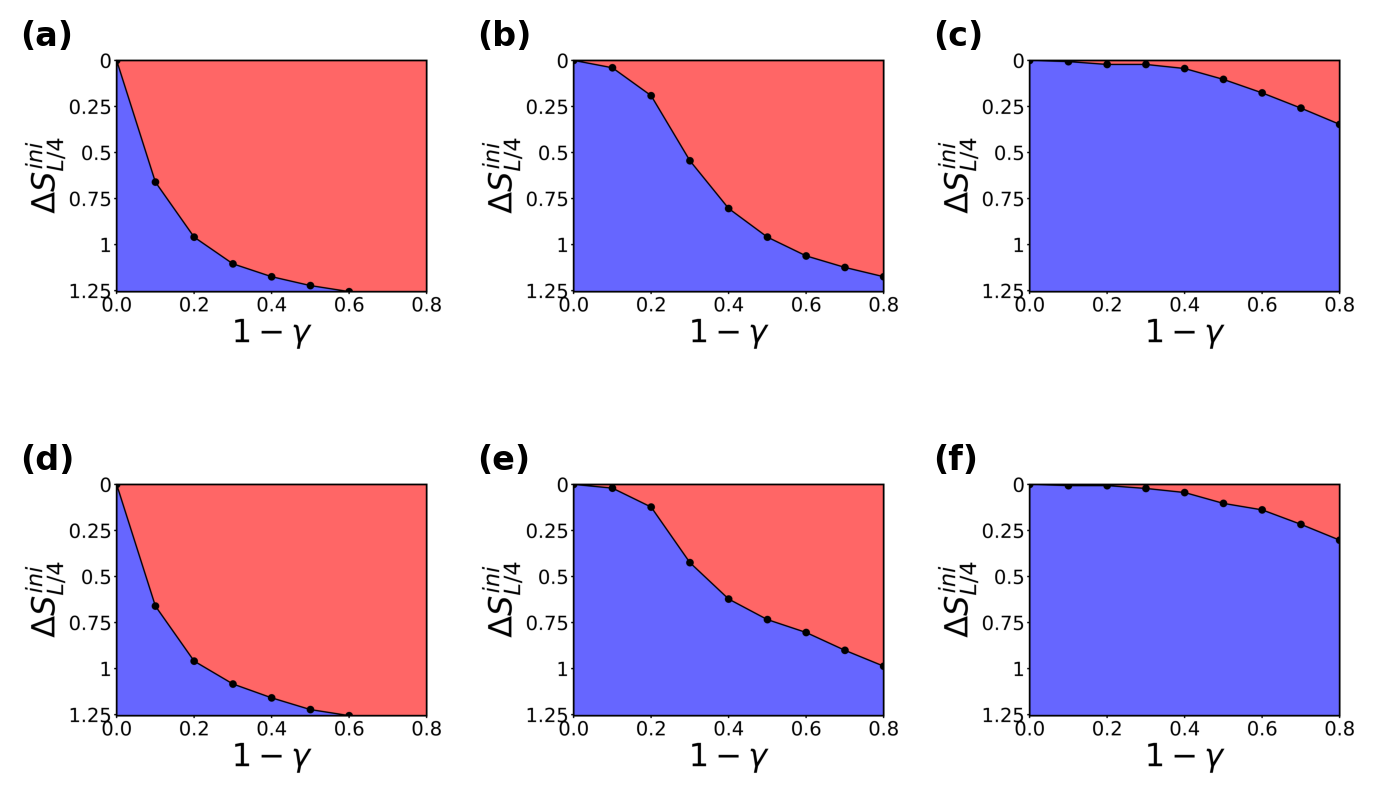}
\caption{Panels (a)-(f) show schematic figures depicting the dependence of EA dynamical patterns on the initial value of EA, $\Delta S_{L/4}^{ini}$ and $1-\gamma$ for different initial states and Hamiltonians. From left to right: initial states are  Ferromagnetic, Domain wall and Antiferromagnetic states, respectively. Panels (a), (b), and (c) are obtained using $H_{1}$, while panels (d), (e), and (f) are based on $H_{2}$.}
\label{fig:Ham_phaseinitialEA}
\end{figure}

As shown in Figs. \ref{fig:Ham_phasediagram_one} and \ref{fig:Ham_phasediagram_two}, the figures for ferromagnetic and antiferromagnetic states obtained through $H_{2}$ are almost identical to those from $H_{1}$. A slight difference is observed in the case of the domain wall state. Since the boundaries are different for different initial states, a natural question arises -- is the boundary only different in terms of $\theta$ or also different when translating $\theta$ to initial EA.  In Fig. \ref{fig:Ham_phaseinitialEA}, we replace $\theta$ with the initial value of the entanglement asymmetry, $\Delta S_{L/4}^{ini}$, on the $y$-axis. We find that the boundary is still different in terms of initial EA for different types of initial states.

\subsection{Late-time entanglement asymmetry}
\begin{figure}[ht]
\centering
\includegraphics[width=0.9\textwidth, keepaspectratio]{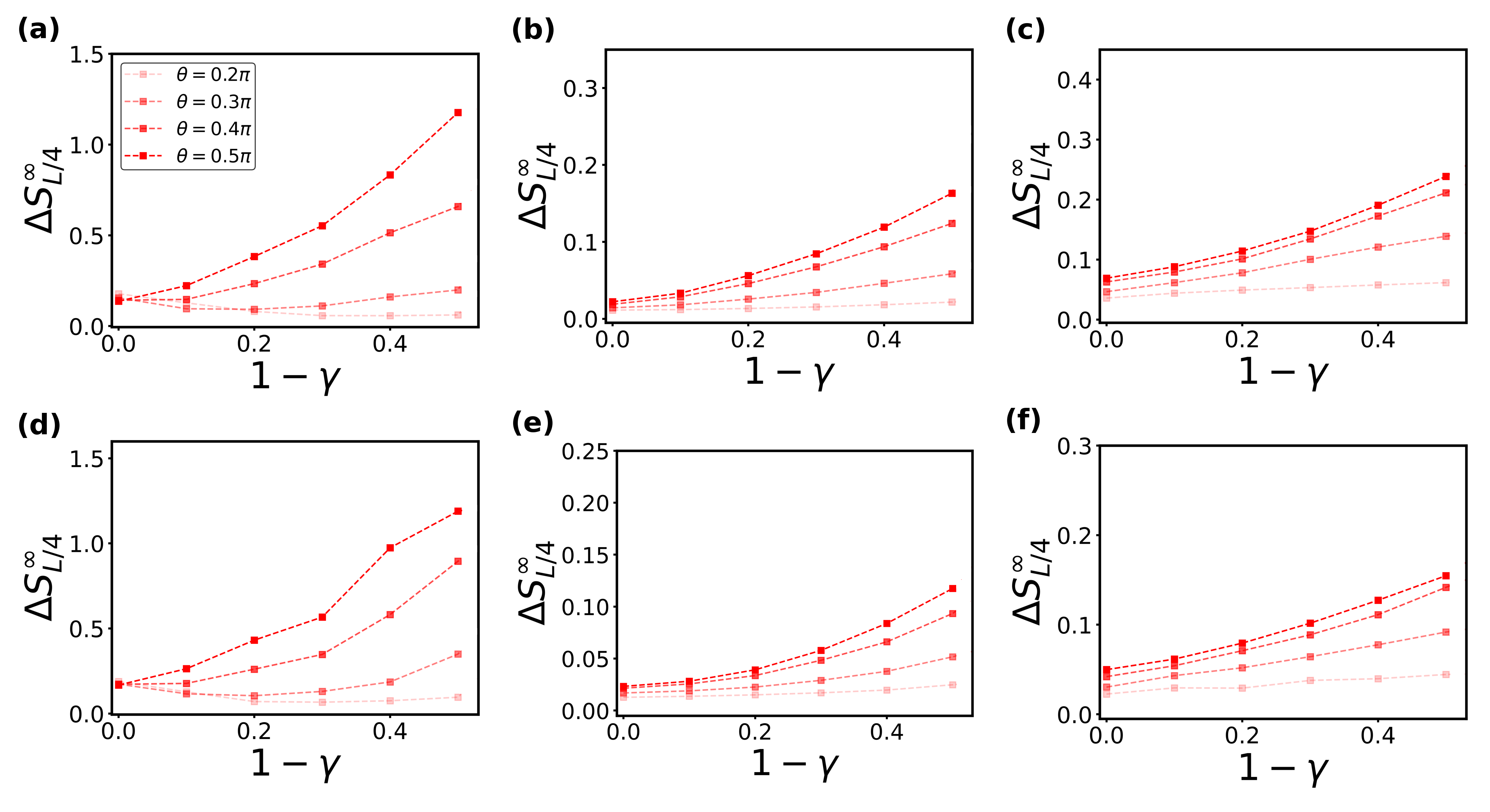}
\caption{The late-time entanglement asymmetry, $\Delta S_{L/4}^{\infty}$, as a function of $1-\gamma$ with $L=12$. From left to right: the initial states are tilted ferromagnetic, tilted domain wall and  tilted antiferromagnetic states, respectively. Panels (a), (b), and (c) correspond to results obtained using $H_{1}$, while panels (d), (e), and (f) are based on $H_{2}$.}
\label{fig:Ham_EAlate}
\end{figure}

The late-time entanglement asymmetry is obtained by  averaging $\Delta S_{L/4}$ over $2000$ time points between $t_{1}=2000$ and $t_{2}=40000$, as described in the main text. In Fig. \ref{fig:Ham_EAlate}, we compute the late-time EA for different initial states under the evolution of $H_{1}$ and $H_{2}$. We find two general trends: (1) The late-time value of EA increases with the strength of symmetry breaking $1-\gamma$. (2) For states with higher initial asymmetry, the entanglement asymmetry at late times is greater and increases more rapidly compared to states with lower initial asymmetry. 

\section{More numerical results of charge variance for U(1) non-symmetric Hamiltonians}
\begin{figure}[ht]
\centering
\includegraphics[width=0.8\textwidth, keepaspectratio]{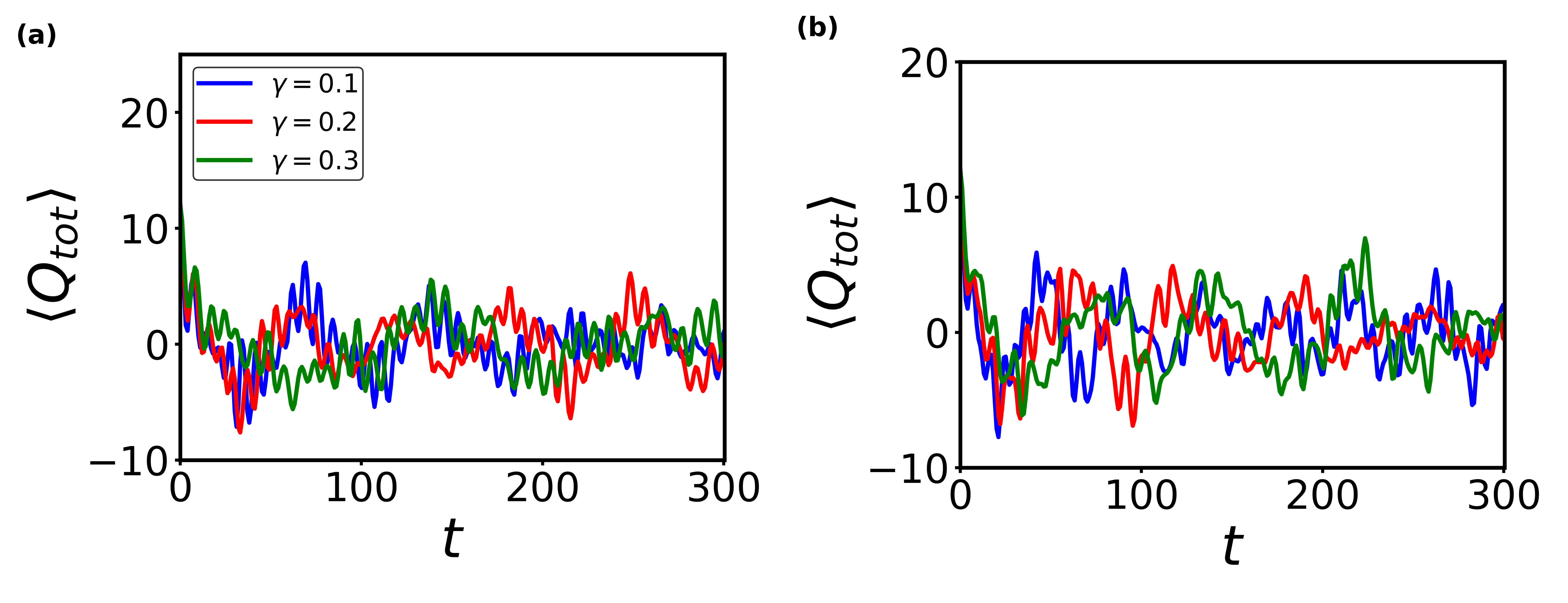}
\caption{The time evolution of the expectation value of the total spin in the $z$-direction, $\langle Q_{tot} \rangle$, is investigated for various values of $\gamma$ with a ferromagnetic initial state, where $L=12$. Here, $\hat{Q}_{tot} = \sum_{i=1}^{L} \sigma_{i}^{z}$. Left: $H_{1}$. Right: $H_{2}$.}
\label{fig:Ham_Q}
\end{figure}

\begin{figure}[ht]
\centering
\includegraphics[width=1\textwidth, keepaspectratio]{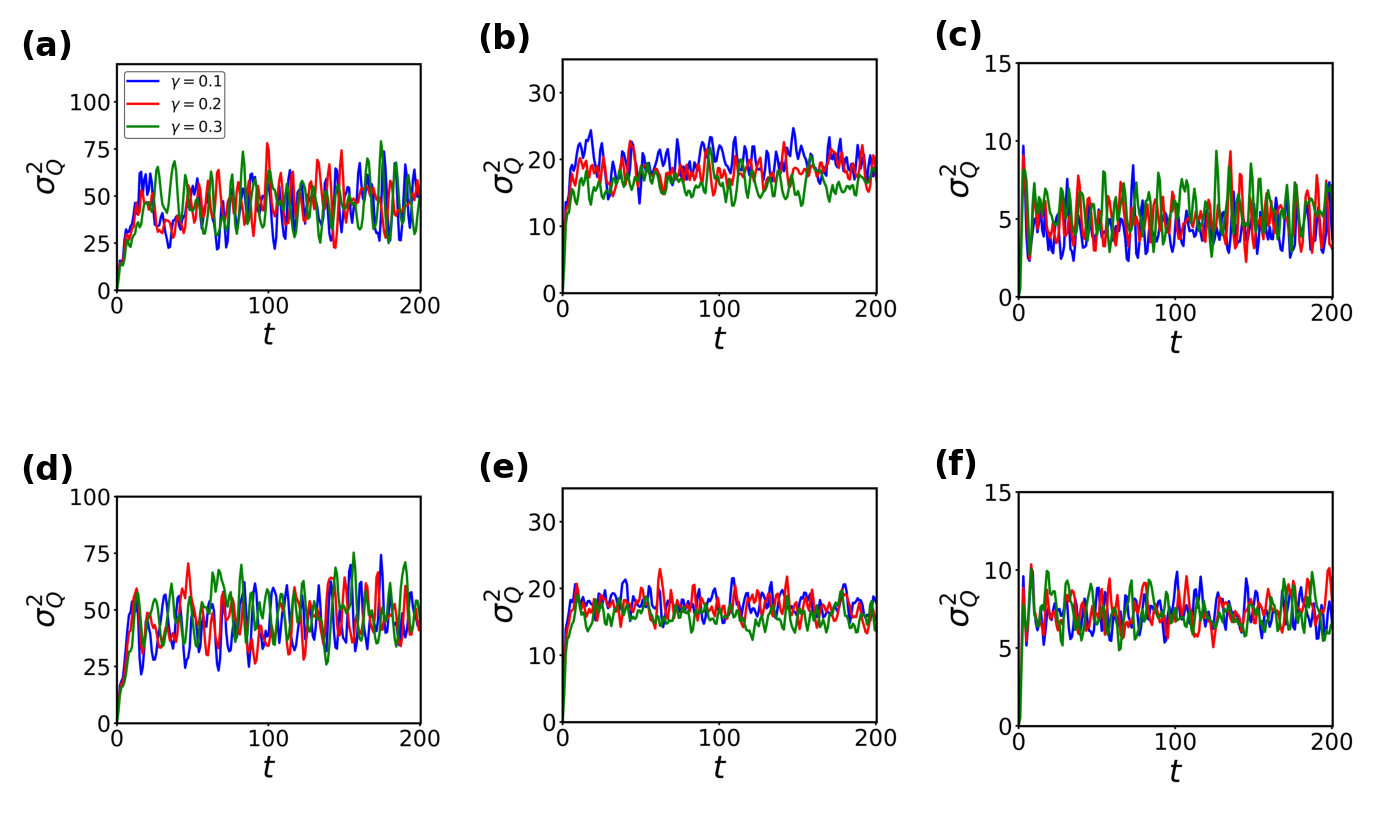}
\caption{The variance of $Q_{tot}$, $\sigma_{Q}^{2}$, as a function of time for different values of $\gamma$ with $L=12$. Top row: $H_{1}$. Bottom row: $H_{2}$. From left to right: Ferromagnetic, Domain Wall, and Antiferromagnetic states.}
\label{fig:Ham_varQ}
\end{figure}

\subsection{Expectation values of $\hat{Q}$ and $\sigma_{Q}^{2}$}
The expectation value of the total spin in the $z$-direction as a function of time is shown in Fig. \ref{fig:Ham_Q} for ferromagnetic states. At the start, $Q_{tot} = L=12$ since all the spins are aligned upward. Over time, the state progressively loses memory of its initial configuration, and $Q_{tot}$ settles into oscillations around $0$.
For domain wall and antiferromagnetic states, $\langle \hat{Q}_{tot}(t) \rangle = 0$ is a constant due to symmetry argument which are elaborated below. 

Suppose we have a transformation $T$:
\begin{align}
T: & \quad \sigma^{z} \rightarrow -\sigma^{z}, \quad \sigma^{y} \rightarrow -\sigma^{y} 
\end{align}
It is straightforward to verify that the Hamiltonians remain invariant under this transformation $T$. Thus, we have
\begin{align}
\langle \hat{Q}_{\text{tot}}(t) \rangle &= \langle \psi(t)| \hat{Q}_{\text{tot}} |\psi(t) \rangle \notag \\
&= \langle \psi(0)|e^{iHt} \hat{Q}_{\text{tot}} e^{-iHt}|\psi(0) \rangle \notag \\
&= \langle \psi(0)|T^{\dagger}Te^{iHt}T^{\dagger}T \hat{Q}_{\text{tot}}T^{\dagger}T e^{-iHt}T^{\dagger}T|\psi(0) \rangle \notag \\
&= \langle \psi(0)|T^{\dagger}e^{iHt} (-\hat{Q}_{\text{tot}}) e^{-iHt}T|\psi(0) \rangle \\
&= \langle \psi_{p}(0)|e^{iHt} (-\hat{Q}_{\text{tot}}) e^{-iHt}|\psi_{p}(0) \rangle \notag\\
&= -\langle \psi_{p}(t)|\hat{Q}_{\text{tot}}|\psi_{p}(t) \rangle \notag\\
&= -\langle \psi(t)|\hat{Q}_{\text{tot}}|\psi(t) \rangle  \notag\\
&= 0  \notag
\end{align}
Here $|\psi_{p}(t)\rangle$ is related to $|\psi(t)\rangle$ by a permutation of site indices. Throughout the derivation, we use the fact that $T \hat{Q}_{\text{tot}}T^{\dagger}=-\hat{Q}_{\text{tot}}$. Therefore, $\langle \hat{Q}_{tot}(t) \rangle$ is strictly zero for domain wall and antiferromagnetic initial states. 

Fig. \ref{fig:Ham_varQ} shows the time dependence of the variance of total charge operator for different initial states under $H_{1}$ and $H_{2}$. Notably, the variance of the charge typically saturates later than the entanglement asymmetry. The charge variance also characterizes the U(1)  symmetry breaking in the evolved state from some aspects, similar as EA explored in the main text. However, the symmetry breaking dynamical behaviors show distinct patterns in the two metrics. EA shows an evident overshooting while the charge variance directly saturates. The differences may provide further insight into the multifaceted physics of symmetry breaking. 

\subsection{Dynamics of charge variance for different U(1)-asymmetric initial states}
\begin{figure}[ht]
\centering
\includegraphics[width=1\textwidth, keepaspectratio]{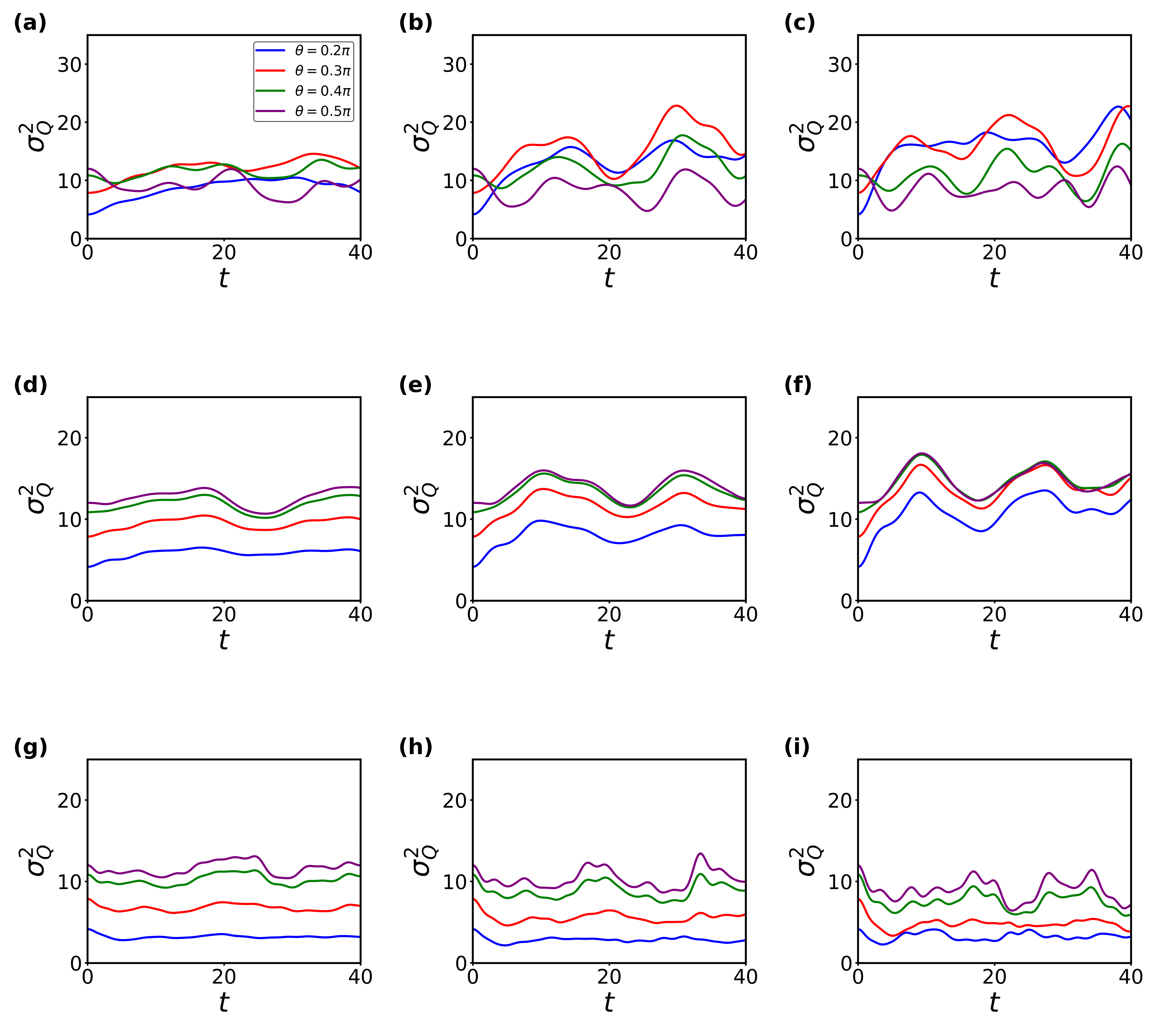}
\caption{Temporal evolution of the charge variance, $\sigma_{Q}^{2}$, under the integrable Hamiltonian $H_{1}$ for a system size $L=12$. The top row corresponds to tilted ferromagnetic states, the middle row to tilted domain wall states, and the bottom row to tilted antiferromagnetic states. From left to right, the columns show results for $\gamma=0.9$, $0.8$, and $0.7$.}
\label{fig:Ham_CVearly_int}
\end{figure}

\begin{figure}[ht]
\centering
\includegraphics[width=1\textwidth, keepaspectratio]{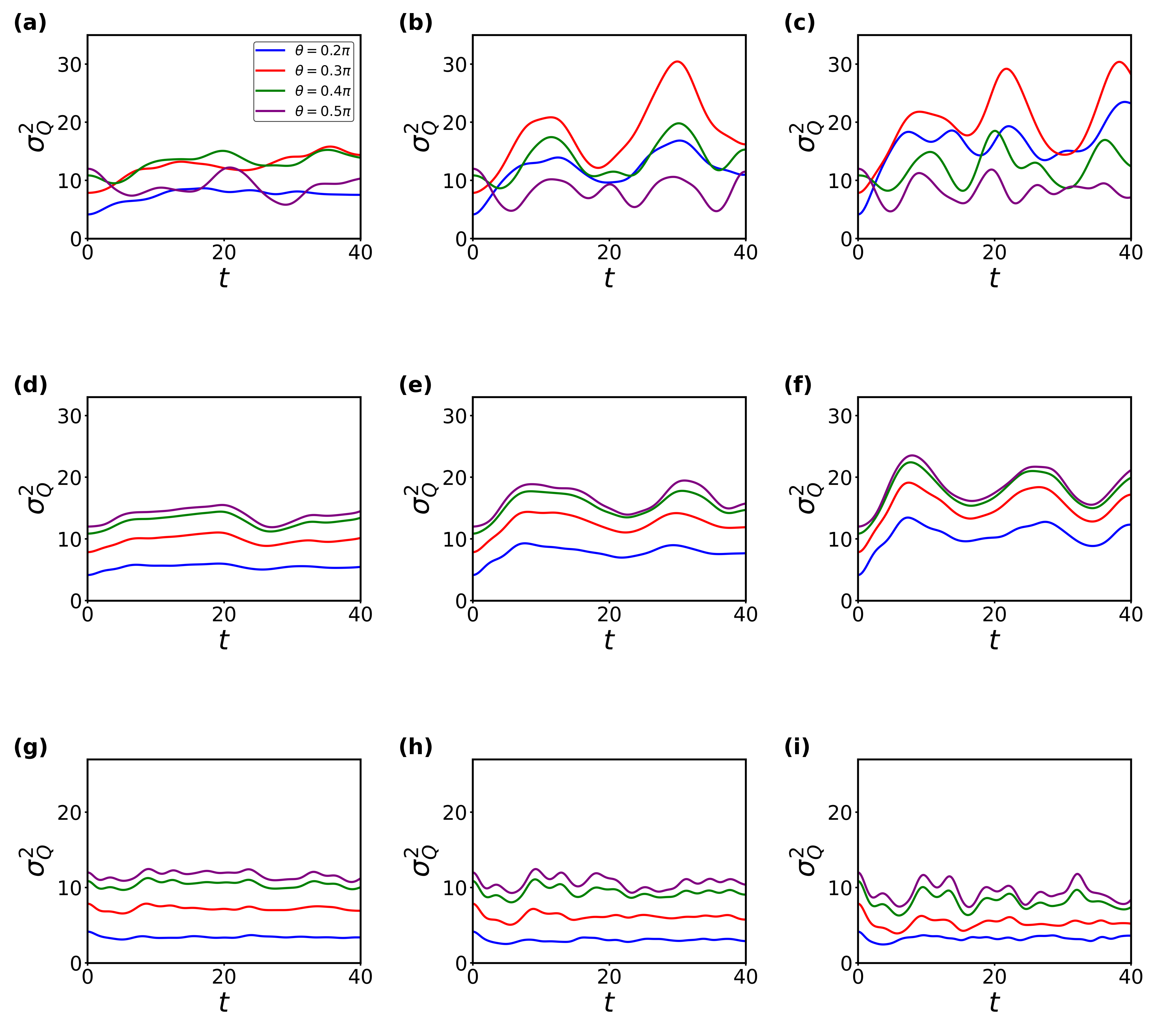}
\caption{Time evolution of the charge variance, $\sigma_{Q}^{2}$, under the non-integrable Hamiltonian $H_{2}$ for a system size $L=12$. The top row corresponds to tilted ferromagnetic states, the middle row to tilted domain wall states, and the bottom row to tilted antiferromagnetic states. From left to right, the columns show results for $\gamma=0.9$, $0.8$, and $0.7$.}
\label{fig:Ham_CVearly_nonint}
\end{figure}

Figs. \ref{fig:Ham_CVearly_int} and \ref{fig:Ham_CVearly_nonint} show that, for charge variance evolved from the tilted domain wall and tilted antiferromagnetic states, the charge variance is consistently larger for more asymmetric states (large $\theta$) than for less asymmetric ones (small $\theta$). However, this trend does not apply to the initial tilted ferromagnetic state, where the monotonic relationship of charge variance with respect to $\theta$ is reversed at early times. This results in an early-time crossing, which becomes more pronounced as $1-\gamma$ increases. These behaviors are observed in both integrable and non-integrable Hamiltonians, highlighting their universality across different dynamical regimes.

\subsection{Dynamics of charge sector probability distributions}
\begin{figure}[ht]
\centering
\includegraphics[width=1\textwidth, keepaspectratio]{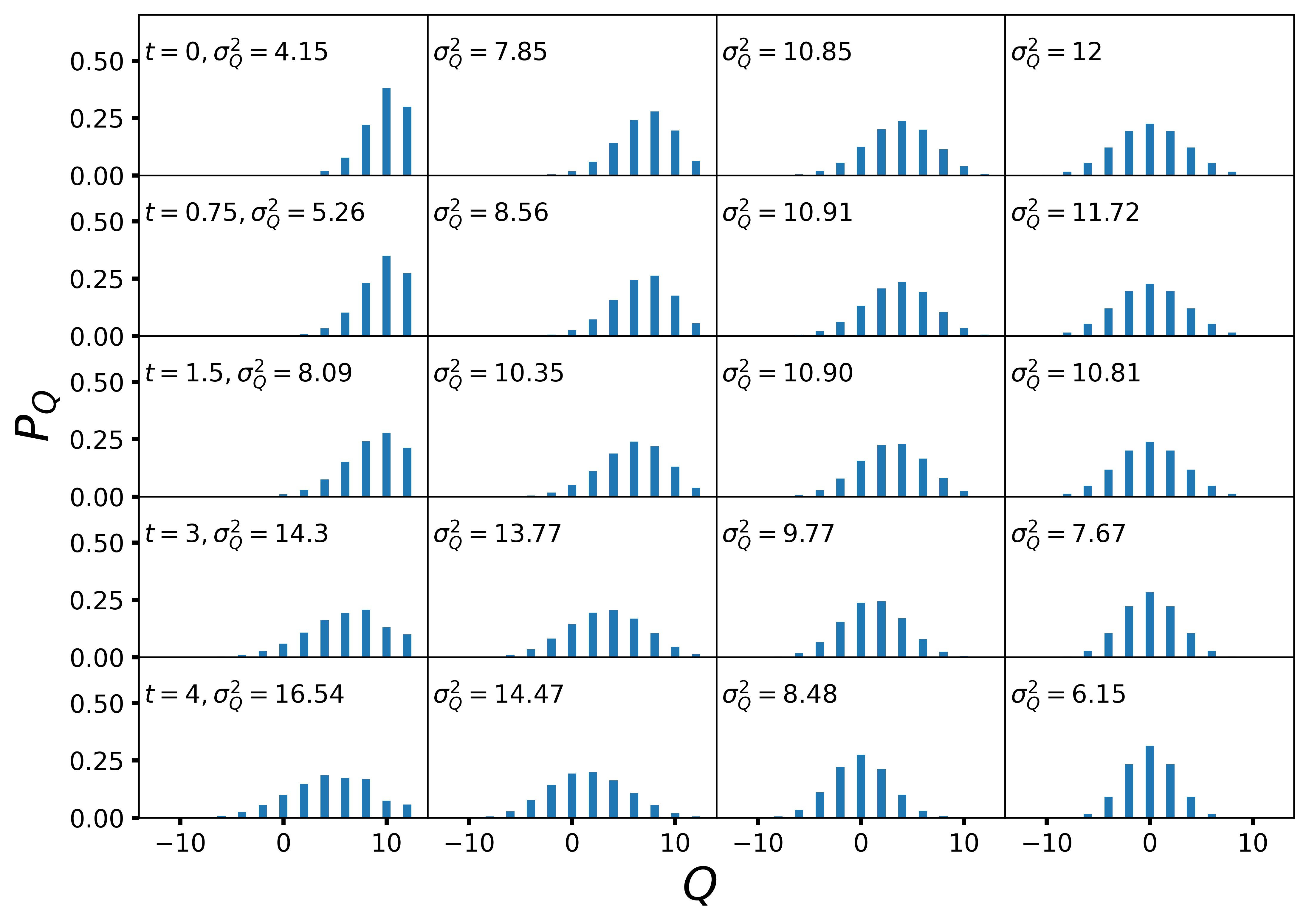}
\caption{Time evolution of the probability distribution, $P_{Q}$, for each charge sector $Q$ under the Hamiltonian $H_{1}$ with $\gamma=0.6$. The charge sectors range from $Q=-12$ to $Q=12$ in increments of $2$. The rows, from top to bottom, correspond to time points $t=0$, $0.75$, $1.5$, $3$, and $4$. The columns, from left to right, represent different tilted ferromagnetic states with $\theta=0.2\pi$, $0.3\pi$, $0.4\pi$, and $0.5\pi$.}
\label{fig:Ham_Chargesector}
\end{figure}

Fig. \ref{fig:Ham_Chargesector} displays the probability distribution $P_Q$ across various charge sectors
$Q$ for the evolved state $|\psi(t)\rangle$ at different times for initial tilted ferromagnetic states. Here, $P_{Q}$ at time $t$ is defined as $\sum_{q=Q}|\langle \psi(t)|\psi_{q}\rangle|^{2}$, where $|\psi_{q}\rangle$ represents the basis wave function corresponding to charge $q$. For more asymmetric initial states (larger $\theta$), the charge distribution shrinks, leading to a decrease in charge variance. In contrast, for less asymmetric states, the distribution spreads across a broader range of charge sectors, increasing the charge variance. This behavior accounts for the crossing phenomenon observed in Figs. \ref{fig:Ham_CVearly_int} and \ref{fig:Ham_CVearly_nonint}.

\subsection{Kinematic behavior of charge variance in U(1) non-symmetric Hamiltonian}
\begin{figure}[ht]
\centering
\includegraphics[width=0.8\textwidth, keepaspectratio]{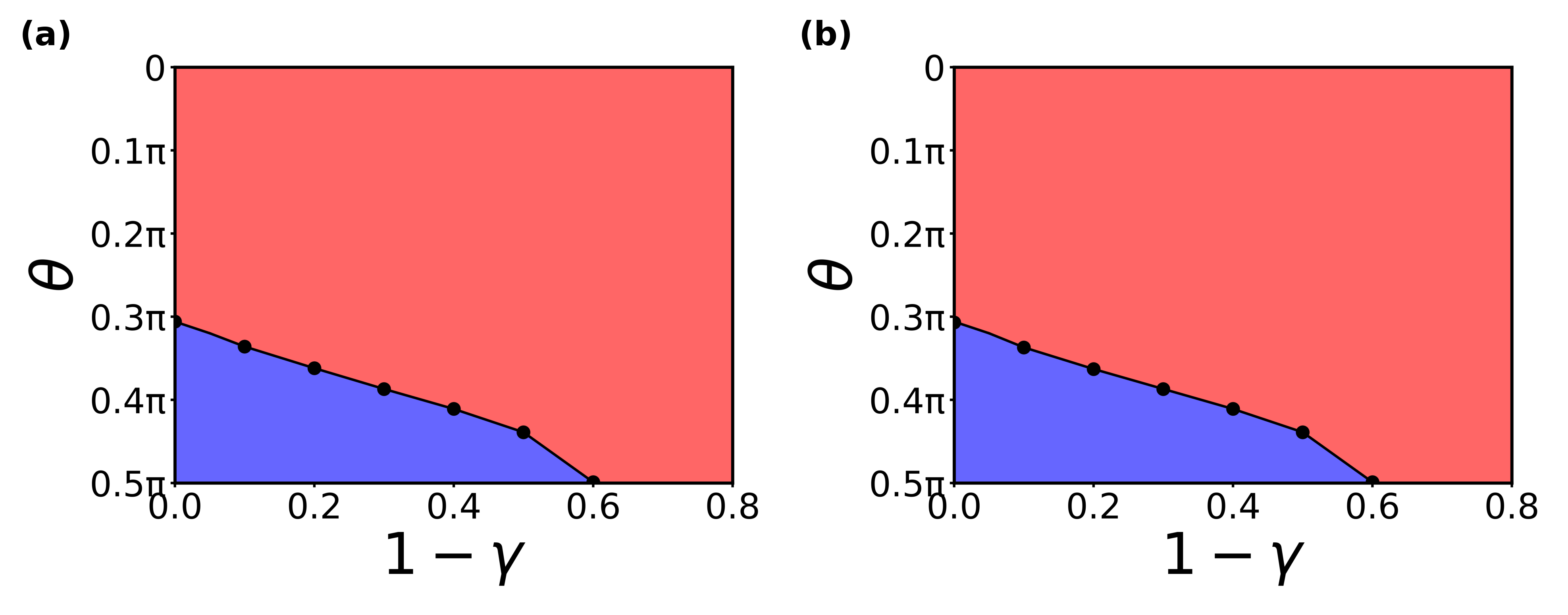}
\caption{Panel (a) and (b) present schematic figures depicting the dependence of charge variance on $\theta$ and $1-\gamma$ with ferromagnetic state evolving under $H_{1}$ (left) and $H_{2}$ (right).}
\label{fig:Ham_Chargephase}
\end{figure}

Fig. \ref{fig:Ham_Chargephase} presents schematic 2-dimensional figures that capture the dependence of early-time CV dynamics on the parameters $\theta$ and $1-\gamma$ for the ferromagnetic state evolving under $H_{1}$ and $H_{2}$. The figure derived from $H_{2}$ closely resembles those obtained from $H_{1}$. The red region indicates where CV initially grows, While the blue region corresponds to an initial decrease in CV. At $\gamma=1$ (left boundary), the charge variance remains constant and equal to its initial value.

\subsection{Late-time charge variance}
\begin{figure}[ht]
\centering
\includegraphics[width=1\textwidth, keepaspectratio]{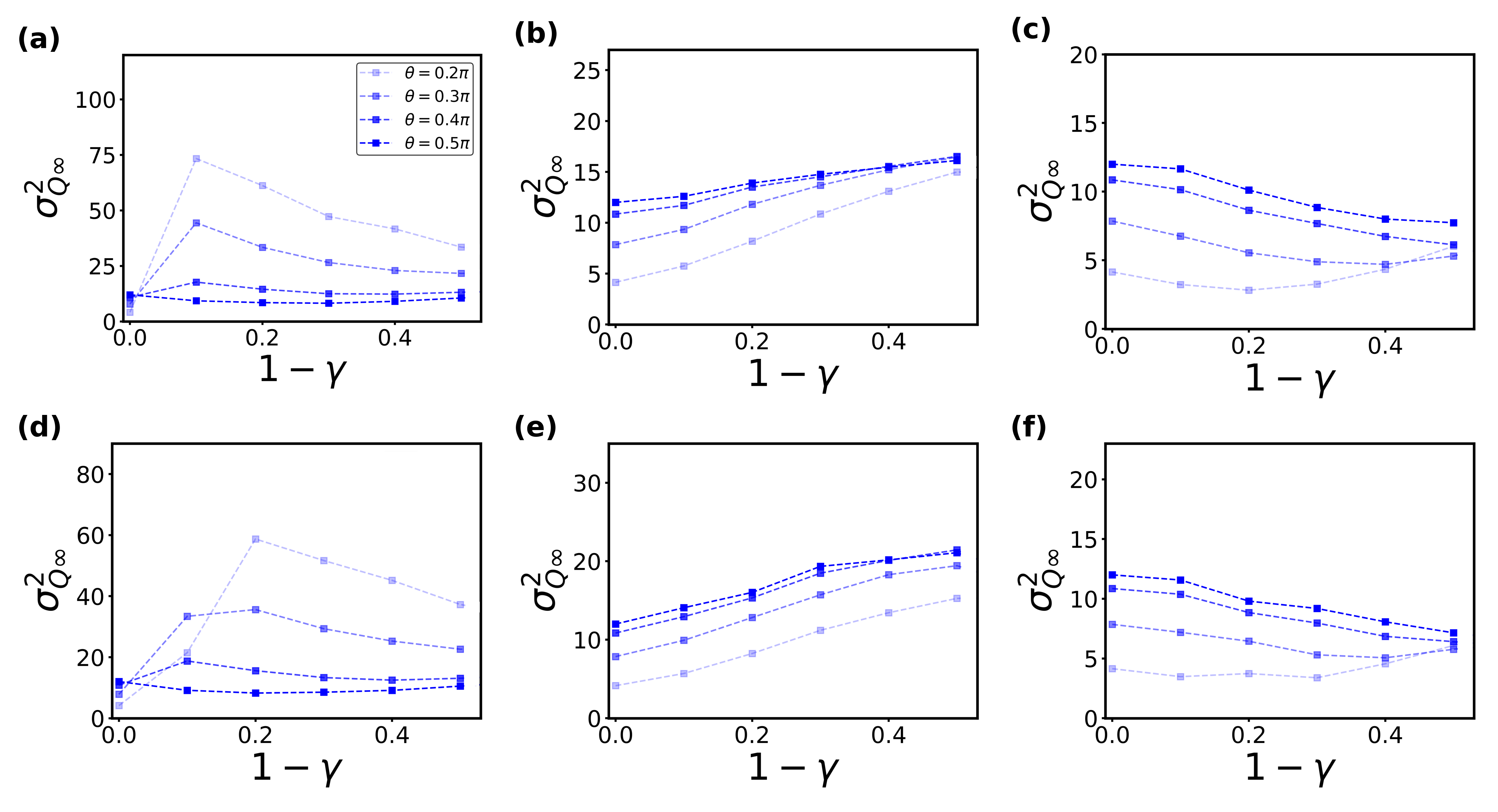}
\caption{The late-time charge variance, $\sigma_{Q\infty}^{2}$, as a function of $1-\gamma$ with $L=12$. From left to right: the initial states are tilted ferromagnetic, tilted domain wall and  tilted antiferromagnetic states, respectively. Panels (a), (b), and (c) correspond to results obtained using $H_{1}$, while panels (d), (e), and (f) are based on $H_{2}$.}
\label{fig:Ham_CvLate}
\end{figure}

The late-time charge variance, $\sigma_{Q}^{2}(t\rightarrow \infty)$, is computed in a manner similar to the late-time entanglement asymmetry. We examine the late-time charge variance for different initial states under the evolution of both $H_{1}$ and $H_{2}$, as illustrated in Fig. \ref{fig:Ham_CvLate}. It is observed that the late-time charge variance increases with the tilted angle $\theta$ for domain wall and antiferromagnetic states at a fixed $\gamma$. In contrast, for ferromagnetic states, the late-time charge variance decreases as $\theta$ increases, which is highly non-trivial. This behavior, also observed at early times (as seen in the crossing in Figs. \ref{fig:Ham_CVearly_int} and \ref{fig:Ham_CVearly_nonint}), continues into the late-time regime.

\section{Analytical results of charge variance at early times}
\begin{figure}[ht]
\centering
\includegraphics[width=0.85\textwidth, keepaspectratio]{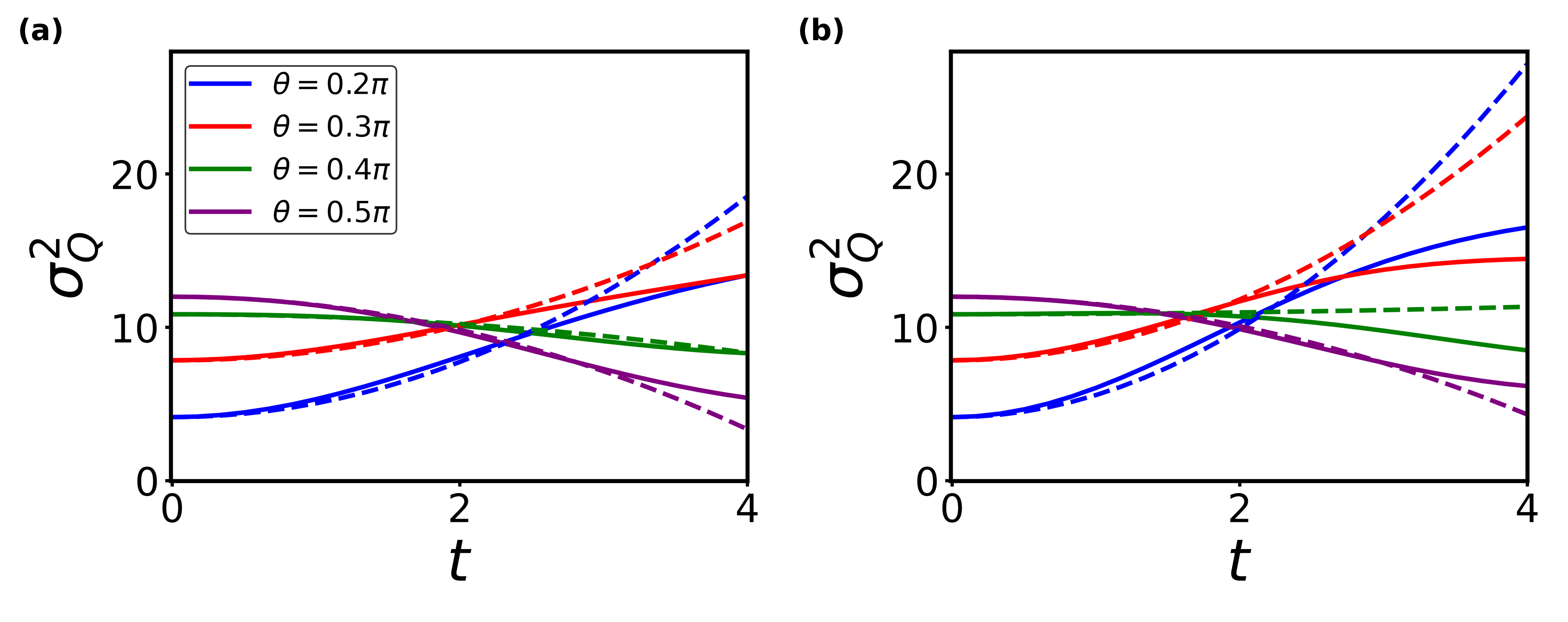}
\caption{Comparison of the early-time behavior of the charge variance, $\sigma_{Q}^{2}$, between analytical solutions (dashed lines) and numerical simulations (solid lines). Results are obtained from tilted ferromagnetic states under the Hamiltonian $H_{1}$ with $L=12$. Panel (a): $\gamma=0.7$. Panel (b): $\gamma=0.6$.}
\label{fig:CVanly_numer}
\end{figure}

The early-time dynamics of charge variance, $\sigma_{Q}^{2} = \langle Q^{2} \rangle -\langle Q \rangle^{2}$, can be expanded as:

\begin{eqnarray}
\sigma_{Q}^{2}(t) \approx \sigma_{Q}^{2}(0)+t\left.\frac{d\sigma_{Q}^{2}}{dt} \right|_{0}+\frac{t^{2}}{2}\left. \frac{d^{2}\sigma_{Q}^{2}}{dt^{2}} \right|_{0}+... \label{Cvt expansion}
\end{eqnarray}
where the ellipsis represents higher order terms that are neglected. Here, $\left.\frac{d\sigma_{Q}^{2}}{dt} \right|_{0}$ and $\left. \frac{d^{2}\sigma_{Q}^{2}}{dt^{2}} \right|_{0}$ denote the first and second derivatives of the charge variance evaluated at $t=0$, respectively. The initial state is chosen to be a tilted ferromagnetic state. 
Using the Heisenberg equation of motion, the first derivative of the charge variance at $t=0$ is given by:

\begin{eqnarray}
\left.\frac{d\sigma_{Q}^{2}}{dt} \right|_{0} = i\langle [H,Q^{2}] \rangle_{0} -2i\langle Q \rangle_{0}\langle[H,Q]\rangle_{0}
\end{eqnarray}
To evaluate $[H,Q]$, where the Hamiltonian is: 
\begin{eqnarray}
H=-\frac{1}{4}\sum_{i=1}^{L}(\sigma_{i}^{x}\sigma_{i+1}^{x}+\gamma\sigma_{i}^{y}\sigma_{i+1}^{y}+\Delta_{1}\sigma_{i}^{z}\sigma_{i+1}^{z})
\end{eqnarray}
and the charge operator is $Q=\sum_{j=1}^{L}\sigma_{j}^{z}$, We compute:

\begin{eqnarray}
[H,Q] &=& [-\frac{1}{4}\sum_{i=1}^{L}(\sigma_{i}^{x}\sigma_{i+1}^{x}+\gamma\sigma_{i}^{y}\sigma_{i+1}^{y}+\Delta_{1}\sigma_{i}^{z}\sigma_{i+1}^{z}),\sum_{j=1}^{L}\sigma_{j}^{z}] \\
&=& -\frac{1}{4}\sum_{j}[\sigma_{j-1}^{x}\sigma_{j}^{x}+\sigma_{j}^{x}\sigma_{j+1}^{x}+\gamma\sigma_{j-1}^{y}\sigma_{j}^{y}+\gamma\sigma_{j}^{y}\sigma_{j+1}^{y},\sigma_{j}^{z}] 
\end{eqnarray}
Simplifying, we obtain:
\begin{eqnarray}
[H,Q]=\frac{i}{2}(1-\gamma)\sum_{j}(\sigma_{j}^{x}\sigma_{j+1}^{y}+\sigma_{j}^{y}\sigma_{j+1}^{x})
\end{eqnarray}
Since $\langle \sigma_{j}^{y} \rangle_{0}=0$, it follows that $\langle [H,Q]\rangle_{0}=0$. Similarly, one can show that $\langle [H,Q^{2}]\rangle_{0}=0$. Consequently, the linear term in the expansion Eq. \eqref{Cvt expansion} vanishes.
The quadratic term in the expansion is determined by the second derivative of the charge variance: 

\begin{eqnarray}
\left.\frac{d^{2}\sigma_{Q}^{2}}{dt^{2}} \right|_{0} = -\langle [H,[H,Q^{2}]] \rangle_{0} + 2\langle Q \rangle_{0} \langle [H,[H,Q]]\rangle_{0} \label{all_2nd}
\end{eqnarray}
To compute $[H,[H,Q]]$, we evaluate:

\begin{eqnarray}
[H,[H,Q]] &=& [-\frac{1}{4}\sum_{i=1}^{L}(\sigma_{i}^{x}\sigma_{i+1}^{x}+\gamma\sigma_{i}^{y}\sigma_{i+1}^{y}+\Delta_{1}\sigma_{i}^{z}\sigma_{i+1}^{z}),\frac{i}{2}(1-\gamma)\sum_{j}(\sigma_{j}^{x}\sigma_{j+1}^{y}+\sigma_{j}^{y}\sigma_{j+1}^{x})] \\
&=& \sum_{j}\frac{1}{2}(1-\gamma)((1-\gamma)\sigma_{j}^{z}+\sigma_{j-1}^{x}\sigma_{j}^{z}\sigma_{j+1}^{x}-\gamma\sigma_{j-1}^{y}\sigma_{j}^{z}\sigma_{j+1}^{y}) \\
&-& \sum_{j}\frac{1}{4}\Delta_{1}(1-\gamma)(\sigma_{j-1}^{x}\sigma_{j}^{x}\sigma_{j+1}^{z}+\sigma_{j-1}^{z}\sigma_{j}^{x}\sigma_{j+1}^{x})
\end{eqnarray}
Using $\langle \sigma_{j}^{x} \rangle_{0}=\sin{\theta}$ and $\langle \sigma_{j}^{z} \rangle_{0}=\cos{\theta}$. We find:

\begin{eqnarray}
\langle[H,[H,Q]]\rangle_{0} = \frac{1}{2}(1-\gamma)((1-\gamma)L\cos{\theta}+L\cos{\theta}\sin^{2}{\theta})-\frac{1}{2}\Delta_{1}(1-\gamma)L\cos{\theta}\sin^{2}{\theta}
\end{eqnarray}
Next, we evaluate the term $[H,[H,Q^{2}]]$. Expanding this expression, we obtain:
\begin{eqnarray}
[H,[H,Q^{2}]] &=& [H,Q[H,Q]] + [H,[H,Q]Q] \\
              &=& Q[H,[H,Q]] + [H,[H,Q]]Q + 2[H,Q]^{2} \label{comu Qsq}
\end{eqnarray}
The expectation value of the third term is:

\begin{eqnarray}
\langle [H,Q]^{2} \rangle_{0} &=& -\frac{1}{4}(1-\gamma)^{2}\langle\sum_{i}(\sigma_{i}^{x}\sigma_{i+1}^{y}+\sigma_{i}^{y}\sigma_{i+1}^{x})\sum_{j}(\sigma_{j}^{x}\sigma_{j+1}^{y}+\sigma_{j}^{y}\sigma_{j+1}^{x})\rangle_{0} \\
              &=& -\frac{1}{2}(1-\gamma)^{2}L(1+\sin^{2}{\theta})
\end{eqnarray}
The first term, $\langle Q[H,[H,Q]] \rangle_{0}$, is computed as:

\begin{eqnarray}
\langle Q[H,[H,Q]] \rangle_{0} &=& \langle \frac{1}{2}(1-\gamma)^2\sum_{i,j}\sigma_{i}^{z}\sigma_{j}^{z}+\frac{1}{2}(1-\gamma)\sum_{i,j}\sigma_{i}^{z}\sigma_{j}^{z}(\sigma_{j-1}^{x}\sigma_{j+1}^{x}-\gamma\sigma_{j-1}^{y}\sigma_{j+1}^{y})\rangle_{0} \label{2nd CV} \\
&-& \langle \sum_{i,j}\frac{1}{4}\Delta_{1}(1-\gamma)(\sigma_{i}^{z}\sigma_{j-1}^{x}\sigma_{j}^{x}\sigma_{j+1}^{z}+\sigma_{i}^{z}\sigma_{j-1}^{z}\sigma_{j}^{x}\sigma_{j+1}^{x})\rangle_{0} \\
&=& \frac{1}{2}(1-\gamma)^{2}(L+L(L-1)\cos^{2}{\theta})+\frac{1}{2}(1-\gamma)L(L-3)\sin^{2}{\theta}\cos^{2}{\theta}+\frac{1}{2}(1-\gamma)L\sin^{2}{\theta} \\
&-& \frac{1}{2}\Delta_{1}(1-\gamma)(L(L-3)\sin^{2}{\theta}\cos^{2}{\theta}+L\sin^{2}{\theta})
\end{eqnarray}
Here, we have used $\langle Q^{2} \rangle_{0}=L+L(L-1)\cos^{2}{\theta}$. The second term in Eq. \eqref{2nd CV} has two contributions from $i=j$ and $i \neq j$, $j-1$, and $j+1$, while the third term vanishes due to $\langle\sigma_{j}^{y}\rangle_{0}=0$. Moreover, $\langle Q[H,[H,Q]] \rangle_{0}=\langle [H,[H,Q]]Q \rangle_{0}$.
Collecting all terms in Eq. \eqref{comu Qsq}, the first term in Eq. \eqref{all_2nd} becomes
\begin{eqnarray}
-\langle [H,[H,Q^{2}]]\rangle_{0} &=& -(1-\gamma)^{2}(L+L(L-1)\cos^{2}{\theta})-(1-\gamma)L(L-3)\sin^{2}{\theta}\cos^{2}{\theta}-(1-\gamma)L\sin^{2}{\theta} \\
&+& \Delta_{1}(1-\gamma)(L(L-3)\sin^{2}{\theta}\cos^{2}{\theta}+L\sin^{2}{\theta}) + (1-\gamma)^{2}L(1+\sin^{2}{\theta}) \\
&=& -(1-\gamma)^{2}(L^{2}\cos^{2}{\theta}-L)-(1-\gamma)(L^{2}-3L)\sin^{2}{\theta}\cos^{2}{\theta}-(1-\gamma)L\sin^{2}{\theta} \\
&+& \Delta_{1}(1-\gamma)(L^{2}-3L)\sin^{2}{\theta}\cos^{2}{\theta}+\Delta_{1}(1-\gamma)L\sin^{2}{\theta}
\end{eqnarray}
The second term in Eq. \eqref{all_2nd} is computed as:
\begin{eqnarray}
+ 2\langle Q \rangle_{0} \langle [H,[H,Q]]\rangle_{0} &=& L\cos{\theta} (1-\gamma)((1-\gamma)L\cos{\theta}+L\cos{\theta}\sin^{2}{\theta}) +2L\cos{\theta}(-\frac{1}{2}\Delta_{1}(1-\gamma)L\cos{\theta}\sin^{2}{\theta})\\
&=& L^{2}(1-\gamma)^{2}\cos^{2}{\theta}+L^{2}(1-\gamma)\cos^{2}{\theta}\sin^{2}{\theta}-\Delta_{1}(1-\gamma)L^{2}\cos^{2}{\theta}\sin^{2}{\theta}
\end{eqnarray}
Combining all terms, the second derivative of the charge variance is:
\begin{eqnarray}
\left.\frac{d^{2}\sigma_{Q}^{2}}{dt^{2}} \right|_{0} &=& -\langle [H,[H,Q^{2}]] \rangle_{0} + 2\langle Q \rangle_{0} \langle [H,[H,Q]]\rangle_{0} \\
&=& -(1-\gamma)^{2}(L^{2}\cos^{2}{\theta}-L))-(1-\gamma)(L^{2}-3L)\sin^{2}{\theta}\cos^{2}{\theta}-(1-\gamma)L\sin^{2}{\theta} \\
&+& \Delta_{1}(1-\gamma)(L^{2}-3L)\sin^{2}{\theta}\cos^{2}{\theta}+\Delta_{1}(1-\gamma)L\sin^{2}{\theta} \\
&+& L^{2}(1-\gamma)^{2}\cos^{2}{\theta}+L^{2}(1-\gamma)\cos^{2}{\theta}\sin^{2}{\theta}-\Delta_{1}(1-\gamma)L^{2}\cos^{2}{\theta}\sin^{2}{\theta} \\
&=& L(1-\gamma)^{2}+3L(1-\gamma)\sin^{2}{\theta}\cos^{2}{\theta}-L(1-\gamma)\sin^{2}{\theta} \\
&-& 3\Delta_{1}L(1-\gamma)\sin^{2}{\theta}\cos^{2}{\theta}+\Delta_{1}L(1-\gamma)\sin^{2}{\theta}
\end{eqnarray}
The early-time growth of the charge variance for different values of $\theta$ is then approximated by:
\begin{eqnarray}
   \sigma_{Q}^{2}(t)/L = \sin^{2}{\theta} + \frac{t^{2}}{2}(1-\gamma)\left(1-\gamma +(1-\Delta_{1}) (3\sin^2 \theta \cos^2\theta-\sin^2\theta\right)) \label{complete eq}
\end{eqnarray}
Here, the initial charge variance, $\sigma_{Q}^{2}(0)$, is replaced with $L(1-\cos^{2}{\theta})$. We can now compare this analytical result with numerical simulations at early times. We find that the charge variance obtained from both methods is in good agreement at early times, as confirmed in Fig. \ref{fig:CVanly_numer}. 

